# Roman Galactic Plane Survey Definition Committee Report

Submitted: Aug 18, 2025
Last Revised: Oct 1, 2025

**Committee Members:**
  Robert Benjamin, University of Wisconsin-Whitewater (ret) [Co-chair]
  Rachel Street, Las Cumbres Observatory [Co-chair]
  Rachael Beaton, Space Telescope Science Institute
  Sean Carey,  Caltech-IPAC
  Kishalay De, Columbia University/Flatiron Institute
  Matthew De Furio, University of Texas at Austin
  Janet Drew, University College London
  Thomas Kupfer, Universität Hamburg
  Dante Minniti, Universidad Andrés Bello
  Roberta Paladini, Caltech-IPAC
  Eddie Schlafly, Space Telescope Science Institute
  Catherine Zucker, CfA Harvard & Smithsonian

**Roman Liaisons to Committee:**
  Andrea Bellini, Space Telescope Science Institute
  Dario Fadda, Space Telescope Science Institute

**Contributors:**
  Numerous colleagues have contributed to this report, as will be noted throughout the document, but the committee would like to specifically highlight three groups of contributors:

  **(1)** the authors of white papers and science pitches described in this report[1],
  **(2)** the participants and Scientific Organizing Committee of the *Roman Galactic Plane Survey Community Workshop*[2] held online on Feb 11-13, 2025, and
  **(3)** the participants, Local Organizing Committee, and  Scientific Organizing Committee of the *Galactic Science with the Nancy Grace Roman Space Telescope* workshop held at Yerkes Observatory (Williams Bay, Wisconsin) on June 13-15, 2024.[3,4]

Copies of contributed presentations, manuscripts, and minutes of numerous discussion groups may be found in the links below. The Definition Committee was extremely grateful for the interest and support of our colleagues and for their many contributions to this survey, and it is our hope that the program presented here will provide data of value for science programs for many decades to come.

---

[1] https://asd.gsfc.nasa.gov/roman/white_papers/,  https://asd.gsfc.nasa.gov/roman/sci_pitch/
[2] https://outerspace.stsci.edu/display/GPSCW/Galactic+Plane+Survey+Community+Workshop+Home
[3] https://sites.google.com/cfa.harvard.edu/romanyerkes/home
[4] https://tinyurl.com/roman-yerkes-discussions

# Executive Summary

The Roman Galactic Plane Survey (RGPS) is a 700-hour program approved for early definition as a community-designed General Astrophysics Survey. It was selected following a proposal call for science programs that would benefit from an early community-based definition ([Sanderson et al 2024](#)). The community was invited to submit white papers and science pitches with a deadline of May 20, 2024; the Roman Galactic Plane Survey Definition Committee (RGPS-DC) first met on Sep 11, 2024. After four months of discussion and reviewing community submissions, a science organizing committee comprised of half RGPS-DC members and half community members organized an on-line workshop held on Feb 11-13, 2025 to inform the community of the status of RGPS planning and solicit feedback and discussion on key points where the Definition Committee (DC) decided that further community input was needed. The RGPS-DC continued to solicit and review science pitches through early March 2025.

Based on the input provided, the RGPS-DC recommends a survey consisting of three elements encapsulated in the figure below: **(1)** a *wide-field science* element (691 deg$^2$, 541 hrs) covering the Galactic plane—Galactic latitude $|b|<2º$ and Galactic longitude $l$=+50º.1 to 281º—in four filters (F129, F159, F184, and F213) with higher latitude extensions for the bulge, the Serpens South/W40 star formation region, and Carina, **(2)** a *time-domain science* element (19 deg$^2$, 130 hrs) of six fields—including the full Nuclear Stellar Disk (NSD) and Central Molecular Zone (CMZ)—with coverage in seven filters and repeat observations in one or more filters with cadences from ~11 minutes to weeks, and **(3)** a *deep-field/spectroscopic science* element (4 deg$^2$, 30 hrs) consisting of fifteen Roman pointings—with a wide range of extinction, diffuse emission, stellar density and populations—using longer exposure times in seven filters in addition to grism and prism observations. A description of each element is given below.

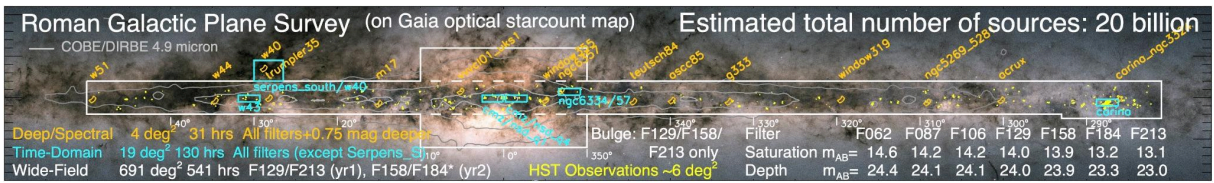

**Figure E.1** Overview of recommended Roman Galactic Plane Survey shown on an Gaia optical starcount map with grey contours for infrared light at 4.9 μm from COBE/DIRBE which show the extent of the stellar disk behind the optical extinction. See text for further detail.

*Wide-field science*: The majority of time allocated for the RGPS (541 hr/77% of the total time) is recommended for a 691 deg$^2$ Wide Field Imaging (WFI) survey of the Galactic plane, $|b|<2º$ and Galactic longitude $l$=+50º.1 to 281º, in four filters—F129 ("J"), F158 ("H"), F184 ("H/K"), and F213 ("K")—with ~sixty-second exposures with a gap-filling dither (LINEGAP2). The recommended wide-field survey also has three vertical extensions: *(a)* higher latitude coverage in the lower extinction region of the Galactic bulge/bar (2º$|b|<6º$) in the three filters F129, F158, and F213, *(b)* coverage of the nearby—$d$=440 pc—high-mass star formation region Serpens South/W40 with $l$=26º.5 to 30 deg and $b$=+2º to +4º.5 deg in five filters (F106 F129, F158, F184, and F213), and *(c)* coverage of the Carina "tangency" direction ($l$=293º to 281º deg) from $b$=−2.5º to +2º to to account for the warping of the Galactic disk in this direction. The single-exposure sensitivity limits range from mag$_{AB}$=23–24, with saturation limits of mag$_{AB}$=13-14. For regions with an extinction of $A_K$=1.4 ($A_V$~18)—the mean level of extinction in the inner Galactic plane—a red clump giant will be detected from 3 to 190 kpc, a solar type star out to 20 kpc, and an M0 dwarf to the distance of Galactic center.

The survey is expected to yield a catalog of approximately 20 billion sources and the highest angular resolution panoramic view of the Galaxy ever obtained. These data will provide unique opportunities for discovery in numerous sub-fields of Galactic astronomy.

The DC requests that the wide-field F129 and F213 filter coverage of the whole wide-field survey footprint be obtained as early as possible in the allotted two-year time frame, with the F106, F158, and F184 to be obtained as late as possible. This time spacing will allow for cross-band proper motion measurements with precision better than 0.5 mas/yr for the brightest sources, sufficient for population selections, cluster identification and characterization, and Galactic dynamics investigations. These data may also be combined with past optical/infrared programs, e.g. *HST*, *Gaia*, *2MASS*, *UKIDSS-GPS*, *VVV*, etc, or future Roman Space Telescope programs (plus JASIMINE and *Gaia NIR*) to allow for long-term proper motion measurements and variability of stellar and diffuse sources.

*Time-Domain Science:* Six subregions of the survey, totalling 19.1 deg$^2$ (130 hr/18.5% of total time), are recommended for all-filter coverage—F062 through F213, excluding the wide F146 filter—and time-domain investigations. These include two 2.1 deg$^2$ regions (six Roman pointings each) on either side of the Galactic Bulge Time-Domain Survey (GBTDS) Galactic Center field, spanning the full Nuclear Stellar Disk and Central Molecular Zone. These fields would be observed in F213 for eight hours at high cadence (11.3 min, 43 visits), an hourly cadence with F213 with increasing intervals (eight visits), and weekly subsequent monitoring (eight visits) in both F129 and F213. The high cadence strategy will be repeated for three regions of similar area containing—but extending beyond—three well-known star formation complexes: Carina ($l$=287º.5), NGC 6334/NGC6357 ($l$=352º.1), and W43 ($l$=30º.6). All but one of these regions will also be monitored by the Rubin Observatory at shorter wavelengths and longer cadences; the northernmost W43 field, which would be monitored in F184 rather than F213, will be covered as part of the Subaru Galactic Plane Survey. Hourly monitoring of the full Serpens South/W40 area with increasing cadence is also recommended. Taken together, these observations are expected to uncover populations of compact binaries, YSO variability, and other classes of variable sources in very different environments. The second year F129 and F213 observations may be combined with early wide-field coverage to constrain longer-term source variability.

*Deep-field/Spectroscopic science*: Fifteen single Roman pointings, totalling 4.2 deg$^2$ (31 hrs/4.4% of total time) are also recommended. These include fourteen deep—four times nominal exposure time—observations in all filters (except F146) as well as two 300 second observations at different roll angles using both the grism and prism. These fields span the full Galactic longitude range of the survey, with extinctions ranging from $A_K$=0.6 to 1.9 ($A_V$=7.6 to 24.3), expected source densities of 8 to 160 million sources/deg$^2$, and a wide range of expected diffuse emission. The data from these observations would provide higher S/N data to validate the wide-field data, would detect an additional ~40% of sources per field (where source crowding allows), would provide full spectral energy distribution information, and would provide data to test the use of the grism and prism in a range of Galactic environments. A fifteenth pointing towards W40 would go even deeper in F129, F158, and F213, and prism observations to probe the substellar content of this nearby star forming region.

# Contents

# 1. Science Motivations

The scientific potential of a Roman survey of the Galactic plane is vast, as evidenced by the breadth of topics covered by the many community white papers and science pitches. A single visit per pointing can yield a wealth of information on Galactic structure, extinction, molecular clouds and stellar characterizations, while proper motions measured from repeated visits will provide valuable insights into dynamical behavior. Time domain observations, over limited but carefully selected regions, will characterize the full range of stellar variability, from young stellar objects through to RR Lyrae and microlensing events caused by stellar remnants. This section provides an overview of the expected scientific return, drawing on contributions from the community that are summarized in Tables A.1 and A.2, including the four-character identifiers used to refer to each submission.

## 1.1 Galactic Structure

The Milky Way is not an unusual galaxy, being a reasonably massive (~L*) barred spiral. But being our own galaxy makes it unique for close-up study at fine-grained scales. The problem that the advanced near infrared (NIR) capability of Roman solves to a considerable extent is extinction – a clear brake on progress until recent times. The proposed wide field NIR survey will operate across the inner Milky Way, capturing (i) the bar (within ±30º in longitude), much of the bulge and the central region interior to the molecular ring , (ii) the disk across nearly all of quadrant 4, and the first, dustier, half of quadrant 1 (see Fig. 1.1).

A feature of the Milky Way is that stellar densities rise and rise as the Galactic Centre is approached along the disk. The ~5-7x increase in angular resolution of Roman, compared to previous NIR surveys from the ground—most notably 2MASS (Skrutskie et al 2006), VVV (Minniti et al 2010), and VVVx (Saito et al 2024)— will be critical for a less-confused view of dense environments such as those existing within the central molecular zone (CMZ, Fig. 1.2). Multi-color information and proper motion constraints from two data-taking epochs will support 3D mapping of regions ranging in angular size from individual star/globular clusters up to the full length of the bar. Ultimate goals will be the characterisation of stellar populations and star formation in the various structures present [WP04, SP17, SP36, SP46].

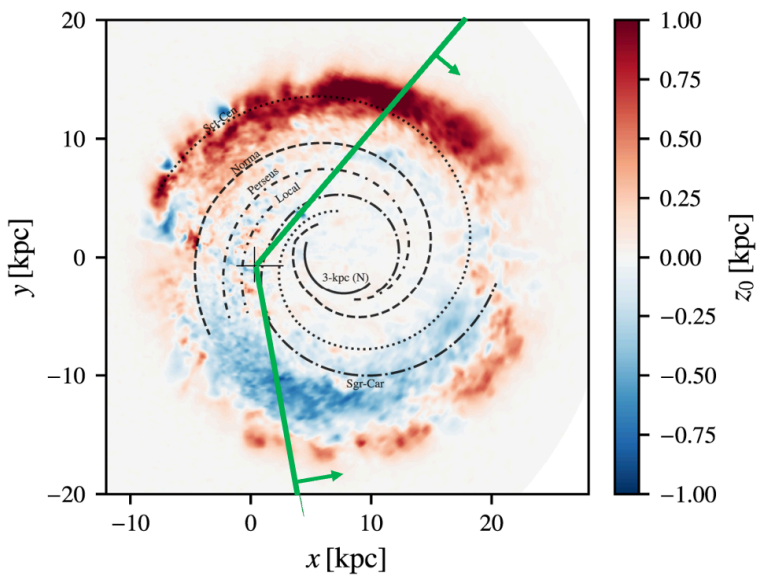

**Figure 1.1** (from Söding et al 2025.) The Galactic disk in face-on view, with the Reid et al 2019 spiral arms drawn in, and colored to identify warping of the plane as deduced from HI and CO gas indicators. The green lines roughly mark the longitude limits for the RGPS, with all the disk to the right included. The tangents of the Sagittarius-Carina Arm nearly coincide with the survey boundaries in longitude.

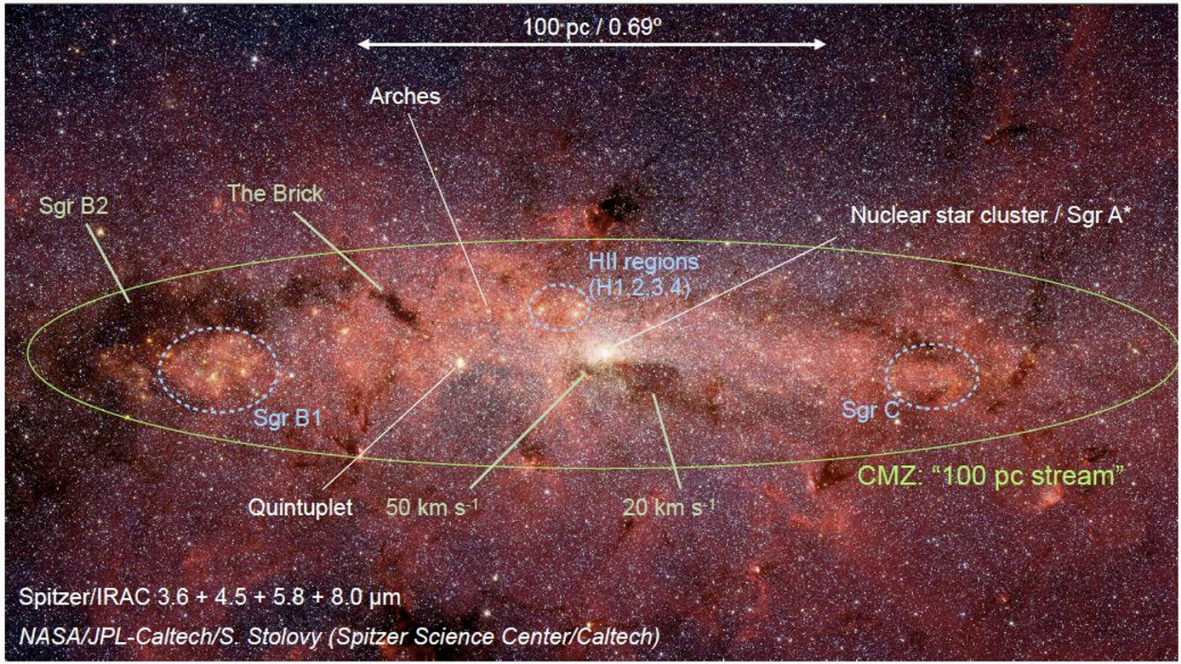

**Figure 1.2** *Spitzer* view of the CMZ with prominent features marked (From Schoedel et al 2023).

Describing and understanding the spiral arms of the Milky Way disk continues: the debate on how many and where they are is settling (see e.g. Reid et al 2019, and Fig. 1.1), but how they shape and are shaped by star formation is still debated. In effect, we do not have a clear idea of how they 'work'. Both the stellar dynamics and the evolution of the interstellar medium within arm/inter-arm ecosystems continue to be explored (e.g. Gaia Collaboration: Drimmel et al 2023, and Bieri et al 2023). A common community interest is in building further on the burgeoning census of star clusters across the disk [SP05, SP24, SP32]. The underlying attraction is that clusters can be aged, chemically-tagged and traced back, kinematically, providing insights into the history of the disk and its spiral structure.

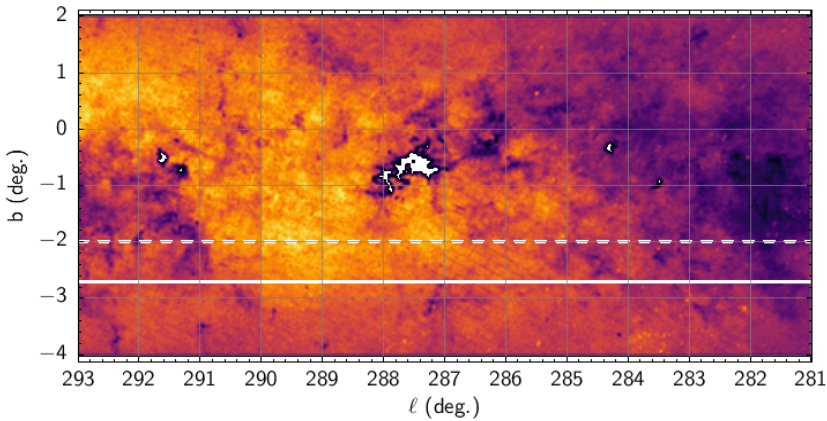

**Figure 1.3** Stellar density map of Carina using DECaPS-2 Y-band data (< 21 mag, Saydjari et al 2023). Density contrast between low (dark colour) and high (bright) is ~10. White patches arise from crowding in NGC 3603, the Carina Nebula, and Westerlund 2. The solid white line marks the lowered bound of the survey strip to better capture the downward-flexing disk.

We do not yet have a complete picture of how and where the stellar disk ends, while the disk behind the bulge/Galactic center remains very hazy. The great opportunity for Roman is that, even in the CMZ near the Galactic center, $A_K$ is commonly less than 3 mag (Gonzalez et al 2012). Roman will also probe small, but valuable, extinction holes [SP33, SP35] that permit multi-color detections of far-side or very distant structures. In part to ensure some access to the far disk beyond the Solar Circle, the proposed disk coverage is asymmetric, enabling the inclusion of Carina ([Fig. 1.3](#)). This region is also important because it brings in a grandstand view of an uninterrupted ~10 kpc section of a moderately obscured ($A_K \sim 1$ mag or less) spiral arm, noted for its intense star-forming activity [SP42] and brilliant nebulosity [SP09]. But is it coherent or broken up ([Drew et al 2021](#))? It has been a magnet for space-based studies, with many Chandra, HST and JWST programmes targeting especially the Carina Nebula (d=2.3 kpc). A possible outer-Galaxy extension of the Perseus Arm behind Carina is well within reach, and we may finally pick up the down-warp of the disk in stars in Carina/Centaurus – known so far only in gas ([Levine, Blitz & Heiles 2006](#), [Soeding et al 2025](#), [Fig. 1.1](#)). At the distance of the Galactic center, Roman will detect main sequence stars down to mid K dwarfs (K5V) even in directions where the extinction is high $A_K$ <2.0, and down to M dwarfs (M0V) in directions where the extinction is lower, $A_K$<1.4. O stars ($M_K$ < -3) will be easy Roman detections to the far side of the disk, and Roman will greatly improve the ability to resolve them in the luminous star-forming clusters known to exist across the Galaxy at distances of d~20 kpc or more ([Chené et al 2021](#)), revealing, for example, the star-forming content of the Scutum-Centaurus Arm extension discovered in CO by [Dame & Thaddeus (2011)](#).

## 1.2 Galactic Dynamics

Roman will obtain proper motions for billions of stars at near-infrared wavelengths. A majority of these stars are inaccessible to Gaia and LSST at optical wavelengths in extinguished or crowded fields. Therefore, these new proper motion maps in the near-infrared will open a powerful window for understanding the dynamics of the Milky Way towards regions of high stellar density and those heavily obscured by dust, including the Galactic bulge, bar, and beyond the Galactic center. While a subset of these regions have previously been targeted astrometrically at near-infrared wavelengths with the VVV survey, Roman is expected to significantly scale up the number of proper motion detections for fainter stars and will be sensitive to Galactic rotation across large swaths of the Galaxy's far side.

With a targeted detection threshold for proper motions of 1 mas/yr, Roman is primed to explore fundamental questions relating to the complex orbital structures of the Galactic bar and nucleus. These observations will offer new insights into the Galaxy's inside-out formation history and the distribution of luminous and dark matter in its central regions [SP31, WP04]. Simultaneously, Roman will also provide constraints on the accretion history of our Galaxy's supermassive black hole by detecting a larger sample of so-called hypervelocity-velocity stars in crowded regions that have been ejected via dynamical interactions with SgrA* [SP17]. Within the disk, Roman will constrain models for radial migration, as it will be sensitive to the relatively unexplored population of metal-rich RR Lyrae stars at low latitudes towards the time domain fields [SP13]. By targeting younger stellar populations, Roman will also characterize non-axisymmetric features, e.g. by better characterizing the processes under which spiral arms form, wind-up, and dissipate over time.

More generally, Roman's proper motions will be able to dynamically distinguish between stellar populations associated with the bulge from those associated with the disk, which is critical for a variety of downstream studies, including constraints on the star formation history of the Milky Way [SP36] and on chemical enrichment and mixing processes. When combined with complementary photometric, astrometric, and spectroscopic observations from other ground- and space-based surveys, Roman's proper-motion-based constraints on Galactic dynamics will provide further insights into the evolution of all the Milky Way's key structural components, including the bulge, bar, nucleus, central supermassive black hole, and spiral structure.

## 1.3 Star Formation and the Interstellar Medium

Star formation is a complex process that arises from the interplay of gravity, turbulence and magnetic fields. The respective role of each of these competing elements is still unclear, as is the dependence from external factors (such as feedback and metallicity). Because of this complexity, studying star formation requires exploring a large parameter space, which can be achieved by observing statistically significant samples of objects in many different environments. Probing different stages of the formation process (Fig. 1.4) is also necessary to understand how the balance between competing forces may change over time.

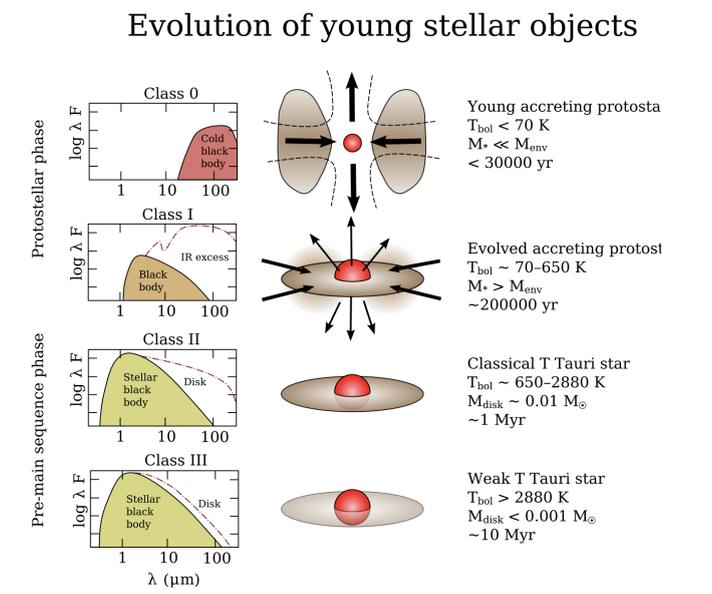

**Figure 1.4** The RGPS filter combination (F129, F158, F184 and F213) will allow placing important data points on the Spectral Energy Distribution (SED) of Class I, II and III YSOs, probing simultaneously the stellar/IR-excess/disk SED components.

The RGPS will cover ~35% of the total surface area of the Galactic plane, and almost the entirety of the inner plane, where the bulk of Galactic star formation occurs [SP21]. This spatial coverage, coupled with high sensitivity and a careful filter selection, will provide access to both obscured and unobscured star formation, and enable the unbiased detection of an unprecedented number (~$10^6$) of pre-main sequence Young Stellar Objects (YSOs) - especially Class I and Class II sources - as well as of young main sequence stars - Class III - [WP08]. In addition, the proper motion measurements, derived from the RGPS multi-epoch observational strategy, will allow membership confirmation for thousands of already known - and relatively close, i.e., 2-3 kpc - young stellar clusters and, even more importantly, to identify new, distant clusters [SP24, WP09].

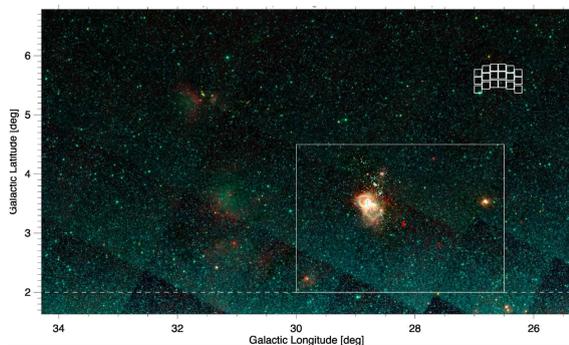

**Figure 1.5** The W40/Serpens South star-forming region: the rectangular area identified by the white solid line denotes the boundaries of the RGPS observations. The bright extended complex on the left is W40, while the smaller region on the right is the HII region Sh 2-62.

Importantly, the measured proper motions of YSOs - typically of the order of ~0.05 arcsec/year for YSOs at ~500 pc [SP25] - will also serve the purpose of investigating potential triggered star formation associated with expanding HII regions, which will be uniquely identified in the F184 band through Paschen alpha emission at 1.84 micron [SP25, SP46, WP09].

As well as encompassing a large continuous area of the Galactic plane, the RGPS will also include targeted observations towards the nearby (d~440 pc) W40/Serpens South star-forming region ([Guttermuth et al 2008](#)) in the Aquila Rift large extinction feature (see [Fig. 1.5](#)). In particular, the Roman observations will encompass an area of roughly 9 deg$^2$ which includes the W40 young star cluster associated with the eponymous HII region, also known as Sharpless 2-64 (e.g., [Smith et al. 1985](#); [Kuhn et al. 2010](#); [Pirogov et al. 2013](#)), and the less prominent HII region Sharpless 2-62 ([Smith et al. 1985](#)). To facilitate YSO detection and characterization, the RGPS observations will be executed using a larger suite of filters compared to the wide-area survey, that is F106, F129, F158, F184 and F213.

The close proximity of this star formation complex will give us the opportunity to take full advantage of both Roman's angular resolution and sensitivity, by resolving cluster members down to ~ 40 au with sensitivity down to Jupiter masses in the background limit [SP19] and allowing us to obtain a complete census of YSOs across all mass regimes. This in turn will enable the accurate measurement of the Galactic star formation rate ([Megeath et al. 2022](#)) and a comparison with extra-galactic estimates which are biased towards high-mass star formation [WP08].

Complementary WFI/prism observations will be performed towards the W40/Serpens South region. These spectroscopy data will provide an invaluable contribution to aid the determination of spectral types of young stars and sub-stellar objects into the planetary mass regime, e.g., from the water and $H_2$ features in the 1 - 2 micron range [WP08, WP09]. In addition, the Paschen beta 1.28 micron line, the [Fe II] 1.257 and 1.644 micron lines, and the H2 (1-0) 2.12 micron line will powerfully probe YSO mass accretion, as signaled by jets and outflows [SP011, SP37].

## 1.4 Low-mass Stars, Brown Dwarfs, and Free-floating Planetary Mass Objects

Roman's WFI is uniquely suited to pursue stellar and sub-stellar science due to its photometric sensitivity, diffraction-limited resolution, spectroscopic modes, and field of view. Recent space-based missions have attempted to characterize the very low-mass end of the initial mass function and its shape in star forming regions by searching for free-floating planetary mass objects down to the opacity limit of fragmentation. While these have successfully identified sources down to 2-3 $M_J$ ([Luhman et al 2024](#), [De Furio et al 2025](#)), statistical analyses exploring a fundamental question of turbulent fragmentation are difficult due to the significant amount of time required to perform mosaic imaging of dozens of fields and obtain follow-up spectroscopy.

With Roman's photometric sensitivity, we can efficiently explore star-forming regions down to the theoretical limit of turbulent fragmentation, 1-10 $M_J$ ([Boyd & Whitworth 2005](#); [Whitworth et al 2024](#)), using deep integrations in nearby regions. Specific observations in W40, a young (~1 Myr) nearby (440 pc) star-forming region, will obtain photometry and spectroscopy for over 1000 cluster members which will be used to both confirm membership and estimate masses of members from spectral types [WP08, WP09, SP16, SP19]. These observations will allow for a robust characterization of the mass function and will serve as a benchmark for future studies in other star-forming environments. More generally, shallower integrations throughout the entire Galactic plane will also allow for an exploration of the stellar and sub-stellar content and characterization of the mass function within many different star-forming regions [WP04, SP19, SP24]. Using the F129,

F158, F184, and F213 filters, these observations will cover near infrared bands where low-mass stars, brown dwarfs, and free floating planets have both significant photospheric emission and atmospheric absorption. The F184 filter covers a broad water absorption feature which can be used to identify low temperature objects and estimate masses in regions with known ages. By understanding these populations, these observations will probe the impact of environment, such as high-mass star UV radiation and stellar density, on the turbulent fragmentation process and the mass function itself. The results will be important inputs to giant planet formation and ejection hypotheses. These observations are crucial examples of the power of Roman, extending the legacy of previous space-based telescopes such as Spitzer, Hubble, and JWST.

Field-aged brown dwarfs are the coldest directly imaged objects outside of the solar system (down to 250K). They are numerous (1 of every 5 stars $< 1\ M_o$), and represent the low mass end of the star formation process. Multiple surveys have characterized the nearby brown dwarf population (Kirkpatrick et al 2024, Best et al 2024) across the MLTY-dwarf sequence, but mostly depend on older, low angular resolution telescopes (e.g. WISE). Given the footprint of the RGPS, brown dwarfs will feature prominently as sources within the observations [WP09]. Brown dwarfs of varying metallicities will also be identified which give insights to the mass function during early stages of the Galaxy [SP03].

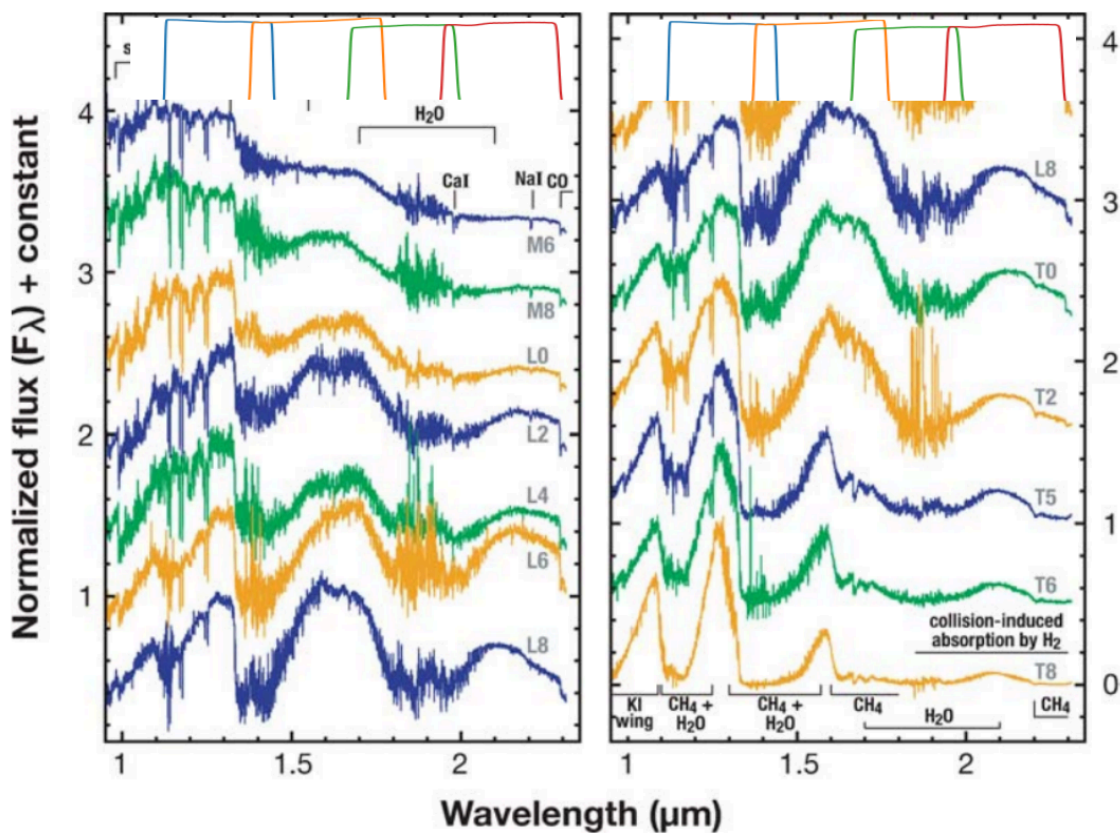

**Figure 1.6** Adapted from Kirkpatrick 2005. Demonstration of the MLT-dwarf sequence with the four Roman filters (F129, F158, F184, and F213) overplotted. F184 overlaps significantly with a strong absorption feature, useful in differentiating between low temperature ($T_{eff} < 3000$ K) objects and reddened background stars.

However, distinguishing between reddened, low-mass stars and unreddened brown dwarfs is difficult with only photometry. The utility of the F184 filter is in part that it mostly covers a water absorption feature in the atmospheres of these cold objects. We can use this filter, in conjunction with the JHK-equivalent filters, to identify candidate low temperature objects throughout the Galactic plane. In Fig. 1.6, we show the evolution of the water band through the mid-late M-dwarf sequence to late T-dwarf sequence, overplotted by the four proposed Roman filters. These data will allow for the identification of candidate brown dwarfs well suited for spectroscopic follow-up and future study which is a frequent science case for JWST and the ELTs. Due to the angular resolution of Roman, we will necessarily identify new brown dwarfs that were unresolved in previous all-sky surveys.

## 1.5 Dust and the Extinction Curve

The RGPS observations are focused on low Galactic latitudes towards the inner Galaxy, the most extinguished region of the sky. Typical extinctions for $|b| < 1°$ and $|l| < 10°$ range from 1 to 2.5 mag in F213 (§C.1). Roman's combination of infrared sensitivity, large field of view, and high spatial resolution will result in detecting many more stars than any other survey in this region (§C.2), allowing for detailed maps of dust extinction. Additionally, the RGPS deep fields will feature slitless spectroscopy from which stellar parameters for relatively bright stars can be robustly extracted, providing access to the dust extinction curve. The community showed extensive interest in studying dust with Roman [WP06, SP04, SP10, SP22, SP23, SP33, SP35, SP39, SP43], using a variety of different techniques.

The RGPS imaging will enable high-resolution, maps of dust reddening and the dust extinction curve via modeling of color-color and color-magnitude diagrams of the stellar populations observed by Roman, e.g., Nataf et al. 2013 and [SP04, SP23, SP33, SP35, SP43]. Three-dimensional dust maps often estimate the reddening and distance to individual stars. Based on expectations from shallower investigations with YJHK photometry, we expect Roman to be able to measure the distance and reddening for billions of stars, inferring detailed distance-reddening posteriors for each (Zucker, Saydjari, & Speagle et al. 2025; see Fig. 1.7). Roman photometry can also be supplemented with Rubin photometry for sufficiently bright, blue stars to improve the accuracy of the inferred stellar parameters. This will lead to significant improvements over Roman-only 3D dust maps in regions of modest extinction outside the inner Galaxy.

Spectroscopy in the RGPS Deep fields will provide an additional means to measure the dust extinction curve [SP39, SP43]. Stellar parameters derived from continuum-normalized spectroscopy are insensitive to dust extinction and can be used to robustly predict the intrinsic colors of stars. Comparison between the observed colors and the intrinsic colors leads to measurements of the dust extinction curve (e.g., Schlafly et al. 2011). RGPS spectroscopy will enable these extinction curve measurements in the Roman filters over a larger range of distances and extinctions than previously possible. This approach is also much less sensitive to the challenges of modeling stellar populations than purely imaging-based techniques because the stellar metallicities and gravities can be directly measured, making it an important cross check on the purely measurements that we will be able to make over the entire RGPS footprint.

Dust is an important component of the interstellar medium, which several science pitches propose to better understand via Roman observations. Roman dust extinction mapping will lead to a better understanding of cosmic ray production in the Galaxy [WP06], star formation, and the interstellar radiation field [SP10]. Combination of RGPS observations with ULTIMATE-Subaru medium and narrow band observations will permit extremely detailed maps of the extinction curve and diagnostics of the ISM [SP22].

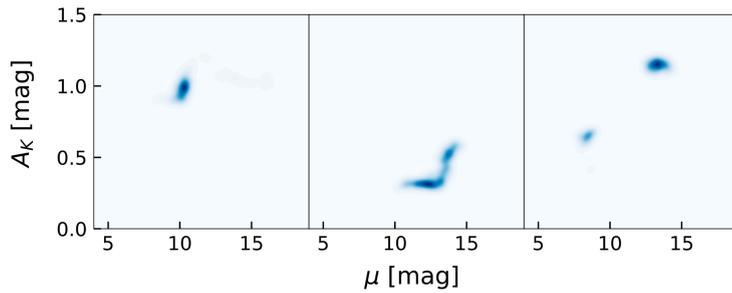

**Figure 1.7** Distance modulus and reddening estimates for stars with 4-band YJHK photometry from Zucker, Saydjari, & Speagle et al. (2025), roughly analogous to the RGPS. Distance and reddening accuracy varies significantly from star to star, from ~10% in distance and ~0.05 mag in $A_K$, to factor-of-two uncertainties, depending on stellar type.

## 1.6 Microlensing

The phenomenon of microlensing offers a unique opportunity to explore populations that are normally too faint to explore, from distant and even free-floating exoplanets to isolated black holes. The gravity of a foreground object will lens the light from a luminous background source lying directly behind it. As their relative proper motions carry the observer, lens and source into and then out of alignment, microlensing can be detected as a time-dependent magnification of the light from the source, and as a displacement of its apparent astrometric photocenter (Paczyński 1986). Since this phenomenon is sensitive to *any* massive object, and does not require light from the lens to confirm a detection, it is a unique tool for detecting distant exoplanets. These lenses include exoplanets, whether bound to host stars or free-floating (FFPs) [SP14, SP26], as well as brown dwarfs, stars of all masses, and evolved objects like white dwarfs, neutron stars, and black holes [WP07, SP09, SP18]. Events range widely in timescale from those for FFPs (~1 day) to compact objects (hundreds of days).

The precise alignment required for a lensing event means that the phenomenon is inherently rare, and most commonly detected in densely populated regions. Roman's high spatial resolution, wide field of view, fast survey capabilities, and infrared sensitivity makes it the ideal survey instrument to deliver the time series photometry and astrometry that can detect and characterize these transient events. This is exemplified by the GBTDS, but several community white papers and science pitches highlighted a number of important ways that the RGPS can extend and complement the science goals of the GBTDS.

Microlensing events are typically identified from time series photometry, but the models suffer several degeneracies and additional data are required to unambiguously determine the mass and distance of the lens. Several community papers highlighted how Roman can deliver these constraints with deep imaging over just a single epoch or few epochs spaced over weeks or months. For example Roman imaging can deliver constraints for thousands of previous events by detecting light from now-separated lenses [SP14], or by identifying the source star in contemporary events, enabled by Roman's high spatial resolution [WP07]. Several authors emphasized synergies between Roman observations and past and future surveys including the VVV, Optical Gravitational Lensing Experiment (OGLE), the Rubin Observatory's Legacy Survey of Space and Time (LSST) and more. Potential science included confirming dozens of candidate black hole events from VVV [SP26] to deblending and characterizing source stars in crowded LSST images across the Galactic plane [WP07]. The value of acquiring high-cadence (~15min) time sampling of selected fields such as stellar clusters was noted [SP26], since this would allow FFP events to be detected.

## 1.7 Star Clusters

Star clusters serve as important anchor points for many astrophysical studies of the Milky Way, providing insights about star formation that benefit all areas of astronomy [SP05][WP07]. They are ideal laboratories containing stars formed at about the same time, from the same parent cloud, with homogeneous chemical composition (with a small spread due to multiple generations of stars). Member stars also evolved in the same environment as they orbited the Galaxy together, lie at the same distance from the sun, and in general contain little dust within the cluster. It is difficult to summarize the revolution that the RGPS will cause in star cluster research; a couple of magnitudes in the distance modulus make an enormous difference for the exploration of the inner Galaxy.

The observed distribution of star clusters shows that the existing samples are very incomplete, especially in Galactic quadrants I and IV [SP49, SP51]. Many clusters remain undetected, as Gaia becomes inefficient at heliocentric distances of a few kiloparsecs, and totally blind beyond the Galactic center. Therefore there are many clusters still to be discovered, especially within the disk and bulge area covered by the RGPS [SP05, SP24, SP32, SP49]. The RGPS depth and resolution will allow one to distinguish clusters from the dense field population, and measure physical parameters like reddening, distance, age, luminosity, size, structures, and mass. For individual clusters, particularly for massive clusters in the inner Galaxy, the study of the low and high ends of the initial mass function will be possible in greater detail than ever before [SP05].

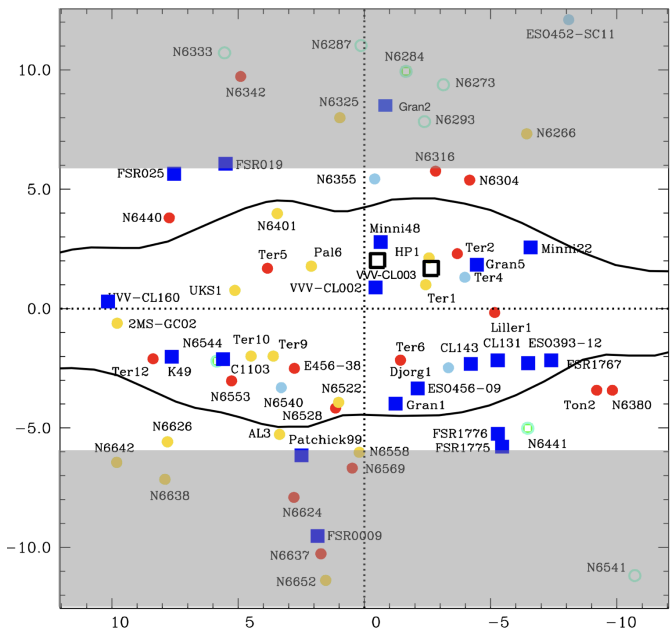

**Figure 1.8** (adapted from Bica et al. 2024, [SP49]): Known globular clusters in the central bulge of the Milky Way, including a couple dozen recently discovered ones. About 10-20 more clusters in this region may still be missing using symmetry arguments, a huge discovery potential for the RGPS. Clusters within the grey area would fall outside of the RGPS footprint.

Globular clusters are critical for the study of the early assembly of the Milky Way because the globular clusters are the oldest known objects in our Galaxy, and constrain the history of the Milky Way. There is a limited sample of Galactic globular clusters known; every new globular cluster discovered is important as they are powerful tools for studying the age of the Galaxy, Galactic chemical evolution, Galactic structure, the distance scale, stellar evolution, the interstellar medium, Galactic dynamics, and collisional systems. Because the RGPS photometry will reach well below the main sequence turn-off (located at K ~ 17 mag at the distance of the Galactic center, beyond the reach of current surveys), data from the RGPS can be used to determine cluster ages homogeneously, even for the innermost reddened reddened globular clusters (Fig. 1.8) [SP49, SP51, WP04].

Many of the innermost globular clusters do not have measured ages; ages that do exist are inhomogeneously obtained. For some globular clusters, it will be possible to measure age ranges to explore multiple stellar populations, as has previously been done in less contaminated globular clusters outside the plane ([Wagner-Kaiser et al 2017](#); [Olivera et al 2020](#)). The RGPS will also enable cluster membership to be confirmed via common proper motions. In addition, proper motions may be combined with radial velocities to yield cluster orbital parameters. These orbital parameters would connect the clusters to different components of the Galaxy (bulge, halo, thin disk, thick disk) or past accretion events. All this will be possible in much more reddened regions and out to much larger distances than have been done to date ([Bica et al 2024](#); [Garro et al 2024](#)). Finally, RGPS time domain observations will provide RR Lyrae detections that can be used to provide accurate distances [WP07] [SP13].

The Milky Way also contains an estimated 20,000 younger clusters, most of which remain unidentified ([Hunt & Reffert 2024](#); [Gupta et al 2024](#)). In the current Gaia era, the vast majority of known clusters are within about 3 kpc of the Sun ([Cantat-Gaudin et al 2018](#); [Hunt & Reffert et al 2023](#)). In the RGPS era, a new atlas of pre-main sequence clusters would map star formation across the Galaxy, illuminating the role of spiral structure and density waves in triggering star formation. With a precision better than 1 mas/yr, it will be possible to distinguish genuine cluster members with a common proper motion from projected members with high confidence. This will allow us to identify the majority of star clusters across the Milky Way, young and old, and build a catalogue of young stellar objects that is 10-100 times larger than present SED-based catalogues, e.g. SPICY ([Kuhn et al 2021](#)), cleaned of contaminants such as dusty asymptotic giant branch stars, and inclusive of pre-main sequence young stellar objects with little or no infrared excess [SP24]. This would enable one to trace sequentially triggered star formation across the different regions of the Galactic plane; deep RGPS multi-epoch imaging will help us understand how the motion of young stellar objects correlates with the expansion of their surrounding shells, probing sequential triggering theory and advancing our knowledge of star formation dynamics in molecular clouds [SP25]. It would also be possible to unveil the multiplicity of young massive stars to understand the early dynamical interactions that shape their brief childhood. This can be done throughout the Galactic plane in clusters, extending to larger distances than those where hundreds of massive stars have been detected using current X-rays and IR surveys [SP15] [SP19].

Finally, for star clusters studies, there is also an important synergy with other surveys like the LSST/Rubin Observatory, that will provide complementary multi-epoch and nearly simultaneous ugrizy optical photometry [WP07] in the region mapped by the RGPS, and the VISTA Variables in the Via Lactea survey that provides JHKs photometry as well as a longer baseline for variability and proper motions [SP16] [SP17] [SP49].

## 1.8 Eruptive Variables

With its wide field of view, mapping speed, spatial resolution and sensitivity, the RGPS will make a major contribution to Galactic time domain science. Not only is it similar in areal coverage to the largest archival Galactic surveys, the RGPS infrared sensitivity is well matched with that of the optical sensitivity of the Rubin observatory time domain surveys, giving it a substantially higher volumetric grasp for dust obscured eruptions. Even in areas that will not be monitored over several epochs, a multi-color Roman map of the Galactic plane would open new avenues for Galactic transient science—enabling the routine identification of the faint progenitors and remnants of transients. This is in addition to the time domain surveys planned to cover a subset of fields at multiple epochs to enable both variability and proper motion studies. Unlike the GBTDS which will cover a minuscule part (< 2 sq. deg.) of the bulge at extremely high cadence (~ 15 min), the RGPS will cover substantially larger areas at a range of cadences (minutes to months) albeit for a much shorter total duration. Three science opportunities that will open with these capabilities are **(1)** routine identification of faint stellar progenitors of eruptive stars, **(2)** resolved structures of Galactic transients in concurrent surveys and **(3)** the serendipitous discovery of a variety of eruptive phenomena during the planned repeated visits of fields in the same or different filters.

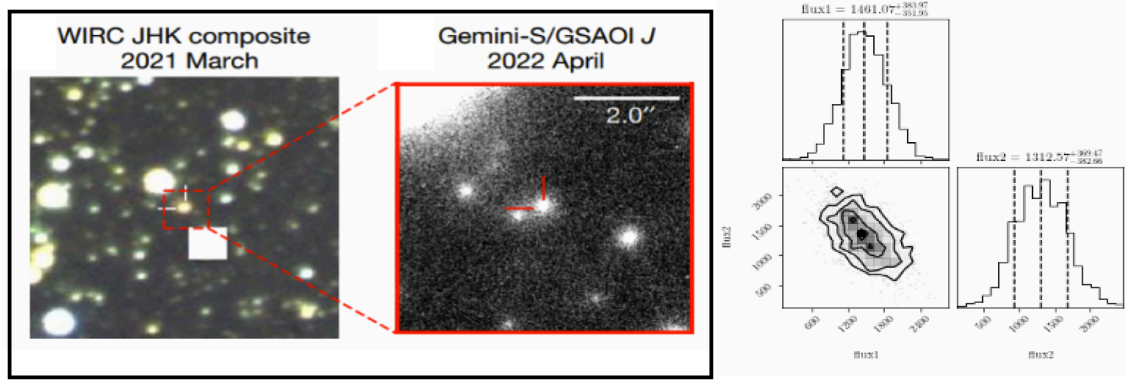

**Figure 1.9** [Left] Demonstration of joint modeling of ground-based seeing-limited imaging (from the Palomar 200-inch telescope) with high resolution post-outburst imaging (here from adaptive optics on Gemini-S) to pinpoint the progenitor (red cross-hair) fluxes (De et al. 2023) of the Galactic transient ZTF-SLRN-2020. [Right] Joint constraints on the fluxes of the blended sources in pre-outburst imaging using joint tractor (Lang et al. 2016) fitting of the blended pre-outburst sources with higher resolution post-outburst images.

First, owing to their wider field of view, ground-based surveys will continue to conduct higher cadence (days – weeks) time domain surveys in the optical bands. With their shallower sensitivity, they will routinely discover thousands of Galactic transients around their peak luminosity -- ranging from dwarf nova outbursts [SP08, SP22], X-ray binaries [SP27], novae [SP40, SP41] and young stars [SP24, SP48, WP08]. RGPS offers the unique capability to constrain the nature of their faint progenitor stars which are commonly inaccessible with ground-based imaging [SP08, SP22, SP27, SP40]. Combining higher resolution space-based imaging of ground-based discoveries will ubiquitously require de-blending techniques for accurate identification. Figure 1.9 demonstrates this method combining high resolution adaptive optics imaging with seeing-limited imaging to measure quiescent counterpart properties: it will be possible to routinely combine imaging from ground-based surveys such as the Rubin observatory and the RGPS.

Second, the spatial resolution of Roman will routinely reveal resolved structures (Fig 1.10) -- opening up synergies with transients discovered in contemporaneous ground-based searches. While previous similar efforts using HST have only been limited to targeted narrow-field observations at single epochs, the RGPS will serendipitously cover active outbursts as part of the WFI survey in at least one epoch, offering the capability to constrain the geometry of outflows [SP24,SP48,WP08] and structures left from past eruptions [SP41] in Galactic eruptive objects. Multiple epochs of imaging will enable time-resolved studies of both evolution in the outflow geometry as well as enable the use of reflections off nearby clouds to ascertain the distance and energetics of the eruptions. And a comparison of RGPS and archival HST will constrain the decades-long evolution of numerous jets and outbursts.

Third, the RGPS itself will be capable of serendipitously discovering outbursts during its planned wide-field and dedicated time-domain surveys. The IR sensitivity will particularly benefit low luminosity Galactic transients such as dwarf novae, helping address the long-standing dearth of cataclysmic variable population invoked to explain the Galactic ridge X-ray emission [SP08,SP22]. When combined with multi-wavelength information, the survey opens up avenues to complete census of outbursting X-ray binaries near the Galactic plane [SP27]. Similarly, the multiple epochs of observations over star forming regions will probe variability and outbursting characteristics of young stars over a range of mass scales [SP48], further benefiting from the long-time baseline of archival surveys such as the Vista Variables in the Via Lactea (VVV) [SP24].

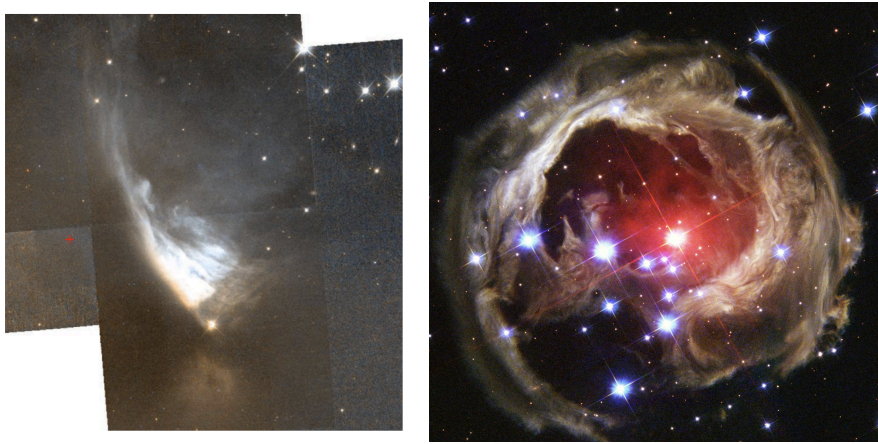

**Figure 1.10** (Left) Hubble Space Telescope image mosaic (taken from STScI/HLA) of the young stellar object PV Cep -- illuminating a large conical scattered light nebula. (Right) An expanding light echo around the stellar merger V838 Mon observed with the Hubble Space Telescope (Bond et al. 2007).

## 1.9 Pulsational Variables and Evolved Stars

Pulsational variables and evolved stars are broad stellar classes that have unique properties in the infrared and, thereby, can have an outsized scientific impact. Pulsational variables like Classical Cepheids, RR Lyrae, and Miras are powerful distance indicators in the Galaxy and beyond (e.g., Subramanian et al. 2017, Beaton et al. 2018). Likewise, evolved long-period variables (a class that includes Miras) are unusually bright in the infrared, and thereby can probe the largest volume of the Galaxy with even modest observations (e.g., Majewski et al. 2003, Auge et al. 2020, among others), but are complex and enigmatic (e.g., Nikolaev & Weinberg 2000, Soszynski et al. 2013, among others). Several white papers and science pitches either directly or indirectly connect to these topics.

There are over 110 unique categories of variable stars (Samus' et al 2017). The pulsational variables form the largest group of variable stars and the sub-classes are shown on a Hertzsprung-Russell diagram in Fig 1.12 (Pollard et al 2016). The RGPS observations will provide measurements of these types of stars across the Galaxy, with the yield depending on the region of the Galaxy being studied. The time-domain observations will have discovery potential for some classes of variables, whereas the large area observations will provide precise photometry that can be used in conjunction with detection and classifications obtained from other observational facilities [SP02, SP06, SP22, SP28, SP30, WP07].

Compact pulsators residing below the main sequence have typical pulsation periods <60min due to their compact nature. Most are driven by the κ mechanism in the partial ionization zones of iron (Charpinet et al. 1997). These consist of different classes including hot subdwarf pulsators (Kilkenny et al. 1997, Green et al. 2003) as well as white dwarf pulsators (ZZ Zeti stars; see Bell et al. 2025 for a recent review). Recently, a new class of radial mode pulsators has been discovered at low Galactic Latitudes; these are Blue Large Amplitude pulsators (BLAPs). They show radial mode pulsations and follow a period-luminosity relation as well as they have only been detected at low Galactic Latitudes suggesting that high metallicity is required to drive the pulsations (Pietrukowicz et al. 2025). The RGPS time-domain part is ideally suited to discover additional systems at much greater distance compared to previous surveys.

The "classical pulsators" are all radial pulsators, and the majority occur in the "classical instability strip" (see Fig 1.11). All components of the Milky Way contain some classical pulsators depending on the stellar populations in that part of the Galaxy. Accordingly, the RGPS will provide data on all types of variables in Fig 1.11 but the classical pulsators are particularly valuable for studies of Galactic structure [SP13, SP35, SP51].

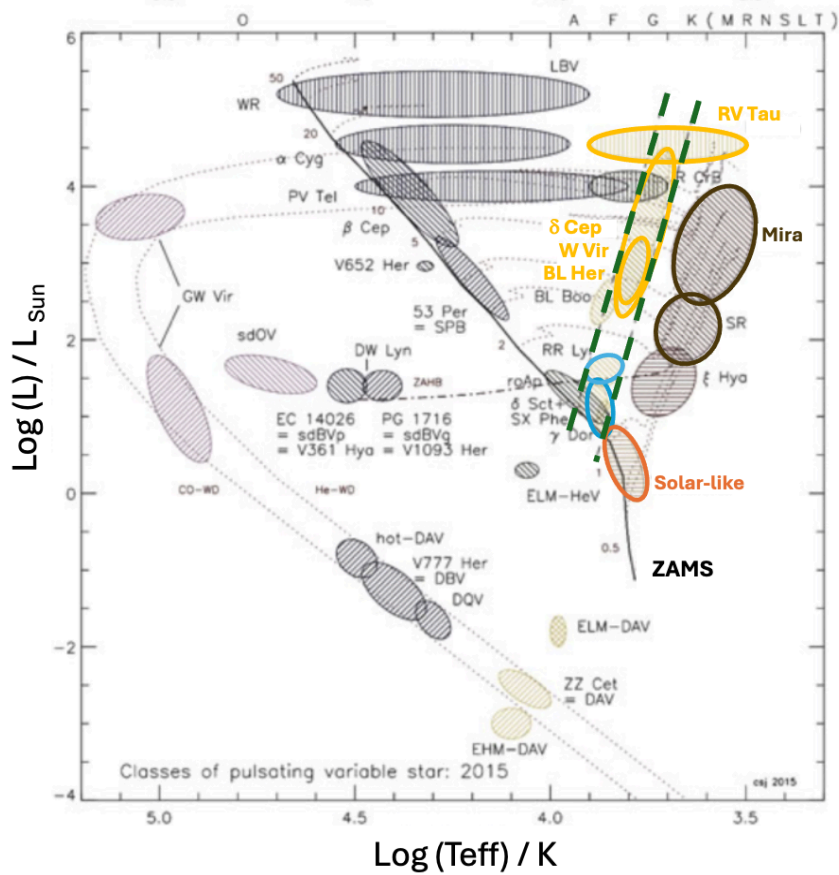

**Figure 1.11** Hertzsprung-Russell diagram showing the approximate locations of the classes of pulsating variable stars adopted from Pollard et al. (2016; their figure 1). The classes of pulsational variable stars discussed in relation to the RGPS are highlighted. The green dashed lines show the region known at the instability strip.

To use classical pulsators for distances, the pulsators need to have enough epochs of observation to reveal periods, and sub-typing is required to place them on the right relationships. Extensive studies on the optimal detection and period identification exist, generally indicating logarithmically spaced observations over the period range of interest. The RR Lyrae variables have periods between 0.5 to 1.5 days, the Type II Cepheid range is from ~2 days to ~60 days, the Classical Cepheids from ~6 days to 120 days, and the Miras from ~200 days to ~1200 days. Optimizing observations to discover and characterize classical pulsators over this broad range is out of scope for the constraints on the RGPS.

The light curve shapes for classical pulsators change with wavelength as is demonstrated for two RR Lyrae stars in Fig 1.12 (Monson et al. 2017). Infrared light curves are not optimal for identification and characterization of classical pulsators, especially where periods for different types of stars overlap (see Fig 1.11). However, these same properties have the positive impact of making the mean magnitudes more precisely derived with sparse data and, additionally, the period-luminosity relationships show overall smaller intrinsic scatter (Monson et al. 2017, Beaton et al. 2018). Together, this means that even single data points, coupled with light curve templates, can lead to extremely precise and accurate distances. As a result, any known RR Lyrae, say from Gaia, VVV, or LSST (among others) could be placed precisely within the Galaxy with "snapshot" style observations [SP02,

SP06, SP22, SP28, SP30, WP07]. Light curve templates can be derived with serendipitous observations in the time domain observations either in the RGPS or in the GBTDS. Thus, even a single Roman measurement can have significant value-add for using classical variables for galactic structure science.

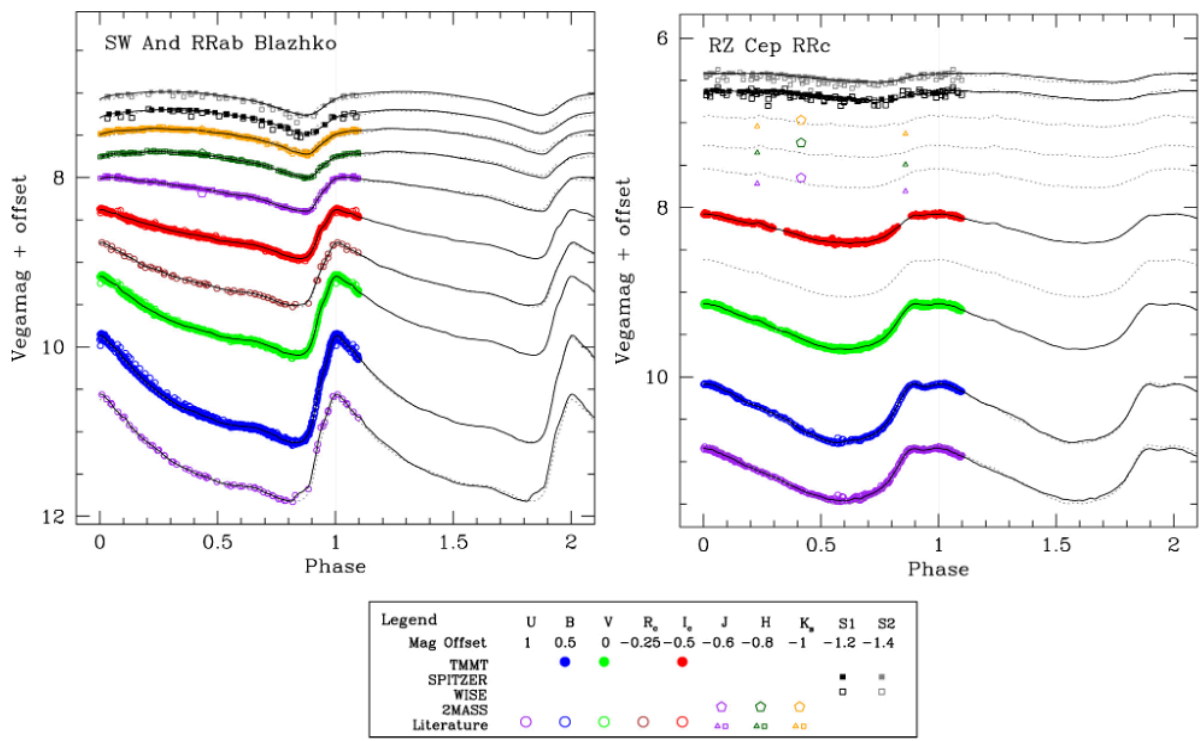

**Figure 1.12** Demonstration of the change in light curve amplitudes and shapes in multiple wavelengths for two RR Lyrae stars. The light curves become more sinusoidal and the amplitudes decrease dramatically; both characteristics lend to more precise mean magnitudes and distances, but with the consequence of being harder to identify and characterize at infrared wavelengths. Adapted from Monson et al. 2017.

## 1.10 Compact binaries

Compact binaries are a class of binary system with periods below several hours (detached or semi-detached) and physical separations between components as small as the Earth-Moon distance, consisting of a neutron star (NS)/white dwarf (WD) primary and a helium star/WD/NS secondary (Postnov & Yungelson 2006 for a review). The study of these short period systems and their subsequent mergers is important to our understanding of such diverse areas as supernova Ia progenitors, binary evolution and they are predicted to be the strongest gravitational wave sources in the *LISA* band (Amaro-Seoane et al. 2023). Compact binaries are believed to form in interacting binaries through various poorly-understood phases of mass transfer. The extreme tightness of the binaries calls for an enormous loss of energy and angular momentum during binary evolution as well as for physical mechanisms to avoid merging in earlier phases of evolution.

The largest uncertainties in predictions for different populations and derived merger rates, and hence SN Ia rates and expected numbers of gravitational wave sources detectable for *LISA*, is the treatment of the common envelope and the onset of mass transfer when the larger object fills its Roche Lobe (e.g. Nelemans et al. 2001, Marsh et al. 2004, Gokhale et al. 2007, Shen 2015). The population of short period binaries consists of several classes, which have been addressed in different white papers and science pitches. Those include accreting (semi-detached) and non-accreting (detached) systems:

Cataclysmic Variables (CVs) are accreting WDs with a main sequence or a red giant donor [SP08, SP40, SP41]. Meanwhile AMCVn type systems are accreting WDs with WD/He-star donors [WP03]. If the accretor is a NS, the system is called an X-ray binary as these systems are strong X-ray sources [SP27]. Among the non-accreting (detached) systems, the RGPS is expected to find several WD/NS + WD/He-star/NS binaries in tight orbits [WP03]. The known sample of compact binaries is still limited and biased due to different detection techniques and selection effects. Although the population is expected to reside mostly in the Galactic plane, previous studies have focused mostly at high Galactic latitudes, avoiding the Galactic plane due to crowding and extinction.

The RGPS will allow the pursuit of compact binaries via two main avenues: first, the identification of new sources in highly obscured low Galactic latitude regions based on proper motions and color selection; second, the identification of new sources based on periodic photometric variability in the dense time-domain fields. Roman's angular resolution is roughly an order of magnitude better than that of seeing-limited ground-based observations, leading to significantly less blending [WP03]. Color information from several filters will provide a way to identify compact binaries in dense stellar environments [SP08, SP40]. Additionally, for accreting systems, the IR bands in particular are useful for classifying the companions in nova systems, where the optical suffers from significant accretion disk flux. The IR is a strong tracer for the brightness of the companions, enabling classification efforts for the companion stars [SP08, WP03]. Of the ~10 detected nova eruptions in the Milky Way each year, approximately 50-70% show evidence for the formation of dust in the IR (especially the K band). Some nova remnants, especially those from recent eruptions in the last ~ 1 – 100 years, may still have a dusty shell of surviving dust formed during the eruption. Observations with the H/K and K bands (F184 and F213) from Roman could potentially constrain the amount of dust surviving in the nova shells, and in some cases even constrain the distribution of dust relative to the gas. Many of these recent nova shells are predicted to be ~0.5 – 2 arcsec in diameter [SP41], easily resolved by Roman.

Photometric light curves of compact binaries show variations on timescales of the orbital period, e.g., due to eclipses or tidal deformation of the components (e.g. Kupfer et al. 2022). Therefore, photometric time-domain surveys are well suited to identify compact binaries in a homogeneous way. The Roman telescope combines unprecedented angular resolution with photometric sensitivity unattainable from the ground, making it ideally suited to identifying compact binaries from their photometric variability. The time-domain part of the RGPS is expected to discover several new *LISA* gravitational wave sources, supernova Ia progenitors [WP03], cataclysmic variables [SP08] as well as compact X-ray binaries [SP27] in Galactic plane regions inaccessible for ground based surveys.

## 1.11 Synergies with other surveys

The Roman Space Telescope will not operate in isolation, and its observations will be highly complementary to those from other facilities, both from archival data and current and future surveys. A number of community contributions highlighted several surveys where RGPS observations can be used to maximize the scientific return of the combined programs.

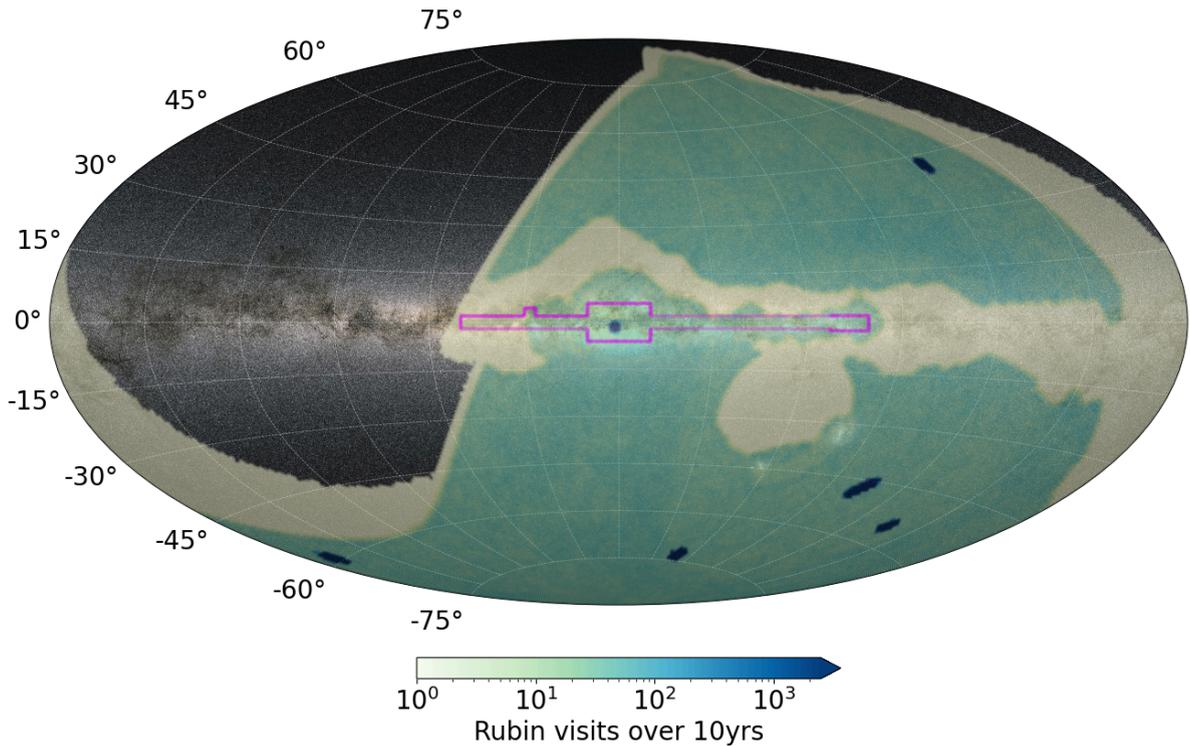

**Figure 1.13** Sky map comparing the regions surveyed by Rubin's LSST (green-blue color map) and the RGPS survey (purple outline). The shading scales in proportion to the number of visits Rubin is expected to make in all filters over 10yrs (based on survey simulation baseline_v4.3.1). The survey footprints are overlaid on an all-sky map of stellar density (in grey). The LSST footprint is highly complementary to that proposed for the RGPS.

While a number of surveys have made extensive observations in the Galactic bulge and plane—including OGLE, MOA, KMTNet, ZTF, Gaia, VVV, PRIME, and more— Roman will be able to probe to much deeper limiting magnitudes with the exquisite spatial resolution possible from space. This will not only dramatically increase the catalog of known stars and variables of all kinds, but also allow far more accurate deblending of objects detected in previous surveys. This in turn enables a significant improvement in the measurements of stellar parameters for objects studied by the other surveys [SP14].

The Roman Mission will operate concurrently with the Rubin Observatory's LSST [WP07]. This ground-based observatory, due to enter science operations in late 2025, will survey the entire Southern sky in 6 optical passbands (SDSS ugrizy) for 10 years (Figure 1.13). Rubin's large 8.4 m aperture and wide (9.6 sq. deg.) field of view make it one of the few facilities capable of probing to limiting magnitudes competitive with that of Roman in fields of low extinction. By combining Roman and Rubin observations we can extend our measurements of object SEDs from ~0.3 to 2.3 micron (F213), enhancing spectral classifications. At the same time, the long time baseline and range of cadences of LSST can complement the more intensive, but shorter duration, time series photometry from Roman. LSST includes a Galactic plane survey region which receives higher cadence than the surrounding "dusty plane" area,. The number of visits per field varies for different filters, with the low-cadence dusty plane receiving ~20-30 visits per year. This includes all filters, though most observations will use (g)rizy filters. Within Rubin's Galactic plane time domain region, the number of visits rises to ~85/year, predominantly in griz. In addition, a single field in the Galactic bulge will receive augmented cadence of up to ~250 visits/yr (again mostly in griz), starting from ~year 4 of LSST. As Figure 1.13 illustrates, the RGPS wide area survey forms a band across the center (in |*b*|) of the Rubin Galactic plane time domain survey region, covering virtually its whole extent in |*l*| but also extending out to *l*>30º in the dusty plane.

In the Galactic plane, the combined data will be highly valuable for characterizing the source stars in microlensing events. Cataclysmic variables benefit from high cadence Roman observations to measure their short orbital periods (~80 min) while Rubin's long baseline will identify outbursts. As another example, YSOs exhibit variability at all timescales and have complex SEDs due to surrounding accreting material and jets, so the complementarity of Roman and Rubin's wavelength coverage and cadence will enable these systems to be characterized in a range of star forming regions. Roman's high spatial resolution and precision astrometry will also be extremely valuable for deblending crowded fields, allowing accurate optical and NIR color measurements. This is particularly valuable for stars in heavily extinguished regions, such as the Galactic Center. Observations of RR Lyrae in the Galactic bulge, plane and Globular Clusters will provide unprecedented insights into the Milky Way structure and kinematics. In addition, Roman will help distinguish the cause of transients detected by LSST such as Tidal Disruption Events, Active Galactic Nuclei flares or supernovae.

The SPHEREx Mission, launched earlier this year, complements Roman by delivering spectroscopy at longer wavelengths (0.75 - 5 micron) with an R~35–130. Over its two-year mission, SPHEREx will produce four all-sky spectroscopic images. Roman observations of the Galactic plane will deliver higher spatial and wavelength resolution, complementing SPHEREx's planned survey studying the growth and evolution of interstellar ices from the diffuse ISM into stars and proto-planetary disks [SP06].

Spanning wavelengths even further into the NIR at 4.6 and 8.0 microns, the NEOSurveyor Mission will deliver time-series observations covering much of the Galactic plane. Its short cadence (hours to days) is well suited to characterizing several forms of stellar variability and short time-scale transients such as planetesimal collisions in moderately young stellar systems. Roman data will enable us to detect these sources in quiescent phases and measure their multiplicity. Observations by Roman of the region where the ecliptic plane intersects the Galactic bulge will help to characterize slow moving (i.e. distant) Solar System objects. Roman will provide high resolution imaging in complementary bandpasses from which a deep source catalog can be built, allowing faint moving objects to be identified. It will also aid the detection of any future interstellar objects like 1I/'Oumuamua, which are likely to approach the Solar System from the Galactic bulge direction due to the motion of the Solar System in the Galaxy and the locations of nearby stellar associations [SP30].

Roman will also complement X-ray and gamma-ray survey facilities such as eROSITA and Fermi-LAT, both of which have a significant number of unidentified sources. For example, these facilities have already provided a large sample of candidate Active Galactic Nuclei (AGN) with jet structures, but time-domain observations from Roman in the NIR will penetrate the extinction in the Galactic plane making it possible to extend this catalog to a previously inaccessible region [SP12].

The ULTIMATE-Subaru project (Ultra-wide Laser Tomographic Imager and MOS with AO for Transcendent Exploration) is developing a novel wide-field adaptive optics system for the 8.2m Subaru Telescope. It is anticipated that it will achieve a FWHM ~ 0.2 arcsec over a 14x14 arcmin field of view in K-band.  Once operational in ~2028, it will perform a Galactic plane survey spanning 60 sq. deg from l=20º-50º in narrowband filters such as Paschen-beta, CN, Brackett-gamma, Fe II and reach comparable limiting magnitudes to Roman.  Roman observations of the same region of the plane would provide highly complementary broadband filter coverage, and could be coordinated to be contemporaneous since both projects will operate at the same time.  Science drivers include the identification of Cataclysmic Variables from transmission in hydrogen recombination lines as well non-variable OH/IR stars [SP22].

It is also valuable to consider archival data from previous surveys in order to optimize the science return from Roman.  The Wide-field Infrared Survey Explorer (WISE) mission, and the Spitzer and Herschel Space Telescopes all made observations in the Galactic plane at longer wavelengths than Roman, though to more shallow limiting magnitudes.  Roman data of the same regions will enable these data to be reprocessed to deblend crowded sources and determine stellar multiplicity [SP28].

Finally, if approved, GaiaNIR will map the entire sky using astrometry, photometry and spectroscopy for radial velocities. The wavelength range will be 800–2300 nm and astrometry will mostly use the full wavelength range but also includes four color filters for photometry. The primary mirror will be at least 1.7 m (0.24" resolution); a larger segmented mirror up to 3.5 m (0.12" resolution) is being considered. (D. Hobbs 2025, priv comm). Thus, GaiaNIR—as well as the Japanese astrometric mission JASMINE ([Kawata et al 2024](Kawata et al 2024))—will be complementary to Roman in terms of resolution and wavelength range.  A combination of Roman and GaiaNIR could be obtained to get improved proper motions for stars not seen by Gaia but measured by two new NIR telescopes with comparable resolution with a time baseline of 30 years.

# 2. Committee Definition Process

This section describes the overall process by which the RGPS Design Committee arrived at the recommendations in this report, including the formation and charge to the committee itself, its review of community contributions and solicitation of further input, and its decision making process.

## 2.1 Charge to the Committee

The Committee was created by the Roman Science Coordination group led by Senior Project Scientist Dr. Julie McEnery and was charged with the following tasks.

- To review existing community science input into the science investigations that could be enabled with the Galactic Plane Survey;
- To prioritize the most compelling scientific investigations for driving the survey design;
- To analyze the impact of different observational strategies on the scientific return;
- To work with representatives from the Science Centers and Roman Project to assess the feasibility of various observational strategies;
- To create a detailed observational strategy and a description of its anticipated scientific yield, for consideration by the Roman Observations Time Allocation Committee;
- To define the final survey strategy at the level required for implementation by the Roman Science Centers;
- To recommend initial survey observations and/or evaluation metrics to assess whether the

survey implementation will meet expectations; and
- To engage openly and transparently with the science community.

The charge emphasized that in undertaking the above tasks, the Committee should maximize the scientific return and legacy value of the survey data within the allotted time allowance and other constraints such as visibility from the spacecraft. The Committee was also required to consider the potential synergies of the Roman survey with existing or concurrent surveys carried out with other observatories.

In the interests of transparency, the Committee was charged to engage the wider community throughout its deliberations and through a variety of mechanisms, such as participation in meetings such as those of the American Astronomical Society (AAS). The products of the Committee, including this report were required to be public and visible to the science community. Specifically, the Committee was charged to provide a written report to the ROTAC, summarizing their recommendations for the survey implementation and its anticipated science yield, by April 2025. This was extended upon request to August 2025.

The ROTAC is responsible for reviewing this resulting report together with recommendations from the other Core Community Surveys, and delivering their final recommendations for Roman Mission operations.

## 2.2 The Design Process and Community Engagement

Scientific input from the wider community formed the essential basis of the committee's work through its deliberations. The committee is greatly indebted to the many community contributors for their advice, recommendations and insights. Links to white papers, science pitches, and minutes of numerous discussion groups are provided on the first page of this document and a table of contributions is provided in an appendix. In this section, we outline the steps that were followed, from the inception of the Early Definition program to the creation of this report.

### 2.2.1 Request for Information

In 2021, the Roman Mission issued a Request for Information to the community, asking whether the mission should undertake an Early-Definition General Astrophysics Survey, and if so, requesting proposals for such a survey. Twenty proposals were received, with contributions from 340 authors, and were reviewed by the Early-Definition Astrophysics Survey Definition Committee[5]. This committee recommended (Sanderson et al. 2024) that a survey of the Galactic plane should be defined early.

### 2.2.2 White Papers and Science Pitches

In response to the Sanderson et al. (2024) report, the Roman project announced its plans for a Roman survey of the Galactic plane, and issued a call for community input with a deadline of May 20, 2024[6]. This call accepted feedback in the form of short (1-2 paragraph) science pitches as well as more extensive white papers, with the goal of lowering the barrier to contribute ideas by reducing the time commitment required to respond.

---
[5] https://roman.gsfc.nasa.gov/science/Early-definition_Astrophysics_Survey_Assessment.html
[6] https://roman.gsfc.nasa.gov/science/galactic_plane_survey_definition.html

In this initial call, nine white papers and thirty-two science pitches were received, spanning a wide range of astrophysics (summarized in the [Science Motivation](#) section). These contributions were reviewed by the RGPS Design Committee and motivated its preliminary survey designs. In many cases, this more detailed information was needed to understand the survey requirements for the science, e.g. specific areas to target, listing of preferred and minimal filter sets, etc. Co-chair R. Street worked with the contributors to clarify what was needed to complete the proposed science programs, (e.g. area to cover, filters needed) and compiled these results as described in §3.3. The committee is grateful to all of our colleagues who provided this feedback. The form for submitting white papers and science pitches was kept open after the formal deadline for contributions, and the community was encouraged to provide additional feedback.

### 2.2.3 Committee process

The committee met for the first time on Sep 11, 2024 together with several members of the Roman Science Coordination Group. Throughout the whole process, minutes and recordings of the meetings were made available to committee members on the RGPS Confluence page, and most communication was on a public channel (#galactic-plane-survey) in the Roman Slack workspace.

The following month was spent reviewing the contributed science pitches and white papers; the full committee read all contributions and one committee member provided a presentation to the group summarizing the key ideas of each contribution. In mid-October, the committee requested further information from the Roman Science group on planning ground rules and formed three working groups on Metrics (De, De Furio, Paladini, Street), Survey Strategies (Beaton, Benjamin, Kupfer, Minniti), and Proper Motions (Carey, Drew, Schlafly, Zucker). Over the next two months, the working groups reported their results. Each committee member was then invited to propose a design for the survey, and the different approaches were compared and debated, and key differences were identified.

Going into the Jan 2025 AAS meeting, the committee ranked eight possible design options for the wide-field component of the survey. These covered the Galaxy in just three filters, just four filters, or mixed coverage with 2/4 filters or 2/4/6 filters; each of these options had a choice of bulge coverage up to $\pm 6^\circ$ or $\pm 10^\circ$ deg. In addition, each option could vary in coverage along the Galactic plane depending on the fraction of time allocated for wide-field science. A four-filter survey with bulge coverage up to $b=\pm 6^\circ$ received the greatest number of $1^{st}$ place votes, but no option clearly stood out. . There was also no consensus on the ideal set of filters. Although the F129, F158, and F213—similar to the ground-based JHK filters—were clear committee (and community) favorites, the committee (and community) was evenly split between F106 and F184 for the fourth filter. Finally, there was a range of opinions on the amount of time to devote to time-domain science (TDS); based on the science requests, the committee agreed to bracket the percentage of time allocated to TDS to be between 15% to 35% of the total time.

The status of committee deliberations were presented at a Jan 14, 2025 Roman Town Hall meeting by co-chair R. Benjamin at the $245^{th}$ meeting of the American Astronomical Society in National Harbor, MD. Several committee members were also present at this Town Hall to answer questions.

## 2.2.4 Galactic Plane Survey Community Workshop

To solicit further community input on the key issues that the committee had identified, we organized a 3-day community workshop on Feb 11-13, 2025. The Science Organizing Committee for this event was split evenly between DC members (†) and invited community members (*) and consisted of Elias Aydi[*], Eric Bellm[*], Jonathan Bland-Hawthorn[*], Matthew De Furio[†], Emily Hunt[*], Thomas Kupfer[†], Rachel Street[†] (chair) and Catherine Zucker[†].

The workshop was conducted entirely through a virtual meeting platform, with two three-hour sessions per day timed to enable researchers to attend from multiple timezones. This was advertised through the standard Roman distribution channels. Over 260 community members from around the world filled out a pre-registration form indicating their interest in attending the workshop. A breakdown of pre-registrants by career stage is provided in Figure 2.1. Although the total number of participants was not tracked, we estimate that the workshop drew between 100 and 125 participants; just under seventy participants filled out the first-day meeting survey.

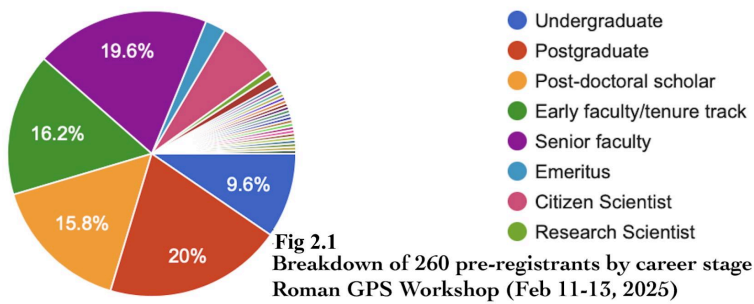

**Fig 2.1** Breakdown of 260 pre-registrants by career stage Roman GPS Workshop (Feb 11-13, 2025)

All content associated with this workshop can be found on the event's website[7]. The first day was dedicated to plenary sessions where members of the Roman Science coordination group and we presented overviews of the context of and motivation for the RGPS, preliminary survey designs, and the metrics used to evaluate them. The remainder of the meeting was dedicated to parallel breakout sessions where participants were invited to give feedback on outstanding questions relating to the survey design. Each breakout session was attended by two or more members of the DC, and each session had a chair and a scribe. A breakout session began with some invited presentations, followed by contributed presentations that community members offered after the first day, and then followed by moderated discussion. The breakout topics were:

- Footprint of wide-area survey and percentage of time allocated to wide-area/Time Domain Astrophysics
- Filter selections
- Proper motion strategies
- Choice of fields and strategy for Time Domain Astrophysics
- Synergies with other surveys
- Science drivers for spectroscopy

Participants were encouraged to submit new science pitches after the workshop; by March 5, 2025, a further nineteen cases were received. These contributions were then reviewed by the Design Committee, and included in the evaluation of the metrics for the survey.

---

[7] https://outerspace.stsci.edu/display/GPSCW/Galactic+Plane+Survey+Community+Workshop+Home

## 2.2.5 Final Design Decisions

Following the Workshop, the committee met weekly to discuss the principal themes that emerged during the workshop and the additional science pitches. With this input, the DC was able to reach consensus for the RGPS design, shown in Figure 2.2 and described in detail in §4. A survey design, implemented in APT (Astronomers Proposal Tool) was delivered to the Roman schedulers on May 1, 2025. The design was slightly modified—to expand spatial and temporal coverage of the Nuclear Stellar Disk/Central Molecular Zone—with the approval of the full committee, on June 18, 2025 and an APT file describing the survey design presented in this report was submitted on July 30, 2025.

A summary of studies, simulations, and metrics that informed this final design is provided in §3. More details on the elements, individual regions, and technical considerations is provided in §4, and constraints on the observations and the need for early science verification observations in §5.

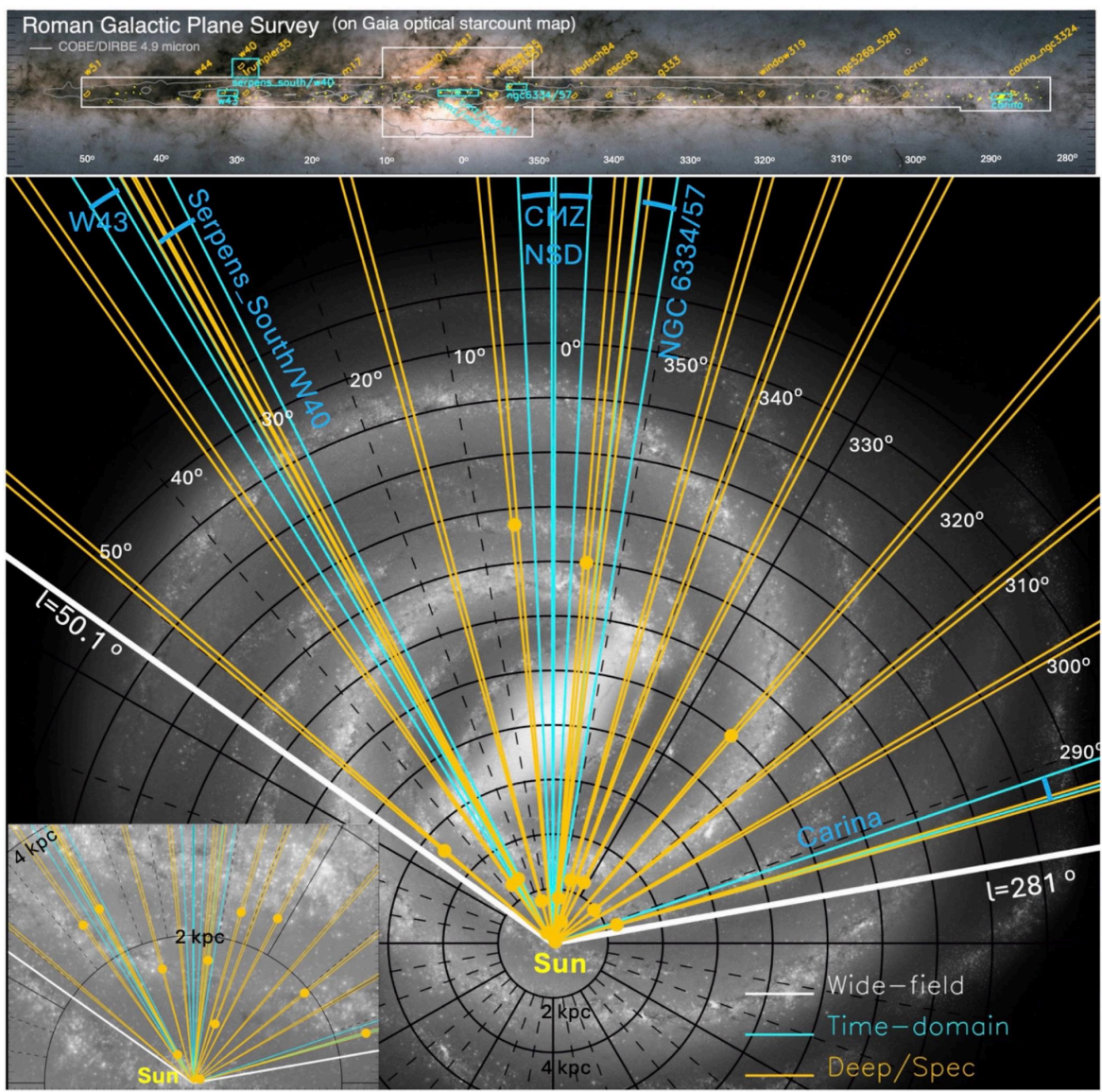

**Figure 2.2** [top] RGPS survey outline showing the wide-field (white), time-domain science (cyan), and deep/spectroscopic science (orange) fields on a Gaia starcount image with COBE/DIRBE 4.9 micron contours showing the stellar disk/bar. [bottom] Survey directions shown on a face-on representation of the Milky Way by R. Hurt. The distance to "objects of interest" (labeled in the upper panel, and listed in Table 4.4) in each Deep/Spectroscopic field are shown. An inset shows a blow-up of these sightlines within 4 kpc of the Sun.

# 3. Discussion of Results

In this section, we describe the general considerations that led to the recommended RGPS design and the information that informed these decisions. In the following sections, we provide an overview of our design considerations (§3.1), the metrics developed and how the survey design satisfies the science cases proposed by the community (§3.2), some calculations and modeling that informed our decision making process (§3.3), and options that were considered but not recommended (§3.4).

## 3.1 Overview of Design Considerations

The construction of the RGPS involves making trade-offs among different survey dimensions: *sky coverage*, *photometric depth*, *number of visits* (and the associated cadences), and the *number, choice, and timing of observations in filters*, including the grism and prism. A breakdown of the recommended elements of the program, together with some useful survey parameters, e.g. saturation and sensitivity limits, extinction values, expected number of sources, Vega to AB magnitude conversions, etc. is given in Table 3.1. In this table, we divide the wide-field survey into three sections: Disk ($|b|<2°$, $l=+50°.1$ to $280°$, with an extension to $b=-2°.7$ for $l=293°$ to $281°$), Bulge ($|l|<10°$ and $|b|=2°$ to $6°$), and Serpens South ($l=26°.5$ to 30 deg and $b=+2°$ to $+4°.5$ deg). There are also six time-domain survey (TDS) regions, and fifteen single pointings for deep and spectroscopic observations (Deep/Spec).

**Table 3.1 Overview Table for Roman Galactic Plane Survey**

| Filter | Wavelength[1] (microns) | PSF FWHM[1] (arcsec) | # sources | Saturation[3] (AB mag) | Depth[4] (AB mag) | Saturation[3] (Vega mag) | Depth[4] (Vega mag) | $A_\lambda$/$A_{5420}$[6] | Vega Zero mag (Jy)[4] | Area (Sq Deg) | Areas |
|---|---|---|---|---|---|---|---|---|---|---|---|
| F213 ("K") | 1.95-2.30 | 0.169 | 14 B | 13.01 (14.21) | 23.01 | 11.18 (12.38) | 21.18 | 0.0745 | 675.5 | 691.2 | Disk+Bulge+Serpens_South+TDS+Deep/Spec |
| F184 ("H/K") | 1.68-2.00 | 0.146 | 11B | 13.16 (14.35) | 23.30 | 11.59 (12.78) | 21.73 | 0.0979 | 854.3 | 531.2 | Disk+Serpens_South+TDS+Deep/Spec |
| F158 ("H") | 1.38-1.77 | 0.128 | 20 B | 13.89 (15.08) | 23.94 | 12.59 (13.78) | 22.64 | 0.1371 | 1091.5 | 691.2 | Disk+Bulge+Serpens_South+TDS+Deep/Spec |
| F129 ("J") | 1.13-1.45 | 0.106 | 14 B | 14.08 (15.27) | 24.04 | 13.11 (14.30) | 23.07 | 0.2102 | 1483.3 | 691.2 | Disk+Bulge+Serpens_South+TDS+Deep/Spec |
| F106 ("Y") | 0.93-1.19 | 0.087 | 68 M | 14.22 (15.41) | 24.10 | 13.56 (14.75) | 23.44 | 0.3106 | 1969.8 | 23.0 | Serpens_South+TDS+Deep/Spec |
| F087 ("Z") | 0.76-0.98 | 0.073 | 21 M | 14.23 (15.43) | 24.06 | 13.73 (14.93) | 23.56 | 0.4709 | 2297.4 | 14.2 | TDS+Deep/Spec |
| F062 ("R") | 0.48-0.76 | 0.058 | 3 M | 14.55 (15.74) | 24.43 | 14.40 (15.59) | 24.28 | 0.8276 | 3162.5 | 14.2 | TDS+Deep/Spec |
| Prism R=80-180 | 0.75-1.80 | — | — | 9.0—11.0 | See text | 7.5—10.5 | See text | — | — | 4.2 | Deep/Spec |
| Grism R=481 λ (μm) | 1.00-1.94 | — | — | 6.5—8.5 | See text | 5.0—8.0 | See text | — | — | 3.9 | Deep/Spec |

[1] Wide Field Instrument Performance webpage (v1.1, Jan 2025) https://roman.gsfc.nasa.gov/science/WFI_technical.html
[2] See Sec 3.2 for description of how number of sources was estimated.
[3] Saturation limits for the first and second resultants using proposed MA table.
[4] Depth for single frame photometry. WF observations will have between 1 and 4 observations per sky position, and deep fields go 0.75 mag deeper.
[5] $m_{AB}-m_{Vega}$ and Vega Zero mag from Durbin et al (2025) except for F062 from Lacaster et al (2022)
[6] Schalfly et al (2016)

**Wide-Field Science:** For most of the science pitches and white papers received, the most important survey dimensions from the list above was *sky coverage*: more area provides larger samples. As Sanderson et al (2024) noted, "Many of the science cases require proper motion information for which the precision is maximized if the plane survey is started early in the Roman lifecycle." Maximizing sky coverage will maximize the number of sources for which one ultimately can measure long time-baseline proper motions. The total amount of sky requested by the community greatly exceeded the amount of sky that could be feasibly covered with a 700 hour program (Figure 3.1). In part, this was driven by requests to observe catalogs of known objects, especially open clusters and star forming regions, some of which subtend large areas often some distance away from the Galactic plane. However, in many cases the authors of these requests advocated for the value of including *some* of the targets in the catalogs, rather than necessarily all of them.

An emphasis on *sky coverage* rather than *photometric depth* was made even more palatable by the sensitivity of the Roman telescope. Even with the shortest recommended exposures of ~60 seconds[8], the photometric depth of the RGPS will still be more than a factor of ten deeper than any previous IR survey of the Galactic plane (Figure 3.2), but with significantly higher angular resolution, opening a wide range of discovery space. Although some science cases required greater depth, the targeted deeper observations recommended in the current program could provide the data needed to justify a future wide-area program of greater photometric depth. In addition, the region interior to $|l|<5º$ and $|b|<2º$ is expected to have greater than one source per every ten pixels, making source confusion—and not sensitivity—the limiting consideration in this region.

**Figure 3.1** Community requests for sky coverage as a function of filter for the six filters with the largest

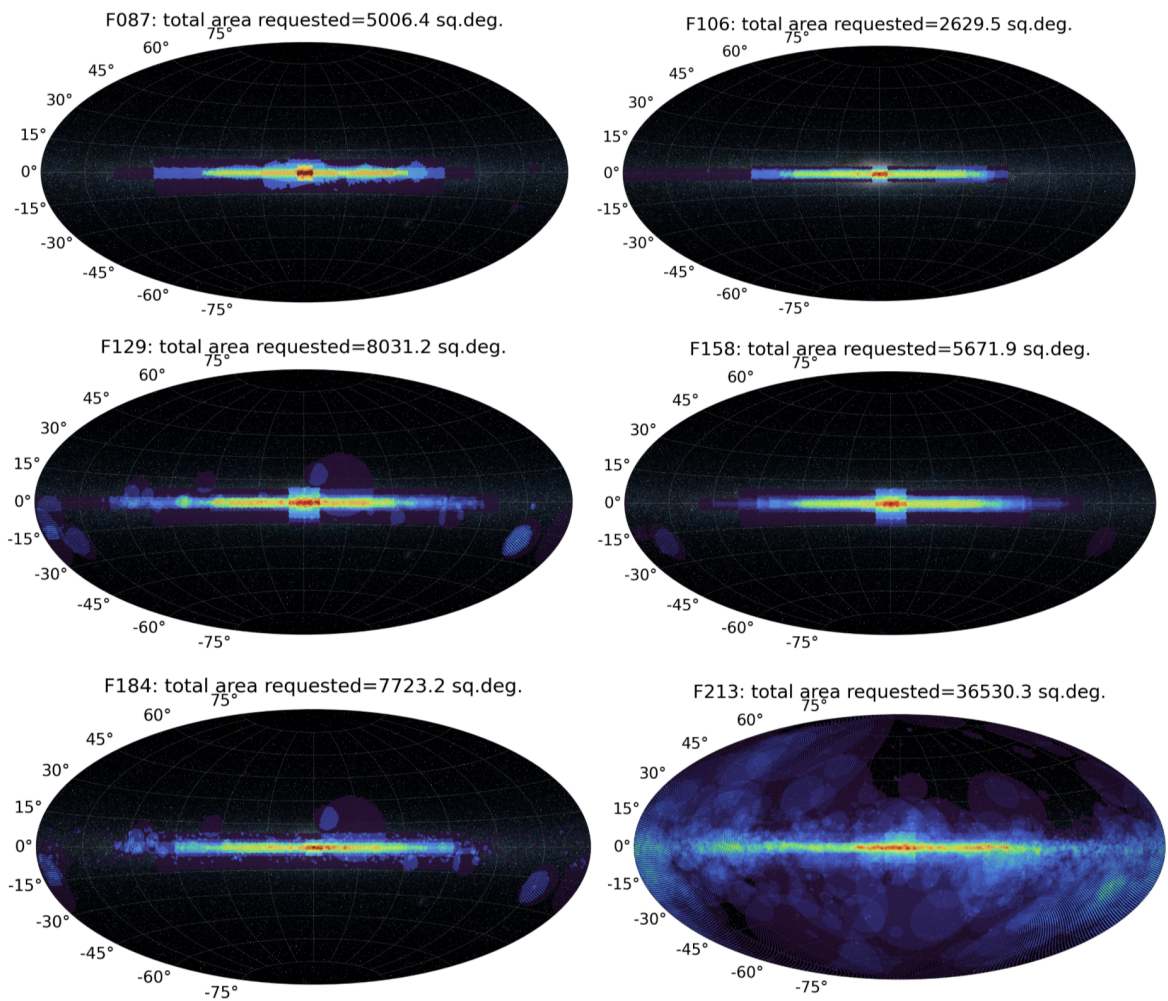

requested area, with rainbow color-table indicating the number of requests. Compare the total area requested, to the area of the recommended wide-field survey: 691 deg$^2$. For all filters, the central ±10º of longitude had the highest demand. Note also the off-plane regions requested, principally the directions of low-mass star formation regions. Total areas requested for other filters were 1908 deg$^2$ (F062), 5177 deg$^2$ (wide F146), 2066 deg$^2$ (grism), and 3.4 deg$^2$ (prism).

A more difficult trade-off to navigate was between *sky coverage* and *number of filters*. For some science cases, particularly ones which were taking advantage of Roman's angular resolution in combination with ancillary data, only a single filter was needed to accomplish the science goals, but the committee agreed that a minimum of two filters should be required to provide color information. However, many other science cases reported the need for (at least) four photometric bands, to break degeneracies between stellar temperature and extinction, for example, and preliminary modeling (§C.3) confirmed the value of adding a fourth filter. We considered two, three, four and six-filter designs with different sky coverage. Following the workshop, and for reasons encapsulated in Figure 3.3, we decided that four-filter coverage of the disk ($|b|<2°$) and three-filter coverage of the bulge/bar ($|l|<10°$ and $|b|=2°$ to $6°$)—where the extinction is much reduced—would strike the best balance between *number of filters* and *sky coverage*, allowing the RGPS to cover ~130° degrees of Galactic longitude. Regarding *filter choices* for the wide-field survey, we recommend that the four filters for the Galactic plane survey should be F129, F158, F184, and F213. The similarity of three of these filters (F129, F158, and F213) to the ground-based JHKs filters will allow for a comparison to previous infrared surveys, and they were the most requested by the community. In §3.4.1 we discuss some of the factors that led us to favor F184 over F106 as the fourth filter.

---

[8] "There is little point considering faster survey speeds, because sensitivity drops rapidly for modest increases in survey speed at yet shorter integrations." (https://roman.gsfc.nasa.gov/science/WFI_technical.html)

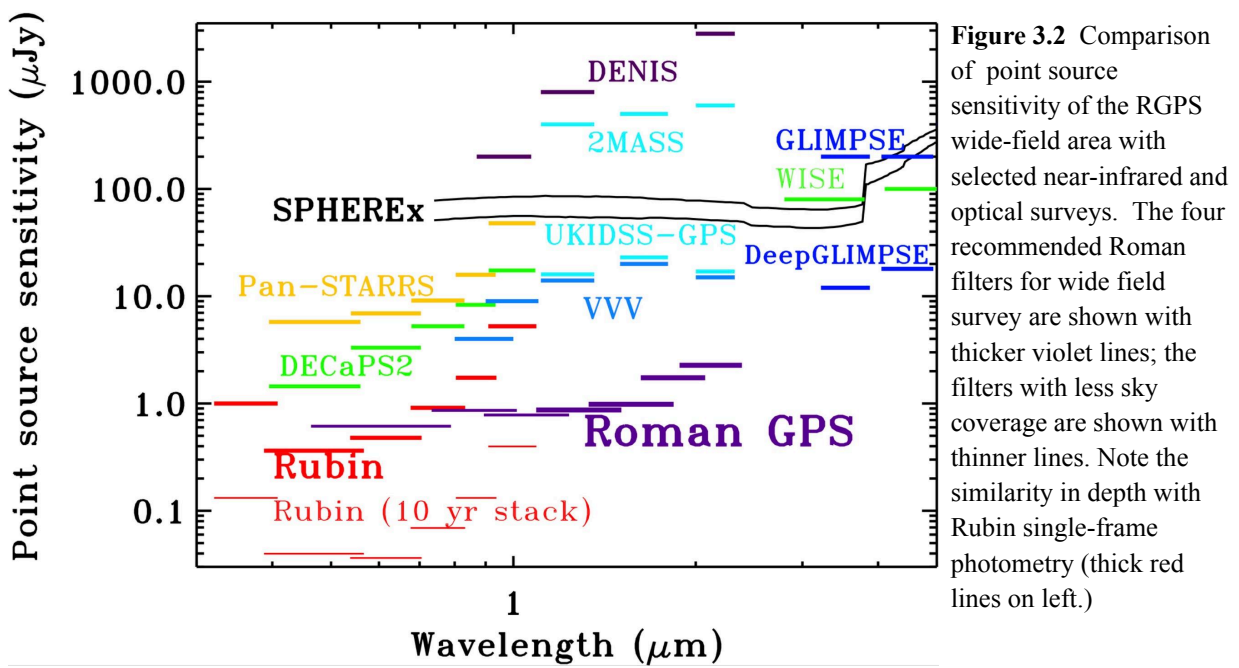

**Figure 3.2** Comparison of point source sensitivity of the RGPS wide-field area with selected near-infrared and optical surveys. The four recommended Roman filters for wide field survey are shown with thicker violet lines; the filters with less sky coverage are shown with thinner lines. Note the similarity in depth with Rubin single-frame photometry (thick red lines on left.)

Another important consideration is the *timing of the observations*. Although the original expectation of the committee had been that the proper motions of the sources detected in the RGPS would have to be obtained with some future epoch of observations, with a ~two year baseline the measurements of proper motions will be good enough for many science goals provided by the community. By separating the observations in the different filters by two years (F129 and F213 as early as possible, and F158 and F184 as late as possible), cross-band measurements of proper motions will provide a down-payment on what we expect will be a long-term program of obtaining high precision proper motions for billions of stars in the Galaxy. Additional issues regarding the timing of the observations in different filters, and implication for studying variable sources, are provided in §3.4.

Regarding the footprint for the *sky coverage*, we focused on three issues: contiguity, Galactic longitude asymmetries, and Galactic latitude extensions. It was agreed that the RGPS should be contiguous, both to allow for the continuous measurement of stellar and star formation properties along the full disk covered, and because of the difficulty of justifying future "gap-filling" programs. Regarding asymmetries in Galactic longitude, there were a significant number of science pitches and papers requesting the Carina star formation and tangency direction out to $l=280°$. (Note also the large number of Hubble observations in this direction shown in Figure E-1.) A symmetric design within |l|< 50º, with an extension to l=-80º, encompasses nearly all of the inner Galaxy star formation within the Sun's orbit, fully spanning the "Sagittarius-Carina" arm, the collection of star forming regions mostly inside the Solar circle.

We also recommend three latitude extensions: *(1)* the extension in the bulge/bar will resolve the stellar population of the bulge/bar down to the confusion limit of the survey, *(2)* the extension to negative latitudes in the Carina region will follow the warp of the gas and stellar layer in this direction/distance, and *(3)* the extension to positive latitudes in the Serpens South/W40 region will take advantage of a uniquely nearby region of star formation region that— unlike Orion, Taurus, Ophiuchus and other nearby star forming regions—could be added as a contiguous section of the Galactic plane coverage.

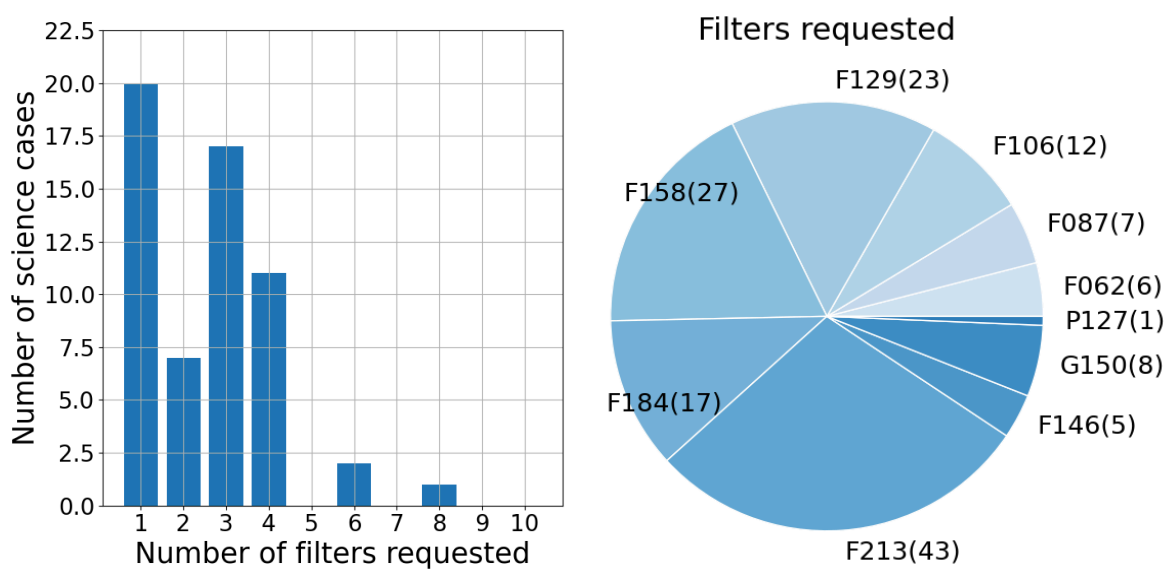

**Figure 3.3** [Left] The number of science cases provided by the community that requested observations in one or more bandpasses. The grism and prism are included in the count of filter choices available. Note the drop in the number of filters requested between 4 and 5. [Right] Breakdown of which filters were most frequently requested. The set F129, F158, F213 (similar to the ground-based JHK) were the most requested; the F106 and F184 filters had a similar number of requests.

**Time-Domain Science (TDS):** The coming decades are certain to be a fruitful time for time-domain science, thanks in large part to the advent of Rubin Observatory operations. The white papers and science pitches reviewed made clear that Roman has unique contributions to make including studying the variability of YSOs in star forming regions, cataclysmic variables, RR Lyrae populations, microlensing events, and X-ray binaries. There were several science cases provided to the committee advocating a large *number of visits* in one or more filters—either for specific regions or Galaxy-wide—in order to identify or better characterize different classes of variable sources. Roman observations would be particularly valuable in crowded fields (which increases the yield) where ground-based observations become compromised due to source confusion, extinction and related photometric uncertainties. Cadences requested ranged from 1 min up to months and durations spanned hours up to the two year limit of the survey. We found that with a judicious choice of field and cadence, the same survey data could serve multiple science cases, motivating the committee to explore a small number of time domain fields. We started with a focus on high cadence (~10 min) observations. This was motivated by community requests, the advanced state of development of this mode for GBTDS observations, and the fact that such observations would fill a niche that the longer cadence Rubin observations would not sample ([Bellm et al 2022](#)) as shown in [Figure 3.4](#). Following the Workshop—where several participants encouraged the DC to consider a broader range of cadences from minutes to months— longer cadence observations (hours to weekly) were examined. Because of the success of the *Spitzer*/YSOvar program ([Rebull et al 2014](#)) in identifying a range of behaviors in YSOs and because of the "intra-night desert" present in the Rubin program—a lack of sampling over timescales of several hours, shown in [Figure 3.4](#)—we favored adding hourly cadences to three of the time domain fields, as well as additional weekly monitoring in both F213 and F129 for the two fields on either side of Galactic center.

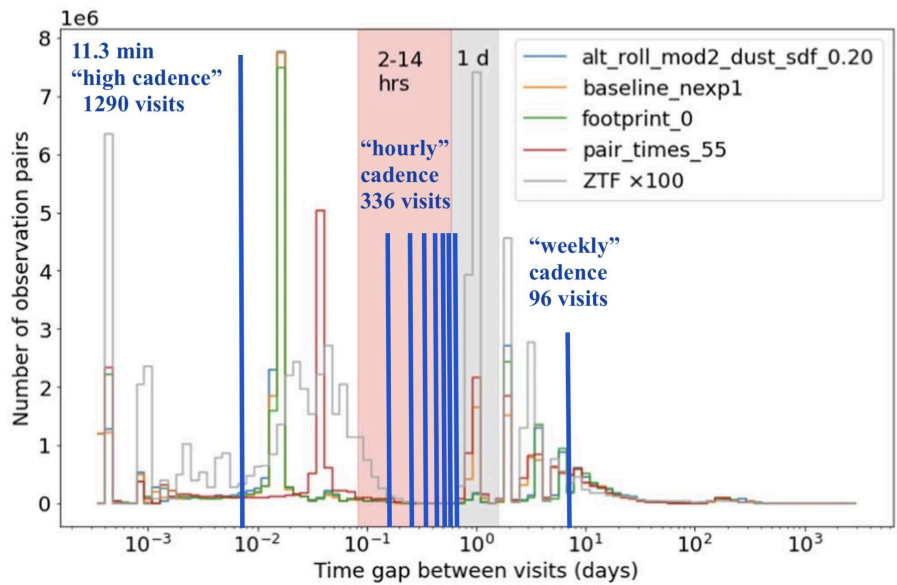

**Figure 3.4** A modification of Figure 1 from [Bellm et al (2022)](#) which shows a histogram of time gaps between visits to the same sky position for several LSST simulations as well as ZTF observations and the time gaps recommended for the RGPS. The five regions of high cadence observations have a total of 1290 visits to 30 Roman pointings.

Three of the recommended regions for monitoring focus on the source-rich inner Galaxy, including two sets of six contiguous Roman pointings that would fully cover the longitude-extent of the *Central Molecular Zone/Nuclear Stellar Disk* with an 11.3 minute cadence over an eight hour duration using F213. These two regions would complement the extensive observations of the GBTDS Galactic Center field, providing wider area coverage,albeit with much sparser temporal sampling. A third inner-Galaxy region, seven degrees from Galactic center, sampling both the nearby (1780±120 pc) *NGC 6334/NGC 6357 star forming regions* and the background Galactic bulge was also selected. All of these regions will also be observed over the same time-frame as Rubin. Because of the geometry of these regions, these high cadence observations are roll-angle constrained.

Further out in the disk, we selected two additional regions for high-cadence-only (11.3 min) observations from community-provided targets. One of the regions, the *Carina star forming region,* also will be monitored by Rubin; the second, the *l=30º direction (including W43)*, overlaps with the planned Subaru Galactic Plane Survey. Unlike the other fields, the DC recommends adopting the Subaru-suggested F184 filter, which is ensitive to variability in the Paschen alpha emission line.

Finally, the RGPS extension to the nearby *Serpens South/W40* is recommended for observations with eight visits spaced with an hourly cadence with increasing intervals between observations, Δt=4, 6, 8, 10, 12, 14, and 16 hrs. This region of the sky strategy was adopted as part of the *YSOvar* program using the *Spitzer Space Telescope,* and the same strategy would also be used to search for variability of YSOs in the Central Molecular Zone.

We recommend imaging these TDS fields in all filters (except the wide F146) with the same ~60 exposures used in the wide-field program, including repeating the F129 and F213 coverage that presumably would be obtained earlier in the RGPS observations. For the two fields on either side of Galactic filter, we recommend deeper observations (four times longer) for the shorter wavelength filters: F067, F082, F106, and F129.

To summarize, these regions will have pan-chromatic coverage, time-domain coverage in one (or in the case of the CMZ/NSD, two filters) at cadences ranging from 11.3 minutes to weeks, and repeat observations in F129/F213 separated by more than a year. It is the expectation that these observations—when combined with the results of ground-based monitoring by Rubin, Subaru, and other telescopes—will provide a large sample of variable sources for joint analysis and a justification for future larger-scale Galactic programs for time-domain science.

**Spectroscopic Science:** The use of the Roman *grism* (wavelength range 1.00-1.94 μm and resolving power R=481 $\lambda$ [μm]) and *prism* (wavelength range 0.75-1.80 and resolving power R=80-180) provide the opportunity to obtain spectral information on an overwhelmingly large number of sources, but also present multiple technical challenges, from guiding in crowded fields to extracting information from overlapping spectral traces. The potential value of these observations in identifying emission line objects and to determine stellar parameters was highlighted by community contributions.

By the time of the Workshop, we had not developed a plan for the use of these spectroscopic capabilities given the technical uncertainties. As a result of the Workshop, several additional science cases and possible approaches were provided to the DC. It was decided to select a set of single Roman pilot regions for spectroscopy that would also double as pan-chromatic deep fields: four times the nominal exposure time or 0.75 mag deeper. Using the sky coverage requests from the community, we selected fourteen directions, spaced roughly every ten degrees. Some of these initial directions were tweaked in order to have regions with a range of stellar density, extinction, and expected level of diffuse emission. A fifteenth direction was added with much greater photometric depth and prism observations in the direction of the nearby star formation region W40 in order to spectrally characterize the sub-stellar content of this region down to one Jupiter mass (photometrically) and three Jupiter masses (spectrally). With the full filter set, spectral information, and greater photometric depth, we expect these fields will be useful in testing different approaches for future coverage of the Galactic plane.

## 3.2 Survey Design and Science Requirements

In order to evaluate how the survey design meets the requirements of the science cases proposed by the community, our metrics working group defined the set of metrics as objective statistical quantities that can be calculated for each science case and survey design. These are summarized in [Table 3.2](#) and described in more detail in [Appendix B](#). The working group concentrated on high-level metrics for the survey's key observational constraints as proxies for how the survey design will enable the science concerned, rather than more specific metrics tailored to each science case.

Every White Paper and Science Pitch submitted was reviewed closely to identify the observational survey requirements necessary to accomplish the science. Specifically, the following observational constraints were identified for each science case:

- The spatial region(s) of interest
- The filters, grism and/or prism requested for each region
- The number of visits requested, the interval(s) between visits and the duration of repeated visits per region

| Table 3.2 Metrics developed for RGPS survey | |
|---|---|
| Metric | Short Description [1] |
| M1: Survey footprint coverage | How well do the survey fields defined cover the survey footprint requested in White Papers/Science Pitches? |
| M2: Star counts | How many stars are included in the survey fields? |
| M3: Extended object count | How many known targets of different types are included in the survey fields? |
| M4: Proper motions | How well can PM be measured ? |
| M5: Color measurements | Sky region observed in multiple bandpasses |
| M6: Sky area number of visits | Sky region to receive the requested number of visits |

[1] A more complete description can be found in Table B.1

Science cases were also categorized by science topic to enable related proposals to be evaluated together. These requirements were captured in machine-readable JSON format so that comparisons could be programmatically made between the requirements of the proposed science and the preliminary and final survey designs. In addition, these science case configurations were used to calculate statistics regarding the science requirements, such as the preferences for the different filters. The results of the analysis were considered during the survey design phase. The software, configuration files and results can be found in the working group's Github repository[9].

The following sections summarize the results of this analysis and discuss what could not be factored into the survey design in order to highlight topics that would benefit from future GA proposals.

## 3.2.1 Survey Footprint

The most numerous group of community proposals recommended a wide-area survey of the Galactic plane with small (1 or 2) numbers of visits to each pointing. Figure 3.5 illustrates how well the wide-area component of the RGPS meets these recommendations.

As discussed in §3.1, one of the major trade-offs the committee had to make was the number and choice of filters used to observe the Galactic plane. The wide-area survey includes the most popular filters, but lacks requested observations in the bluer filters (F062, F087, F106) as well as spectroscopic coverage. The exceptions are those regions, such as Serpens South, where observations using these filters are made as part of a combined time-domain or deep/spectroscopic strategy. Although the total area of sky covered by these elements of the survey is small compared with the wide-area footprint, the fields were chosen to maximize the science potential. This process included choosing pointings that lie within the survey regions desired by community science proposals. Tables B2 and B3 provide a summary of the community proposals served by each deep/spectroscopic and time domain field.

The wide-area survey also confines its observations to a contiguous footprint in the central Galactic plane, with a higher latitude extension in the bulge region. By concentrating towards the central plane, the survey design satisfied a wide range of science goals, and maximized the number of objects observed in regions that are not easily accessible to other surveys. While this strategy serves a large fraction of the community science cases, it unavoidably excluded regions that lie further from the central plane that are nevertheless of interest for Galactic science. Other Roman Core Community Surveys will observe large areas away from the Galactic plane, so some of the proposed science may be achieved with those data.

---

[9] https://github.com/rachel3834/rgps/releases/tag/RGPS_report_metrics_v1.0

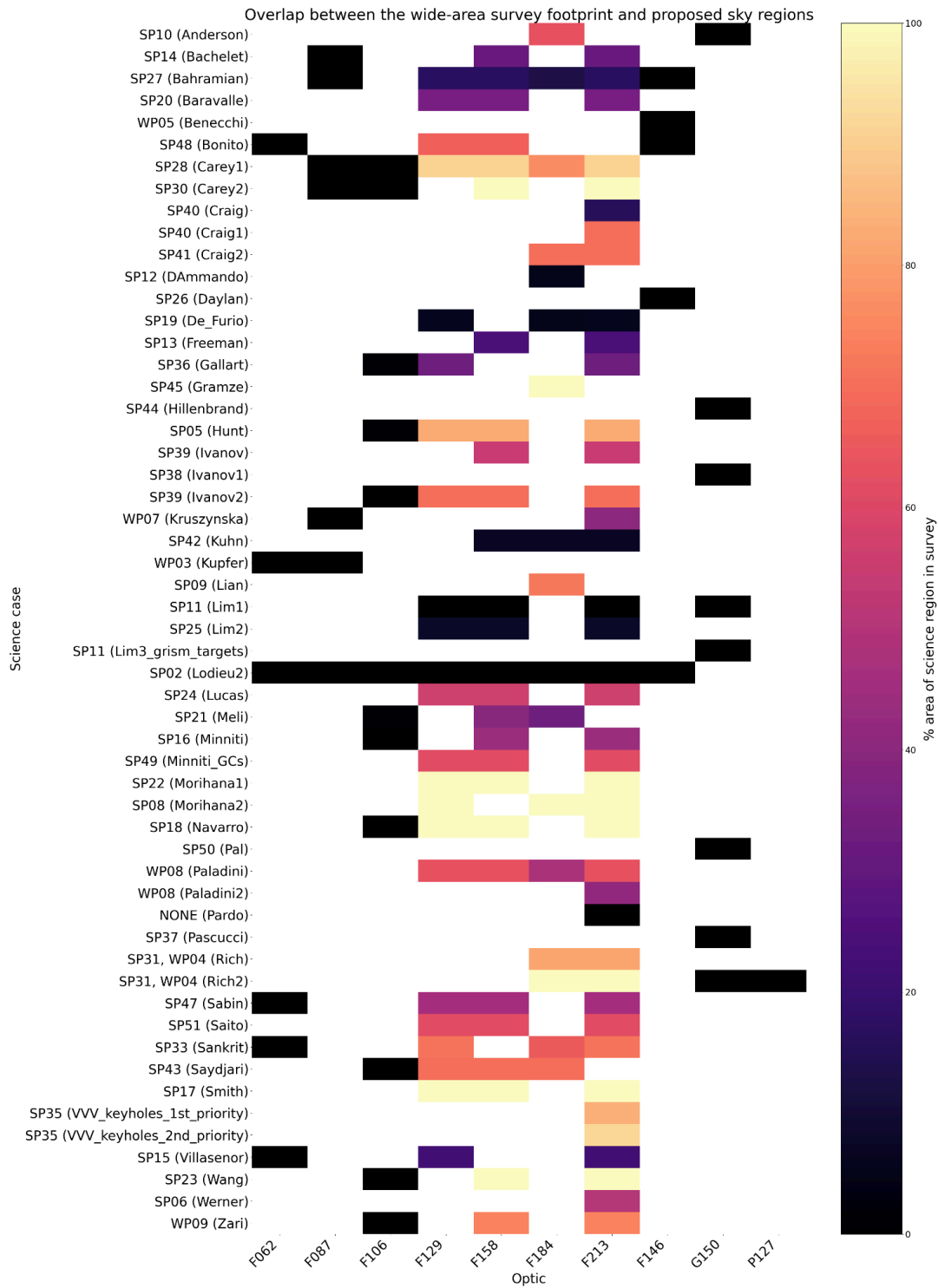

**Figure 3.5** Percentage of survey regions desired by each science case included within the wide-area RGPS survey footprint. White spaces indicate filters that were not requested and not included in the design. Note that for some programs, the percentage of requested area that is incorporated in the RGPS is low because the requested area greatly exceeded 700 deg$^2$.

It should be noted that while many proposals advocated for observations of very large areas of sky, they often indicated that *any* observations within that region would be beneficial, rather than necessarily expecting that their entire area would be included. Feedback was solicited from the community regarding the minimum useful area for each science case, but in the vast majority of cases such thresholds proved very difficult to establish. For example, the expected number of RR Lyrae will scale in rough proportion to the number of stars observed for fields in the Galactic plane, and authors indicated to us that maximizing the latter number was most important.

Many studies in Galactic science concentrate on specific types of non-point-source objects, such as clusters and star-forming regions, for which catalogs of the known examples of the type are available. Some science cases advocated for the RGPS to include as many known targets as possible from large catalogs of open and globular clusters, star forming regions, active galactic nuclei, molecular clouds and HII regions. Others proposed to focus on a subset of targets selected to address particular science questions, such as a samples spanning a range in age or metallicity. In some cases, the full catalog of known objects is quite large, with many entries far from the plane and thus outside the RGPS footprint. Some objects, particularly star formation regions also subtend large areas on the sky, putting some of the area well outside the Galactic plane.

These catalogs were used to maximize as much as possible the number of these objects included within the wide area survey boundaries and to optimize the placement of the deep/spectroscopic and time domain fields. The survey will provide observations of a significant sample of several classes community-supplied objects, as shown in Table 3.3. Although there are numerous catalogs of different classes of objects now available, we only used those supplied to us by the community for the analyses here.

| Object type | Number of targets within footprint | Percentage of catalog within footprint | Source |
|---|---|---|---|
| Star Forming Regions | 8 | 14 | De Furio [priv. com.] |
| Globular Clusters | 39 | 24.5 | Baumgardt-Harris catalog [Harris 1996, Baumgart et al. 2019] |
| Open Clusters | 1829 | 25.5 | Hunt [priv. com.] |
| Active Galactic Nuclei | 24 | 5.9 | D'Ammando, 4LAC DR3 [Ajello et al. 2023] |
| Molecular Clouds | 25 | 56.8 | Villaseñor [priv. com.] |

**Table 3.3** The number of targets of interest (and percentage of total catalog) for which some or all of the target receives some observations in the RGPS wide-area survey.

## 3.2.2 Proper Motions

In order to obtain proper motions, we propose splitting the RGPS observations by filter into an early epoch and a late epoch, separated by roughly two years. This strategy is defined for the bulge and disk fields, and analysis of the expected proper motion precision indicates that the threshold precision of one milli-arcsecond will be reached for all pointings within these regions that receive two visits.

While time series observations are obtained for all other fields (both deep and TDS), the maximum interval between visits are too short (typically a few days-weeks at most) for proper motions to have a measurable effect on target locations. This may be extended if the F213 observations are completed for the full wide-area footprint in Year 1, allowing time to revisit the Serpens South and CMZ/NSD fields.

### 3.2.3 Filter Coverage

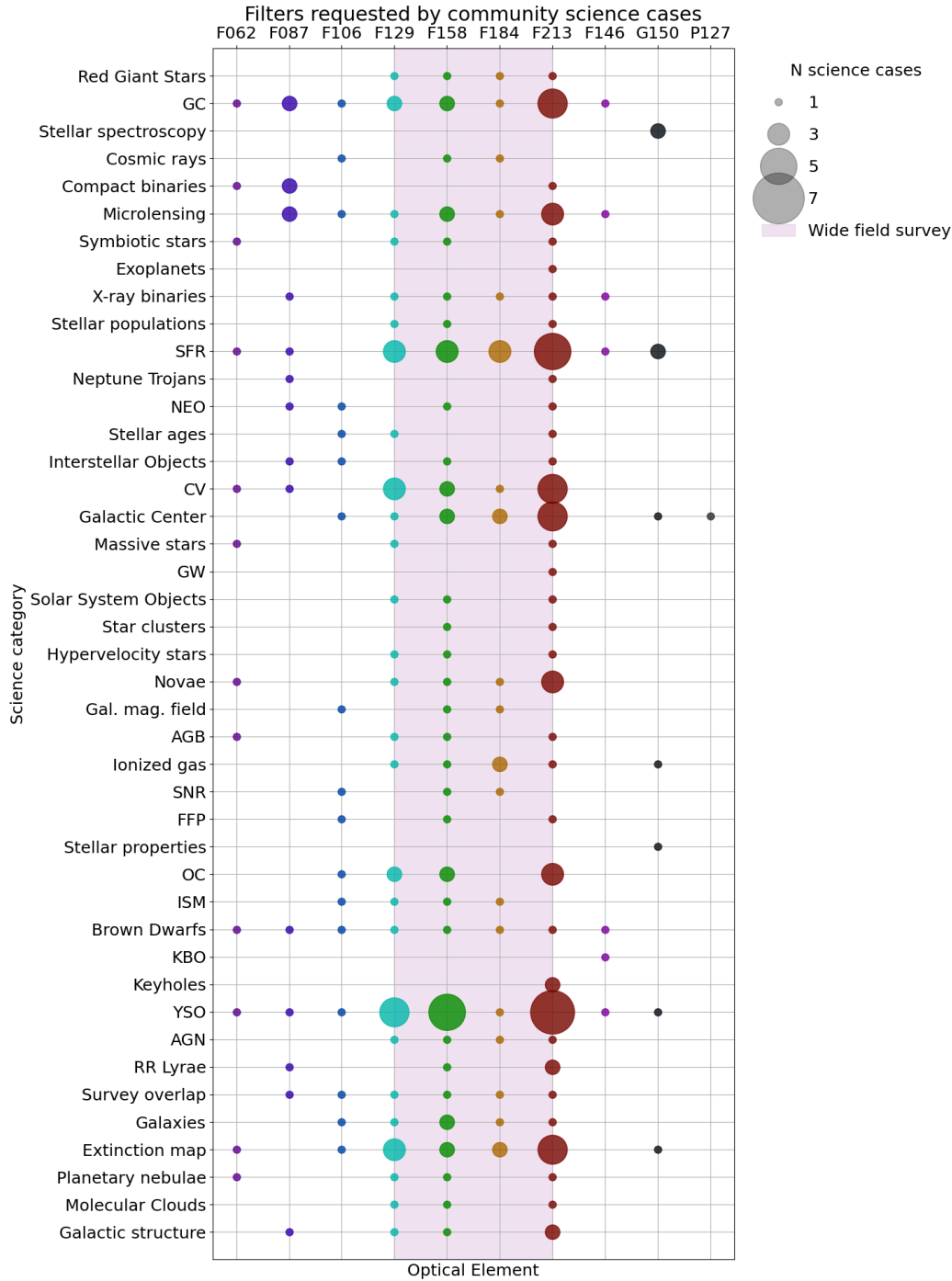

**Figure 3.6** Bandpasses of the observations requested by community proposals for different science topics, compared with those provided by the RGPS wide field survey. In addition, the deep/spectroscopic fields provide observations in the full range of filters in selected areas.

The RGPS addresses the inherent tension between observing large sky areas and obtaining data in multiple bandpasses with a two-fold strategy, as discussed in Section 3.1. The wide-area survey element covers ~700 sq.deg. by adopting a limited subset of filters. These were selected to suit the needs of the majority of community proposals, and to sample stellar SEDs (Figure 3.6) However, strong arguments were made for observations in a wider range of filters, such as extinction measurement, the need with other surveys particularly Rubin for cross-calibration, and the value of bluer wavelengths for characterizing variable sources such as YSOs and CVs. To address these concerns, the survey includes deep/spectroscopic fields for selected regions which will receive observations in all bandpasses.Most time domain science proposals advocated for the majority of observations to be obtained in a single filter, often with observations in one or more secondary filters to be made at less frequent intervals. A more restricted filter set was therefore adopted for the time domain fields, illustrated in Figure B.1, where single observations will be obtained in multiple bandpasses for stellar characterization but the majority of data will be obtained with a single filter, either F184 or F213.

The wide filter, F146, was requested by a number of community proposals but is not included in any element of the RGPS. Most science cases stressed the value of wide area survey or time domain observations over reaching deep limiting magnitudes, and in the vast majority of cases the need for color information was considered to outweigh the need to maximize throughput. This was reinforced by the fact that all elements of the RGPS were designed to serve multiple science cases, and feedback from the community indicated that the wide passband was less important in most cases if data were obtained in a full set of narrower passbands, since the SEDs would be well sampled.

## 3.2.4 Time Domain Science

Metrics were used to examine whether regions of interest for time domain science receive the cadenced observations needed for their goals, and the cadences requested. Since a science case can have multiple desired regions, each with a bespoke filter set and timeseries cadence, it is relatively complex to compare these specifications with all regions covered by the RGPS strategy. We limit the discussion in this section to the survey regions within the time domain element of the RGPS, and to those community proposals that requested time domain observations.

The first comparison is to evaluate the overlap between those fields and those requested. The time domain fields, while limited in extent, were selected to lie within the regions requested by as many science cases as possible. Table B3 summarizes in detail the science cases served by each time domain field.

The temporal strategy of RGPS time domain observations is well suited to the needs of a range of science cases that require high cadence over a finite duration, including YSO and SFR studies, short duration microlensing events, CVs and for characterizing a range of variability in the Galactic Center. This is an important but underserved temporal regime since the typical interval between LSST observations is of order of days for most of the sky. Variables with intermediate timescales (~days – weeks) will also be sampled by the adoption of the increasing-interval time-series in some fields. Feedback from the community indicated a degree of flexibility in many science cases regarding the filter choice(s) for time domain science, so preference was given to achieving good temporal coverage in one, frequently-requested passband rather than several.

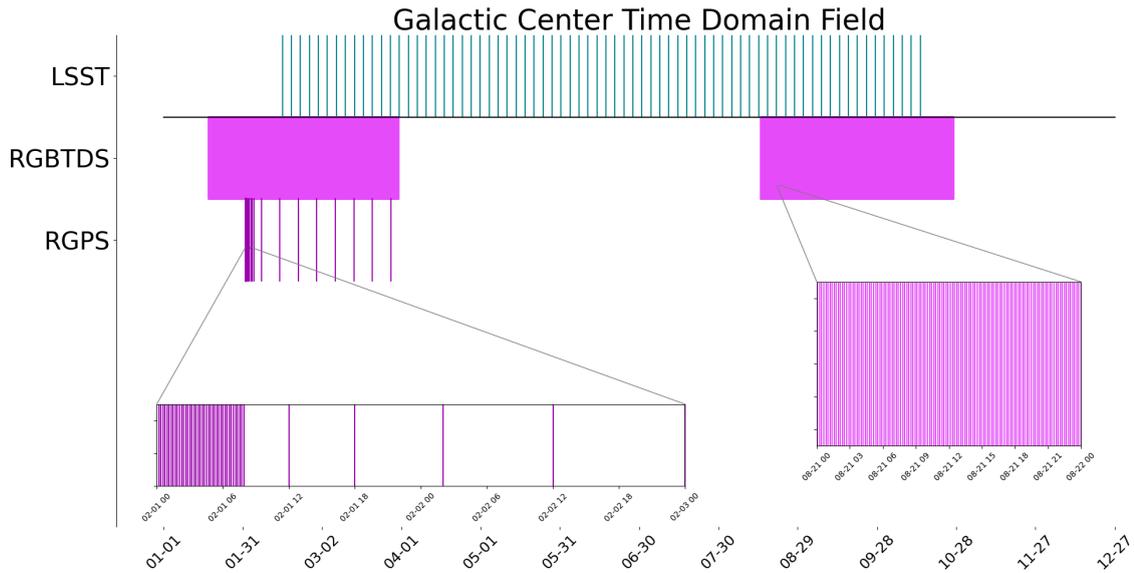

**Figure 3.7** Comparison of the time-domain survey strategies of the Roman GBTDS and RGPS surveys and Rubin Observatory's LSST for one year. The RGPS will extend the spatial area of the Galactic Center to receive cadenced observations complementary to those of the single pointing included by the RGBTDS. Both Roman surveys are complemented over longer timescales by regular observations from Rubin from Year 2 of that survey onwards. Note the minimum LSST cadence is shown here, not the augmented cadence for the Galactic bulge field.

A number of the RGPS time domain fields overlap the Galactic plane survey region of LSST, most notably the Galactic Center fields. Figure 3.7 illustrates how the contemporaneous observations by both surveys would be highly complementary, taking into account the estimated visibility windows of both observatories. The RGPS will extend the spatial area of the Galactic Center to receive cadenced observations complementary to those of the single pointing included by the RGBTDS. Both Roman surveys will be complemented over longer timescales by regular observations from Rubin from Year 2 of that survey onwards. It should be noted that a single 9.6 sq deg. Rubin pointing in the Galactic bulge will receive significantly augmented cadence from Year 4 onwards. A wealth of stellar variability across the Galactic Center will be characterized if the scheduling of these observations is coordinated by both the Roman Mission and Rubin Observatory.

## 3.3 Trade Studies, Modeling and Simulations

Unlike the Core-Community Surveys, there were no Project Infrastructure Teams supporting the RGPS design. However, several calculations of interest to the community were performed both by committee and community members to quantify some of the expectations for the RGPS. The results shared here are preliminary, and should be considered illustrative as opposed to definitive. Future work in the areas described below, both to inform any modifications to survey design and to prepare for the analysis of the results, is highly encouraged. The details of these studies are provided in Appendix C. Here, we summarize some principal results:

*Extinction modeling* (§C.1): For a typical mid-plane ($|b|<1°$) value of extinction of $A_{F213}=1.5$ (or $A_{5240A}=20$), a source suffers a factor of 0.25 (or $10^{-8}$) diminution in brightness. In estimating the effects of dust on source counts, we have chosen to rely on maps based on infrared data only— taken from analyses of 2MASS or VVV data—since maps that require the detection of one or more optical bands are severely limited by the amount of dust in the inner Galaxy. As the extinction structure is very patchy and complex on the sky, we use 2D maps that characterize the integrated extinction out to the distance of the bulge/bar, but note that different maps differ in estimates of $A_K$ by a factor of 1.25, e.g. Figure 16 of Zhang & Kainulainen (2023). We use a composite extinction map to estimate the star counts derived in the following section.

*Star-count modeling* (§C.2): Since many of the community science cases scale with the number of sources, it is valuable to characterize the total number and classes of sources expected from the Galactic plane survey. Using the TRILEGAL model (Girardi et al 2005) with the extinction documented in §C.1, we expect the RGPS to yield 20 billion stellar sources; this number is uncertain to a factor of a few.

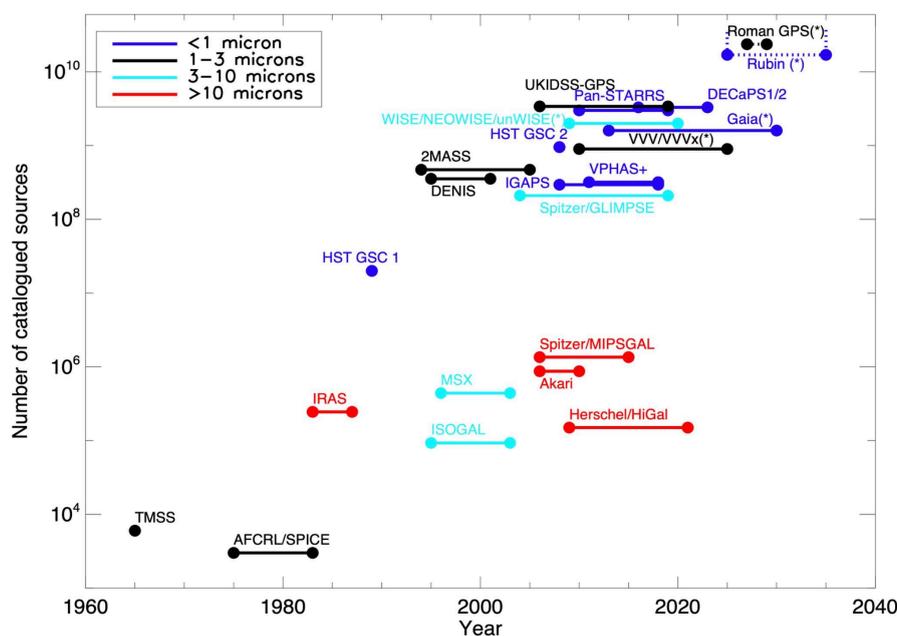

**Figure 3.8** Number of catalogued sources as a function of year for optical (dark blue), NIR (black), mid-IR (cyan) and far-IR (red) showing the years that observing and catalog development were in operation. Programs marked with a (*) are either operational or planned.

This estimate is comparable to the number of stellar sources (17 billion) expected to be in the full-southern-sky 10-year LSST catalog from the Rubin Observatory[10] and an order of magnitude larger than previous surveys (Figure 3.8). It would be valuable to update the different available Galaxy star-count models, together with an improved treatment of the 3D extinction, to prepare for the enormous catalogue of sources that Rubin and the RGPS will produce.

*Measurements of stellar parameters* (§C.3): Many science programs will use the combination of the RGPS photometric measurements, time-domain information, and ancillary information to identify numerous classes of sources for characterization. Figure 3.9 shows an infrared color-*absolute* magnitude diagram, generated using the TRILEGAL model (Girardi et al 2005), that contains all sources brighter than the RGPS sensitivity limit in the direction $(l,b)=(5°,0°)$. This shows some of the principal features of the CMD, including the main sequence both for binaries (upper branch) and single stars (lower branch), the red giant branch, red clump giants, the main sequence turnoff at $M_{F213}=2.50$, and the location of selected spectral classes and the range over which they can be detected for two different values of extinction. In the inner Galaxy, approximately 60 to 70% of the detectable sources are expected to be K and early M dwarf stars, but the fraction of M stars in a given direction will depend sensitively on the extinction. In §C.3, we also provide an example of how one might recover multiple-age stellar populations in the direction of the NSD/CMZ using a combination of

---

[10] https://rubinobservatory.org/for-scientists/rubin-101/key-numbers

Roman and JWST data (Schödel, priv comm) and preliminary results on how the recovery of stellar parameters for individual sources depends on the filter set used (Zucker et al, priv comm).

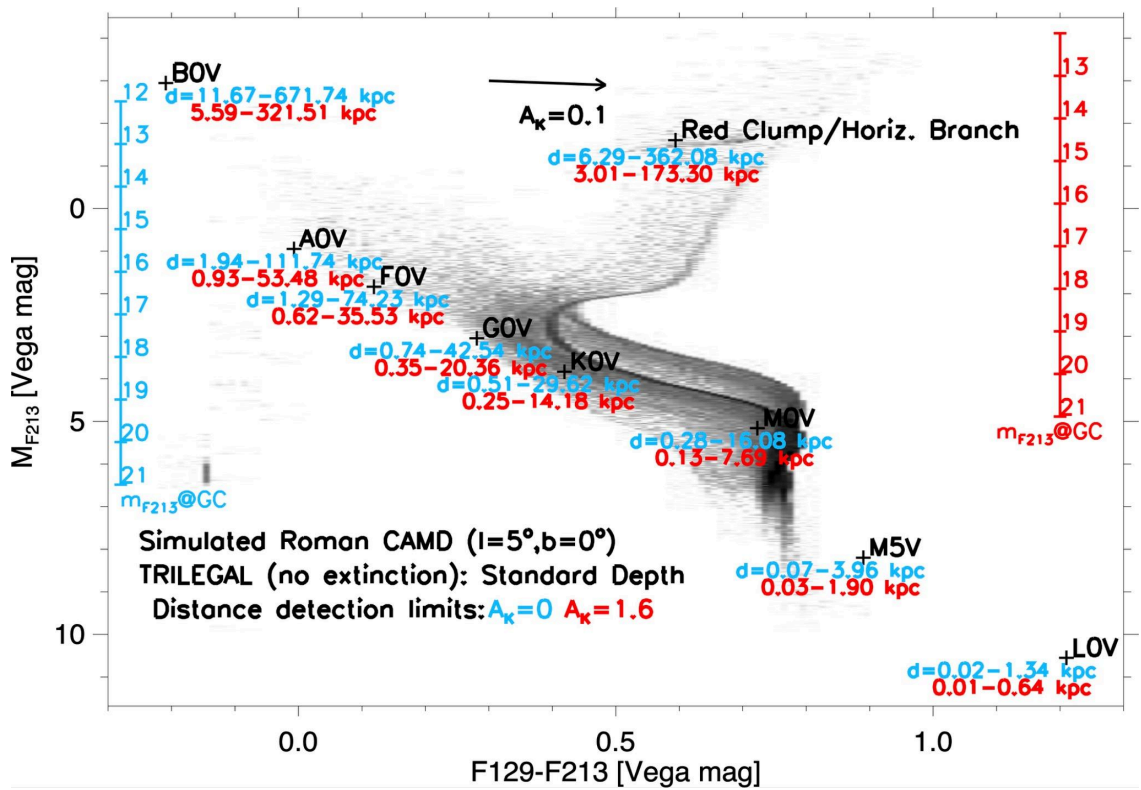

**Figure 3.9** Absolute magnitudes and colors for selected spectral classes, using the updated[11] J and K absolute magnitudes of Pecaut al (2013) and Hawkins et al (2017) converted to Roman filter Vega magnitudes via Durbin et al (2025). Only sources detectable with the photometric depth of Table 3.1 are shown. The distance range over which these sources will be seen are shown for two different values of extinction $A_K$=0 (blue) and $A_K$=1.6 (red). The apparent magnitude of sources at the distance of the Galactic center—using 8 kpc (m-M=14.52)—is also shown.

*Proper motion measurements* (§C.4): With its combination of survey speed, near-infrared sensitivity, and high resolution, over time Roman has the potential to measure proper motions for an unprecedented number of stars. Even the two year time-baseline of the RGPS will allow for the measurement of scientifically valuable proper motions.

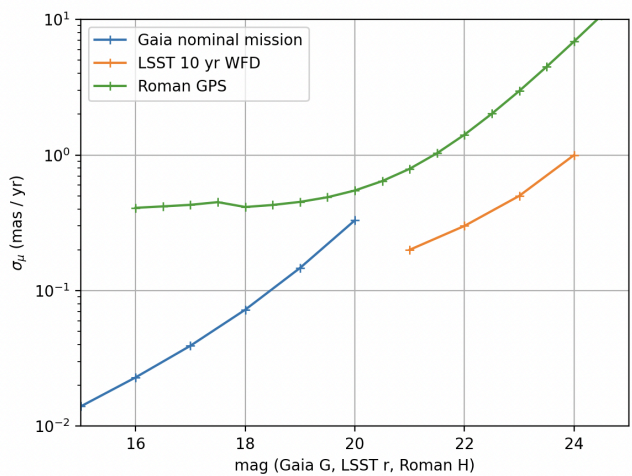

**Figure 3.10** Comparison of uncertainties in proper motion for RGPS with a two year baseline to the Gaia mission and the LSST ten-year WFD catalog. Note that RGPS can achieve uncertainties of 0.4 mas/yr at the bright end.

Figure 3.10 provides a comparison of the estimated RGPS proper motion uncertainties with both Gaia and the ten-year Rubin data as a function of source magnitude. This shows that the RGPS proper motion measurements should be similar to Gaia at the faint end of

---
[11] https://www.pas.rochester.edu/~emamajek/EEM_dwarf_UBVIJHK_colors_Teff.txt

Gaia's measurements ([de Bruijne et al. 2014](#)), while extending to much fainter magnitudes than Gaia at wavelengths much less affected by reddening. LSST performance is expected to be a factor of three better than the RGPS after ten years over its nominal high latitude footprint ([LSST Science Book 2009](#)), but the RGPS performance should be competitive in the crowded and high extinction Galactic plane and will be available before the LSST ten-year WFD catalog. The RGPS performance is compared to different science programs in [§C.4](#).

*Calculating yields for time-domain science* ([§C.5](#)): The Rubin LSST is certain to uncover large numbers of different classes of variable sources with a wide range of behaviours. The time-domain science recommended as part of the RGPS focuses on regions—like the Galactic center— and cadences that will be less well covered by Rubin ([Figure 3.4](#)). It will cover the "intranight desert" ([Bellm et al 2022](#)) of a few hours to a day which can probe stellar flares and short-time-scale accretion events in YSOs, as well as short duration dwarf novae of compact accreting binaries. Calculations provided in [§C.5](#) provide information on the expected yields for different classes of objects using the strategies proposed here.

## 3.4 Other Options Considered

This section documents some of some of the more difficult trade-offs made in the developing the recommended RGPS design. It is our hope that future Galactic plane surveys with Roman (which would improve the precision of proper motion measurements) will also enable some of the additional science that the current design was not able to include.

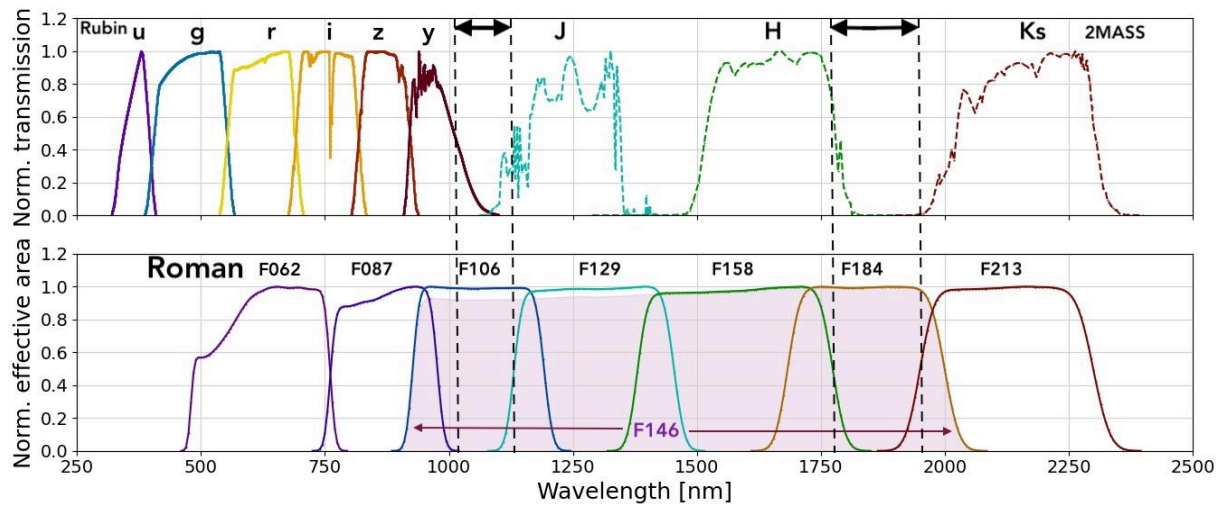

**Figure 3.11** Comparison of Rubin and 2MASS filter bandpasses (top panel) with the Roman bandpasses below; the F146 bandpass is indicated with a shaded region. With a combination of Rubin and the recommended four filters for the coverage of the Galactic disk (F129, F158, F184, and F213), there will be full wavelength coverage of the Galactic plane, with a region where the transmission drops below 50% between 1.05 and 1.15 microns. Note the overlap between ground-based *y* and Roman F106 filter and that the F184 filter covers a section of the IR spectral range unobservable from the ground.

### 3.4.1 The fourth filter: F106 or F184?

Based on community feedback, we were comfortable in recommending the F129 ("J"), F158 ("H") and F213 ("Ks") as the top three filters for the wide-field survey ([Fig 3.3](#), also see [Fig 3.11](#) above). But in early deliberations, both the DC and the community contributions were evenly split on the issue of the fourth most useful filter.

The F184 filter has several benefits. For example, it provides access to spectral features like Paschen alpha emission and water absorption. It is also more sensitive than F213 in regions of moderate to low extinction ($A_K$<1.7, see §C.1), resulting in more detected sources with better SNR. Finally, the F184 bandpass is largely inaccessible from the ground (Fig 3.11), potentially enabling new discoveries.

The F106 filter is another compelling option. The F106 filter provides a longer baseline wavelength coverage, allowing better measurement of sources' SEDs. It also provides a space-based, high angular resolution data set that would reasonably closely match the ground-based *y* filter observations of the LSST. Finally, it would fill the wavelength range 1.05 – 1.15 µm, where both the ground-based *y* and the space-based F129 have very low transmission (Fig 3.11) .

Preliminary simulations to explore which filter provided the most diagnostic power for different classes of sources were inconclusive. We were asked to provide a final recommendation on this issue, noting that a Change Board existed if further investigations provided definitive evidence on this question. A majority of the DC was persuaded by the arguments in favor of F184, but we would be grateful for further community investigation of the relative value of F106 vs F184 for different scientific investigations.

Note that by using F184, the combination of Rubin and Roman data will provide an almost full optical-to-IR coverage for billions of sources, with a low sensitivity area in the 1.05 – 1.15 µm range (Fig 3.11). The high-resolution Roman measurements can be used to deblend the Rubin images, enabling improved photometry across the optical through near-infrared. There will also be 23 square degrees of the sky covered in F106—in Serpens South, and the time-domain and deep/spectroscopic fields— which could motivate a future Galactic plane survey using this filter over a wider area.

### 3.4.2 Going beyond the Galactic plane

As shown in Figure 3.1, the community had a fairly liberal definition of "Galactic plane", with contributors telling us that over 36,000 deg$^2$ (compared to the planned ~700 deg$^2$) would be valuable for their science programs. The decision to make the wide-field survey contiguous made it necessary to exclude directions that were too far from the Galactic plane. This decision had a particular impact on the value of the RGPS for investigations of *(1)* the Solar System, *(2)* extra-galactic studies in the "Zone-of-Avoidance" ([SP12], [SP20]) , and *(3)* nearby star-forming regions, which can be significantly offset from the Galactic plane—for example, Orion, Taurus, and Perseus.

Regarding Solar System science, although the ecliptic plane is tilted by 63 degrees from the Galactic plane, it passes through the Galactic mid-plane near $l$=6º.5, a region where it can be difficult to find solar system objects due to stellar crowding. [WP05] suggested deep and repeated monitoring of a region in this direction to search for Kuiper Belt objects in the vicinity of the New Horizons spacecraft. Unfortunately, the position of this field at ($l,b$)= (17º.4, -14º.6) made these observations unfeasible for the RGPS. Additional solar system science was mentioned in §1.11.

The Serpens South star formation region—which includes W40 at almost the same distance as Orion, but with one-fifth star formation rate—presented the best opportunity to reconcile the significant community demand for data on nearby-star formation regions with the desire to fully cover the inner disk of stars and star formation.

### 3.4.3 Symmetric vs. asymmetric design

A symmetric design for the RGPS, extending to a Galactic longitude of approximately ±90º, would have been valuable for testing the symmetry of Galactic stellar and interstellar structure . It would have also included the impressive Cygnus-X star formation region ([Beerer et al 2010](#)) towards l=81º.5, which has long been a major element of most models of Milky Way spiral structure ([Bok 1937](#)). But this extended wide-field coverage would have had to come at the expense of time-domain investigations. Because of community demand, and because of the possibility of synergy with Rubin, we decided to extend the wide-field coverage as far as possible in the Carina (l=280º) direction, but not the Cygnus direction. An additional factor supporting this decision was the fact that a northern extension towards Cygnus would have required either an expanded latitude range or an offset of the midplane, both because the distant Galaxy warps upwards in this direction ([Levine et al 2006](#)), and the nearby Cygnus X region extends to high latitude. Modern 3D dust maps [(Edenhofer et al 2024)](#) demonstrate that Cygnus-X is part of an extended local dust structure, the Radcliffe Wave ([Alves et al 2020](#)), containing many of the star formation regions of the "Gould's Belt" that subtends a huge fraction of the Galactic plane ($l$=75º to 225º, $b$=-30º to +25º), almost twelve times larger than the RGPS area.

### 3.4.4 Concerns about dithering strategies

One of the principal goals of the RGPS is to allow for the proper motion measurements for as many sources as possible. This requires short integration times and a minimal gap-filling/dithering strategy. We chose a two-dither pattern in order to fill most of the gaps between the Roman detectors, while not spending too much time reobserving the same patch of sky. An ideal two-dither pattern would reduce the area covered by zero exposures as much as possible while introducing little overhead and covering most of the sky one or two times. We currently favor a specific implementation of the LINEGAP2 pattern, LINEGAP2_5, which will cover each part of the survey region thrice (3.7%), twice (67.2%), once (27.9%), or no times (1.2%). The variation in coverage makes for a non-uniform data product, but the inclusion of selection functions in modern analyses has become routine, and since most programs are statistical in nature, the lack of any coverage for 1.2% of the survey area (which will occur in regions around very bright stars anyway) will not compromise any programs. Potentially more serious is the 27.9% of the sky that is only covered once. This could be compromised by persistence, scattered light, cosmic ray removal, and other artifacts. No significant issues with these single-pass observations are expected—and experience with JWST supports this —but it is a concern and is discussed further in §5.

### 3.4.5 How deep is deep enough?

Given the sensitivity of Roman, it was generally agreed that the ~60 second exposure time was satisfactory for most science cases, and necessary to cover a large section of the sky. However, increasing the depth of the survey would benefit some science cases—allowing, for example, higher S/N observations of the MSTO in the bulge [SP36]. The bluer bands in particular would benefit from longer exposures, where high extinction limits their value in much of the inner Galaxy. But going four times deeper for the full widefield survey by increasing the number of visits per sky position from two to eight would cut down the survey area to roughly 170 deg$^2$, smaller than the bulge section of the current program! The scientific value of future deep observations of the Galactic plane, when more time will be available for General Astrophysics Survey programs, can be tested with the data obtained in the time-domain and deep/spectroscopic fields for the current RGPS, as well as with the GBTDS data.

### 3.4.6 The ideal cadence range (and targets) for time-domain science

We agreed early on to bracket the fraction of the RGPS devoted to time-domain science between 15% and 35% of the total time, ending up with a recommendation of 18.5% of the total time, which breaks down into 12.7% monitoring and 5.8% supporting imaging. In terms of choosing the fields, we principally looked for community-requested regions that could satisfy multiple science goals (§3.2.4). Initially, the committee focused on high cadence observations. As a result of the GBTDS development, Roman is specifically well-equipped to carry out these observations, and even short eight hour monitoring programs in high density fields (but without atmospheric blurring) are expected to produce a high return fraction of sources that vary on these time scales.

One field that was initially considered was a *Baade's window* field containing a pair of globular clusters: NGC 6522 and NGC 6528. The goal of this field was to look at the differences in the nature of variable sources in field populations vs. a dense stellar environment. Ultimately, this field was dropped in favor of doubling the coverage of NSD/CMZ, which had a higher overall source density and would benefit more from Roman's infrared capabilities. However, the *l=30º direction (including W43)* also contains the enigmatic GLIMPSE C01 (Kobulnicky et al 2005) which is either a globular cluster or a very high density young massive cluster. This object has been observed with Chandra and in the UV, optical and near-IR with HST (Hare et al 2018). The Roman re-observations will thus allow for the measurement of long-term variability and proper motion as well as the short time-scale monitoring in F184.

During the Workshop, there was discussion of using an unbiased logarithmically increasing time-separation from 10 minutes to two years, but this produced another trade-off between the number of visits per area versus the number of areas visited. (Also, only a small longitude range, $l$=280º to 295º, of the Galactic plane in the RGPS is continuously observable; see Fig 5.1) In the end, only the Galactic Center time-domain fields were recommended for different sets of cadences, with time scales of minutes, hours, and weeks. The combination of Rubin observations, the GBTDS, and these pilot observations will be valuable in defining future Galactic time-domain science programs.

### 3.4.7 How often to change filters?

During the Workshop, some community members expressed concern that the lack of contemporaneous color information would compromise the identification and characterization of variable sources, advocating for frequent filter changes so that observations in different filters would be taken as close together in time as possible. Frequent filter changes would both reduce the mapping speed, and is not advisable for instrument reasons. The re-observations of the time-domain fields in both F129 and F213 (except Serpens South which will only be re-imaged in F213) will help to quantify this concern. We did not explore the cost of breaking the wide-field survey into smaller chunks with more frequent filter changes.

### 3.4.8 Going beyond two years

By design, the call for proposals for Roman early definition science was limited to a two year program of observations, but as noted in §3.3, even a two-year program could yield scientifically valuable proper motions. We debated whether to support a program that goes beyond the two year window, deciding that the timely completion of the survey outweighed the advantage of going an extra six months to a year. The real advantage of these initial observations could come a decade or two down the line when the uncertainties in the proper motions are greatly reduced. However, as discussed in §5.1, there is real benefit to separating the two RGPS epochs as far as possible in time ideally by at least two years. That said, we agreed that a time separation of 1.5 years or more would provide a long enough baseline for most applications.

# 4. Survey Design and Technical Implementation

The recommended RGPS consists of **(1)** 691 deg$^2$ of wide-field (WF) imaging with all areas mapped in F129, F158, and F213, with F184 coverage added for the Galactic plane and Serpens South, and F106 added for Serpens South only (77% of time), **(2)** six time-domain science (TDS) regions totalling 19°.1 deg$^2$ that will be monitored in one filter and observed at least once in the remaining filters (19% of time), and **(3)** fifteen individual fields (Deep/Spec) totalling 4.2 deg$^2$ with deep (4x nominal) observations in all filters along with prism and grism spectroscopic observations (4% of time). A sky view and Galactic face-on view is given in Figure 2.2. Table 4.1 provides a breakdown of the full recommended RGPS program, including the total number of hours (and percentage of the total time), the area covered, and the bounding box of the regions covered for WF or TDS fields, or the pointing center for the Deep/Spec observations.

| Table 4.1 Overview of the elements of the Roman Galactic Plane Survey | | | | | | | | | |
|---|---|---|---|---|---|---|---|---|---|
| | Hours | Percent | Area(deg$^2$) | $l_{min}$ | $l_{max}$ | $b_{min}$ | $b_{max}$ | $l_{cen}$ | $b_{cen}$ | Filters/Cadence* |
| **Wide-Field Science** | 540.65 | 77.2% | 691.15 | | | | | | | |
| Disk | 382.18 | 54.6% | 468.40 | -67.00 | 50.10 | -2.00 | 2.00 | — | — | **F129**,F158,F184,**F213** |
| Disk_Carina | 44.43 | 6.3% | 54.00 | -79.00 | -67.00 | -2.50 | 2.00 | — | — | **F129**,F158,F184,**F213** |
| Bulge_Bpos | 52.27 | 7.5% | 80.00 | -10.00 | 10.00 | 2.00 | 6.00 | — | — | **F129**,F158,**F213** |
| Bulge_Bneg | 51.94 | 7.4% | 80.00 | -10.00 | 10.00 | -6.00 | -2.00 | — | — | **F129**,F158,**F213** |
| Serpens_South | 9.83 | 1.4% | 8.75 | 26.50 | 30.00 | 2.00 | 4.50 | — | — | F106,**F129**,F158,F184,**F213** |
| **Time-Domain Science** | 129.52 | 18.5% | 19.06 | | | | | | | |
| TDS_Carina | 11.89 | 1.7% | 2.06 | -73.77 | -71.19 | -1.05 | -0.25 | — | — | F213 (min)+F062,F087,F106,F129 |
| TDS_NGC6334_6357 | 11.89 | 1.7% | 2.06 | -9.15 | -6.57 | 0.35 | 1.15 | — | — | F213 (min)+F062,F087,F106,F129 |
| TDS_Galactic_Center_Q4 | 39.04 | 5.6% | 2.06 | -2.80 | -0.22 | -0.53 | 0.28 | — | — | F213 (min/hrs/wks)+F129 (wks) + F062,F087,F106,F129 |
| TDS_Galactic_Center_Q1 | 39.04 | 5.6% | 2.06 | 0.06 | 2.64 | -0.53 | 0.28 | — | — | F213 (min/hrs/wks)+F129 (wks) + F062,F087,F106,F129 |
| TDS_Serpens_South_W40 | 14.86 | 2.1% | 8.75 | 26.50 | 30.00 | 2.00 | 4.50 | — | — | F213 (hrs) (+F106,F129,F158,F184,F213 from Wide Field Science) |
| TDS_W43 | 12.80 | 1.8% | 2.06 | 29.31 | 31.89 | -0.49 | 0.31 | — | — | F184 (min)+ F062,F087,F106,F129,F213 |
| **Deep/Spectroscopic Science**¶ | 30.80 | 4.4% | 4.22 | | | | | | | |
| Deep_NGC_3324_Carina | 1.79 | 0.3% | 0.28 | — | — | — | — | -73.80 | -0.20 | All filters deep, grism,prism |
| Deep_Acrux | 1.79 | 0.3% | 0.28 | — | — | — | — | -59.75 | -0.29 | All filters deep, grism,prism |
| Deep_NGC_5269_5281 | 1.79 | 0.3% | 0.28 | — | — | — | — | -50.86 | -0.55 | All filters deep, grism,prism |
| Deep_Window_319.5-0.2 | 1.79 | 0.3% | 0.28 | — | — | — | — | -40.50 | -0.21 | All filters deep, grism,prism |
| Deep_G333 | 1.79 | 0.3% | 0.28 | — | — | — | — | -27.00 | -0.48 | All filters deep, grism,prism |
| Deep_ASCC_85 | 1.79 | 0.3% | 0.28 | — | — | — | — | -20.17 | -0.32 | All filters deep, grism,prism |
| Deep_Teutsch_84 | 1.79 | 0.3% | 0.28 | — | — | — | — | -15.56 | -0.46 | All filters deep, grism,prism |
| Deep_NGC_6357_Lobster | 1.79 | 0.3% | 0.28 | — | — | — | — | -6.86 | 0.85 | All filters deep, grism,prism |
| Deep_Window_355.0-0.3 | 1.79 | 0.3% | 0.28 | — | — | — | — | -5.21 | -0.37 | All filters deep, grism,prism |
| Deep_VVV-CL001_UKS_1 | 1.79 | 0.3% | 0.28 | — | — | — | — | 5.20 | 0.77 | All filters deep, grism,prism |
| Deep_M17_Omega | 1.79 | 0.3% | 0.28 | — | — | — | — | 15.04 | -0.73 | All filters deep, grism,prism |
| Deep_Trumpler_35 | 1.79 | 0.3% | 0.28 | — | — | — | — | 28.21 | -0.06 | All filters deep, grism,prism |
| Deep_W40 | 5.74 | 0.8% | 0.28 | — | — | — | — | 28.83 | 3.54 | Very Deep F129,F158,F213, prism |
| Deep_W44 | 1.79 | 0.3% | 0.28 | — | — | — | — | 34.70 | -0.40 | All filters deep, grism,prism |
| Deep_W51 | 1.79 | 0.3% | 0.28 | — | — | — | — | 49.40 | -0.20 | All filters deep, grism,prism |
| **TOTAL** | 700.97 | 100.1% | | | | | | | | |

\* "All Filters" excludes the wide F146 filter, grism, and prism. *Wide-Field Science*: Filters in bold recommended for first epoch observations.
  *Time Domain Science*: Fields covered with all filters to "standard" observing time and single filter for repeat visits.
  Note that all Time Domain fields will have F129 and F213 observations separated by > 1 year to assess variability.
¶ Single Roman pointings with x4 standard observing time (0.75 mag deeper). If grism/prism observations are not feasible, 0.4 hrs per field will be
  allocated to longer integrations. Pointing centers may be adjusted depending on roll angle for scheduled observations.

In this section, we provide an overview of the observing strategies (§4.1), more details of the wide-field mapping (§4.2), the TDS regions (§4.3), the Deep/Spec targets (§4.4), and estimates for the sensitivity, saturation, and uncertainties in the proper motion measurements (§4.5).

## 4.1 Observing strategies

Different elements of the programs will use different readout patterns (MA Tables) and dithering patterns. The observing plan described here was generated using APT version 2025.4.1 A summary table for all of these program elements is given in Table 4.2.

*Wide-field:* Most of the RGPS will be in a wide / fast mode. We recommend two 60 second exposures per pointing using MA table `IM_60_6_S` with 6 resultants and $t_{exp}$=60.09 s, and a LINEGAP2_5 dither pattern with a slew along the WFI short axis to the next field. Using slew times for the dither and offset to the next field[12], the mapping speed can be estimated as (0.8 x 0.4 deg$^2$) / (2 (60 s) + 18.9 s + 39.3 s), or 6.5 deg$^2$/hr. Using APT yields a mapping speed of 4.9 deg$^2$/hr per filter. The MA table enables a large dynamic range by beginning with an initial single read resultant followed by a

**Table 4.2 Dithering patterns [1]/Observing strategies**

| | F062 | F087 | F106 | F129 [2] | F158 [2] | F184 [2] | F213 [2] | Grism | Prism | High Cadence [3] | Hourly Cadence [4] | Weekly Cadence [5] | Roll-angle constraints [6] |
|---|---|---|---|---|---|---|---|---|---|---|---|---|---|
| **Wide-Field Science** | | | | | | | | | | | | | |
| Disk | — | — | — | LG2 | LG2 | LG2 | LG2 | — | — | — | — | — | — |
| Disk_Carina | — | — | — | LG2 | LG2 | LG2 | LG2 | — | — | — | — | — | — |
| Bulge_Bpos | — | — | — | LG2 | LG2 | LG2 | LG2 | — | — | — | — | — | — |
| Bulge_Bneg | — | — | — | LG2 | LG2 | LG2 | LG2 | — | — | — | — | — | — |
| Serpens_South | — | — | LG2 | LG2 | LG2 | LG2 | LG2 | — | — | — | — | — | — |
| **Time-Domain Science** | | | | | | | | | | | | | |
| TDS_Carina | LG2 | LG2 | LG2 | LG2 | — | — | → | — | — | F213 | — | — | N |
| TDS_NGC6334_6357 | LG2 | LG2 | LG2 | LG2 | — | — | → | — | — | F213 | — | — | Y |
| TDS_Galactic_Center_Q4 | BG8 | BG8 | BG8 | BG8 | — | — | → | — | — | F213 | F213 | F129,F213 | Y |
| TDS_Galactic_Center_Q1 | BG8 | BG8 | BG8 | BG8 | — | — | → | — | — | F213 | F213 | F129,F213 | Y |
| TDS_Serpens_South_W40 | — | — | — | — | — | — | → | — | — | — | F213 | — | N |
| TDS_W43 | LG2 | LG2 | LG2 | LG2 | — | → | LG2 | — | — | F184 | — | — | N |
| **Deep/Spectroscopic Science** | | | | | | | | | | | | | |
| Deep_NGC_3324_Carina | BG8 | BG8 | BG8 | BG8 | BG8 | BG8 | BG8 | * | * | — | — | — | Y |
| Deep_Acrux | BG8 | BG8 | BG8 | BG8 | BG8 | BG8 | BG8 | * | * | — | — | — | Y |
| Deep_NGC_5269_5281 | BG8 | BG8 | BG8 | BG8 | BG8 | BG8 | BG8 | * | * | — | — | — | Y |
| Deep_Window_319.5-0.2 | BG8 | BG8 | BG8 | BG8 | BG8 | BG8 | BG8 | * | * | — | — | — | Y |
| Deep_G333 | BG8 | BG8 | BG8 | BG8 | BG8 | BG8 | BG8 | * | * | — | — | — | Y |
| Deep_ASCC_85 | BG8 | BG8 | BG8 | BG8 | BG8 | BG8 | BG8 | * | * | — | — | — | Y |
| Deep_Teutsch_84 | BG8 | BG8 | BG8 | BG8 | BG8 | BG8 | BG8 | * | * | — | — | — | Y |
| Deep_NGC_6357_Lobster | BG8 | BG8 | BG8 | BG8 | BG8 | BG8 | BG8 | * | * | — | — | — | Y |
| Deep_Window_355.0-0.3 | BG8 | BG8 | BG8 | BG8 | BG8 | BG8 | BG8 | * | * | — | — | — | Y |
| Deep_VVV-CL001_UKS_1 | BG8 | BG8 | BG8 | BG8 | BG8 | BG8 | BG8 | * | * | — | — | — | Y |
| Deep_M17_Omega | BG8 | BG8 | BG8 | BG8 | BG8 | BG8 | BG8 | * | * | — | — | — | Y |
| Deep_Trumpler_35 | BG8 | BG8 | BG8 | BG8 | BG8 | BG8 | BG8 | * | * | — | — | — | Y |
| Deep_W40 | — | — | — | BG4 | BG4 | — | BG4 | — | ** | — | — | — | Y |
| Deep_W44 | BG8 | BG8 | BG8 | BG8 | BG8 | BG8 | BG8 | * | * | — | — | — | Y |
| Deep_W51 | BG8 | BG8 | BG8 | BG8 | BG8 | BG8 | BG8 | * | * | — | — | — | Y |

(1) Dithering patterns LG2=LINEGAP2_5 with 60s per exposure, BG4=BOXGAP4 with 977s per exposure, BG8=BOXGAP8_2 with 60s per exposures
  *=298 sec exposures at two different roll angles, 9.93 min total integration, **A set of four 997 sec exposures at two different roll angles, 133 min total integration
(2) **Boldface**=as early as possible in 2-yr window, **Red**=as late as possible in 2yr window. TDS fields will be either re-imaged or monitored in selected filters.
(3) "High cadence"=43 visits @ Δt=11.3 min [Duration: 0.36 days]
(4) "Hourly cadence"= 8 visits @ Δt=4, 6, 8, 10, 12,14, 16 hrs [Duration: 2.92 days]
(5) "Weekly cadence"= 8 visits @ Δt=1 week [Duration: 56 days]
(6) See text for further discussion

two-read resultant, so that bright stars can use an effective exposure time of only a few seconds. The last read will also be downloaded as a single resultant without on-board averaging to provide the best sensitivity to cosmic rays and to obtain the best S/N for pixels with high count rates (Sharma & Casertano 2025). Both the MA tables and dithering patterns are under active development; we will continue to examine which options would provide the best data quality for the RGPS. The fact that approximately 28% of the sky would be covered only once per filter, is discussed further §5.3.

---

[12] https://roman-docs.stsci.edu/roman-instruments-home/wfi-imaging-mode-user-guide/introduction-to-the-wfi/wfi-quick-reference

*Time-domain fields*:  The regions selected for time-domain science investigation will be imaged in all of the Roman filters but the exposure times, dithering strategy, and monitoring cadence will depend on the field and filter;  this is summarized in Table 4.2 and explained further in §4.3.

*Deep/Spec pointings*:   With the exception of the W40 pointing, all the regions recommended for deep imaging and grism/prism spectroscopy will be observed uniformly with the same MA table, number of resultants, and exposure time as above. However, these observations would use the BOXGAP8 dithering pattern, yielding (on average) a four-fold increase in integration time and going about 0.75 magnitudes deeper than the wide-field survey. This strategy would be used for all of the filters: F062, F087, F106, F129, F158, F184, and F213 filters, leaving out only the wide filter (W146).

For spectroscopy in the same fields, we recommend the use of MA table `SP_300_16` (Resultant=16, $t_{exp}$=289.19 s ) for both the prism and grism, to be repeated at a different roll angle. The two spectroscopic exposures should be performed at a modestly different roll angle, so that different sources' spectra overlap in the different images.  This roll can be rather small; e.g., a roll of 6 degrees would separate 95% of the sources overlapping in prism images and 98% in grism images. This will yield a total of 10 minutes of data for the grism and for the prism in fields with different stellar densities, extinction, and diffuse emission to explore the value of the prism and grism for science in the Galactic plane. Using the most recent version of APT, the combination of imaging and spectroscopy would take 1.79 hours (or 0.3% of the total program) for each of the fourteen recommended fields, totaling 25.06 hours (3.6%).

The W40 pointing, intended to test the value of very deep photometric and spectroscopic observations in a nearby star formation region, will also be a single pointing but will only use three filters F129, F158, F213 to obtain 66.7 min exposures using MA table `IM_1000_16`   (16 resultant, $t_{exp}$=999.34 s), with dithering pattern BOXGAP4. The prism spectroscopy will use MA table `SP_1000_16`  (16 resultants, $t_{exp}$=1002.53 s), with dithering pattern BOXGAP4.

## 4.2 Wide-Field Science programs

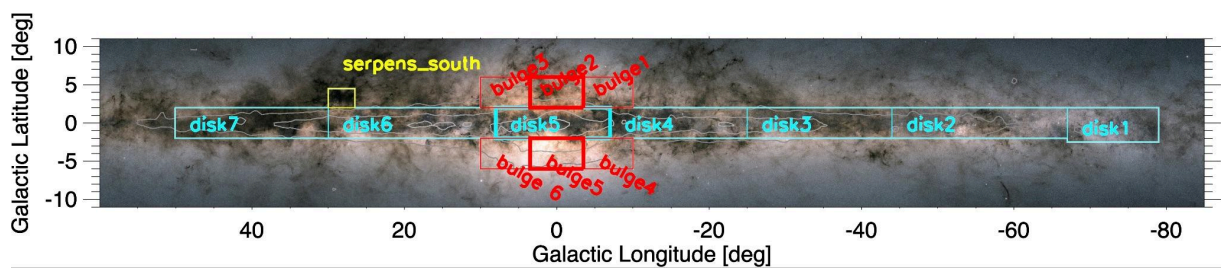

**Figure 4.1**: Observing segments for wide-field observations. Disk 1/Carina ($l$=-79º to -67º),  Disk 2 ($l$=-67º to -44º), Disk 3 ($l$=-44º to -25º), Disk 4 ($l$=- 25º to -7º), Disk 5 ($l$=-7º to +8º),  Disk 6 ($l$=+8º to +30º), and Disk 7 ($l$=+30º to +50.1º). All segments have a latitude width of ±2º except Disk 1/Carina with b=-2º.5 to 2º.0. In the second pass, bulge 1,2, 3=bulge 7  and bulge 4, 5, 6=bulge 8. Two-filter coverage of Disk5+bulge2+bulge 5 (thick lines) can be obtained in a two-day window during the first GBTDS observing season; the positive longitude segments can be covered in four days, and the negative longitude segments in six days.

A total of 541 hours, or 77.2% of the total time,  is recommended for wide-field coverage using three filters—F129, F158, and F213—for the bulge regions, four filters—F129, F158, F184, and F213— for the disk, including an extension in the negative latitude direction, and five filters—F106, F129, F158, F184, and F213—for the nearby star forming region Serpens South and environs. The longitude and latitude extent of each of these regions is defined in Table 4.1. For scheduling, the Disk/Disk_Carina fields are broken into seven segments; see Figure 4.1. In order to be able to measure (cross-band) proper motions, we recommend observing F129/F213 as early as possible during the two year program and the remaining filters as late as possible. The filter pair for the first pass provides a "J-K" color for sources early in the mission, which should provide diagnostic power for source classification early on in the program.

## 4.3 Time-Domain Science fields

The recommended time-domain science element of the RGPS consists of monitoring six different fields and obtaining supporting imaging in all of the filters except the wide filter (and the grism and prism). Table 4.3 provides a summary of the six areas, boundaries, monitoring time and imaging time, filters to be observed. It also includes some notes on targets of interest in each of the fields. With an area of 2.1 deg$^2$ per region, the targets do not completely fill the area surveyed, so that the time-domain science fields will have both *on*-target and *off*-target regions and will probe foreground/background populations. The time-domain science fields break down into three groups, which we describe in turn.

| Fields | Total Time (hrs) | Percent (of total) | Area (deg$^2$) | Monitoring Time (hrs) | Imaging Time (hrs) | $l_{min}$ | $l_{max}$ | $b_{min}$ | $b_{max}$ | Filters/Cadence(s)* | Synergies | Notes |
|---|---|---|---|---|---|---|---|---|---|---|---|---|
| Carina | 11.9 | 1.7% | 2.1 | 8.3 | 3.6 | -73.8 | -71.2 | -1.0 | -0.2 | **F213**: 11.3 min cadence (43 visits over 8.0 hours) + Imaging in F062, F087, F106, F129 | Rubin | Highly requested target (d=2350±150 pc), heavily studied with HST. High cadence observations probe short-term accretion in YSOs, e.g. [SP48] |
| NGC6334_6357 | 11.9 | 1.7% | 2.1 | 8.3 | 3.6 | -9.2 | -6.6 | 0.3 | 1.1 | **F213**: 11.3 min cadence (43 visits over 8.0 hours) + Imaging in F062, F087, F106, F129 | Rubin | Two heavily studied star formation regions (d=1780±120 kpc) with several YSO groups projected against the high density background of the bulge/bar. High cadence observations probe short-term accretion in YSOs and foreground/background compact binaries [WP03] and other high cadence sources. |
| CMZ_NSD_Q4 | 39.0 | 5.6% | 2.1 | 24.6 | 14.4 | -2.8 | -0.2 | -0.5 | 0.3 | **F213**: 11.3 min cadence (43 visits over 8.0 hours) + Increasing hourly visits**(8 visits over 3 days) +Weekly visits over 2 mo (with both **F213** and **F129**) +Imaging in F062, F087, F106, F129 | Rubin | Two sets of six Roman pointings flanking the GBTDS Galactic Center field in the direction of the highest stellar density of the entire GPS. These fields will cover most of the Nuclear Stellar Disk and star-forming Central Molecular Zone and environs [WP04], and the cadences span 11 minutes to weeks, along with reobservations separated by >1 year. See text for more details. |
| CMZ_NSD_Q1 | 39.0 | 5.6% | 2.1 | 24.6 | 14.4 | 0.1 | 2.6 | -0.5 | 0.3 | | | |
| Serpens_South_W40 | 14.9 | 2.1% | 8.8 | 14.9 | — | 26.5 | 30.0 | 2.0 | 4.5 | **F213**: Increasing hourly visits**+ (8 visits over 3 days) | — | One of the closest star formation regions to the Sun containing O stars (d=436±9 pc) and the closest such region to the Galactic plane, the increasing hourly cadence will extend previous Spitzer/YSOvar characterization of YSO and stellar variability. |
| W43 | 12.8 | 1.8% | 2.1 | 8.3 | 4.5 | 29.3 | 31.9 | -0.5 | 0.3 | **F184**: 11.3 min cadence (43 visits over 8.0 hours) + Imaging in F062, F087, F106, F129, F213 | Subaru GPS | With both high stellar and star formation density, this field overlaps with the planned Subaru GPS. The W43 star formation (d=5490±390 pc) (which fills a single SCA) has the highest measured star formation rate in the Galaxy (SFR=287x Orion), and the full field covers the intersection of the end of the bar with the "Scutum Spiral Arm". High cadence observations are expected to yield a population of faint cataclysmic variables, using the accretion dependent Pa-alpha line in the F184 filter, and constrain Galactic diffuse X-ray emission. |
| **Time Domain Science TOTAL** | 129.5 | 18.5% | 19.1 | 89.0 | 40.6 | | | | | | | |

<div align="center">Table 4.3 Overview of Roman GPS Time Domain Science Fields</div>

\* All fields also be observed in F129 and F213 twice with Δt > 1 yr. Note that "all filters" excludes the wide F146 filter, the grism and prism.
\*\* Increasing spacing for hourly visits is Δt=+4, +6, +8, +10, +12,+14, +16 hours

### 4.3.1 The Central Molecular Zone/Nuclear Stellar Disk fields

The most extensive TDS observations will be of the two central Galaxy fields covering the Central Molecular Zone and Nuclear Stellar Disk, CMZ_NSD_Q1, and CMZ_NSD_Q4, which contain six Roman pointings for a total of 2.1 deg$^2$ each on either side of the GBTDS Galactic center pointing. (Figure 4.2). These will be observed in the fifth bulge season (Feb-Apr 2029) or whenever the wide-field mapping in F158 and F184 takes place and their positions may be modified to be consistent with the GBTDS plans. The relatively small size, high-extinction, and dense stellar environment make this region a natural fit to the capabilities of the Roman telescope.

These two regions will be imaged in F129 and F213 early in the RGPS as part of the standard wide-field program. In the fifth bulge season, it will be monitored in F213 with a high (11.3 min) cadence over 8.05 hrs followed by eight "hourly" observations with an increasing time interval (Δt=4, 6, 8, 10, 12, 14, and 16 hrs) for a total of 2.92 days. It will also be monitored in both F129 and F213 with a weekly cadence (with a tolerance ±2 days). Depending upon the timing of other observations during this bulge season, this weekly monitoring may take place either before or after the high-cadence and hourly observations. Finally, we recommend reobserving these fields in two filters between year one and year two in order to constrain variability on longer time scales.

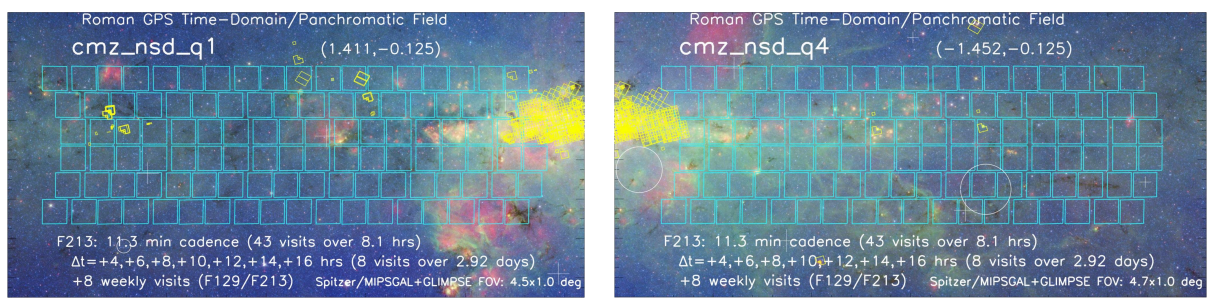

**Figure 4.2** The two Central Molecular Zone/Nuclear Stellar Disk Time Domain Science fields. This (and subsequent figures) show Roman pointings superimposed on either a *Spitzer*/MIPSGAL+ GLIMPSE or *Spitzer*/GLIMPSE-only image of the field, which allow one to see the direction of star formation regions. [Red=emission from warm dust at 22 μm, green=emission from PAHs at 8 μm, blue=stellar emission at 3.6 μm.] Regions previously observed by the Hubble Space Telescope are shown in yellow; the positions and radii ($R_{50}$) of Gaia-characterized stellar clusters (Hunt & Reffert 2023) are shown with white circles and the location of mid-IR extinction windows (Saito et al 2020; Zhang & Kainulainen 2022) are shown with red circles. The size of the crosses is proportional to the number of science cases requesting observations in a given direction. These observations will be matched in roll angle to the GBTDS Galactic center field. See Table 4.2 and text to determine whether the Roman observations have any roll angle constraints.

Because of the exceptionally high extinction in this direction, these regions will be imaged using the BOXGAP8 dither pattern in F062, F087, F106, and F129, with four times the "standard" depth of the wide-field survey. *When combined with the monitoring in F129 and F213, not only will this region be the most extensively surveyed in time domain cadences, it will also have the deepest photometric coverage of the RGPS.*

### 4.3.2 High cadence fields: *l*=30º/W43, NGC 6334/57, and Carina

Three directions in the first quadrant disk (*l=30º/W43*), the bulge direction (*NGC 6334/57*), and the fourth quadrant disk (*Carina*) were selected from community input for high cadence observations with a strategy almost identical to what will be used for the CMZ/NSD (Figure 4.3).

This will allow a comparison of the yields in these different environments with variables detected in both the star formation regions and the foreground/background stellar disk/bulge. Of these three fields, the *l*=30º direction (which contains W43 in a single SCA) will have one small modification to the strategy described above, using F184 for monitoring rather than F213. This field will overlap with the Subaru Galactic plane survey and it was argued that the F184 may be more sensitive to stellar accretion due to flaring in the Paschen alpha line. These observations will test that proposition.

In terms of imaging, these fields will be imaged at "standard" depth in F062, F087, and F106. Since these fields will be imaged in F158 and F184 near the same time as part of the wide-field mapping, they will be only be re-observed in the F129 (all three fields) and F213 (W43 field only) filters. All imaging will be done using LINEGAP2 to the standard depth.

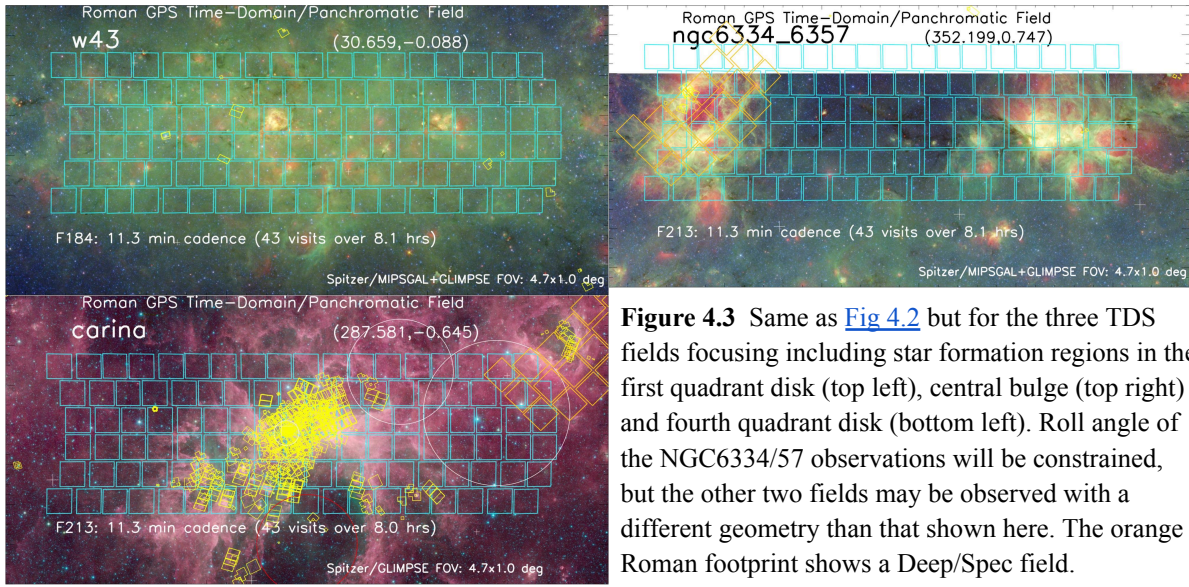

**Figure 4.3**  Same as Fig 4.2 but for the three TDS fields focusing including star formation regions in the first quadrant disk (top left), central bulge (top right) and fourth quadrant disk (bottom left). Roll angle of the NGC6334/57 observations will be constrained, but the other two fields may be observed with a different geometry than that shown here. The orange Roman footprint shows a Deep/Spec field.

### 4.3.3. Serpens South and W40

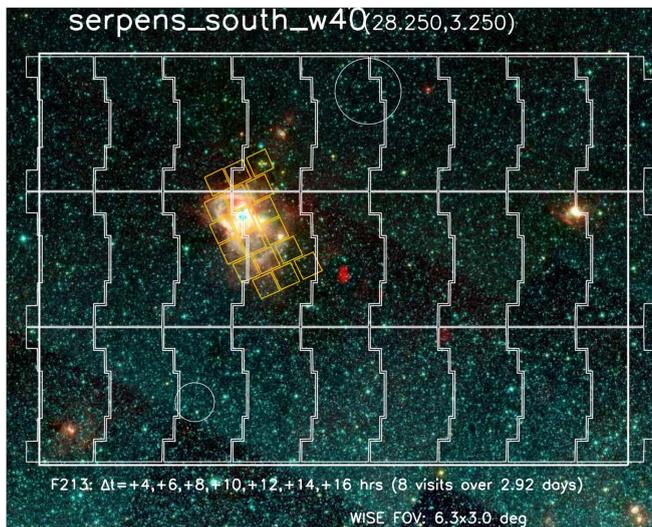

Unlike the previous five time-domain science programs which have identical footprints and always include high-cadence monitoring, the Serpens South field is larger (8.8 deg$^2$), will have only hourly cadence (with increasing time interval) and will not have any imaging beyond what is done for the wide-field survey.

**Figure 4.4**  Same as Fig 4.2 but for the Serpens South star forming region containing the W40 star formation region and shown against a WISE background. The Roman footprints are for scale; observed roll angle is likely to be different.

Note that this field is unique for its larger area F106 coverage, in order to better constrain the SEDs of YSO populations in this nearby star forming environment. This field will be revisited in F213 with eight additional visits spanning 2.92 days, with an increasing hourly time separation, principally to constrain YSO variability. Because of the larger area and depth, other classes of variable sources, e.g. stellar flaring, can be expected to be detected.

## 4.4 Deep Photometric/Spectroscopic Science programs

We recommend fifteen individual fields for deep multiband imaging and spectroscopy with the grism and prism. These fields were chosen using the data and procedures documented in §3.1 and, with the exception of the W40 direction, will be observed using the procedure given in §4.1. Two examples of these targets are shown in Fig 4.5, with a gallery showing the remaining images in Fig 4.6. Table 4.4 provides a summary of the directions, including the mean extinction, level of nebulosity, predicted stellar density, and "objects of interest" that lie along each sightline. These objects of interest include six fields with massive star formation regions, four fields containing one or more open clusters, two fields with IR extinction windows, a field with a large supernova remnant (W44), a field with a large stellar wind bowshock (Alpha Crux) plus the greatest midplane concentration of Sco-Cen members, and a field with two globular clusters.

One challenge of these observations is that very different regions of the sky can be covered depending on the roll angle of the observations. In Figure 4.5, the aperture (pointing) center is shown; the overlap in area between two observations with different roll angles can be quite low. As Figure 4.5 and 4.6 make clear, different directions on the sky have different arrangements of interesting features. With a random roll angle, some of these interesting features could be missed. On the other hand, prescribing the roll-angle of these fields makes them more difficult to schedule, and the principal goal of these observations is a technical demonstration. This is discussed further in §5.2.

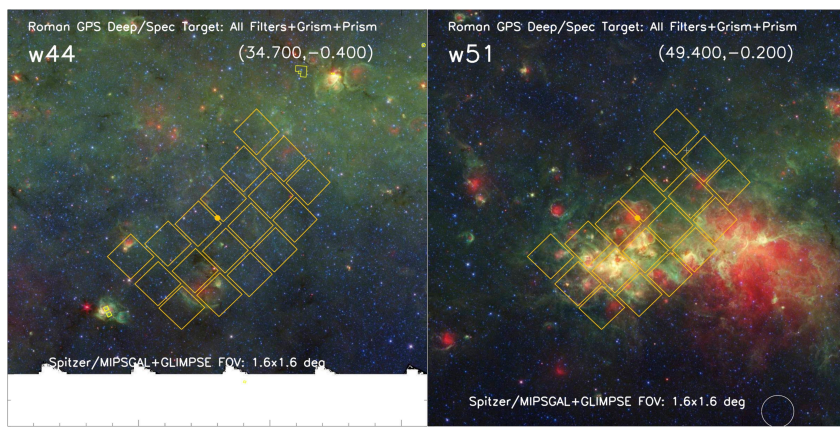

**Figure 4.5** Same as Fig 4.2, but showing single Roman pointings recommended for deep imaging in all filters plus grism and prism observations. The left panel containing a large angular supernova remnant (W44) and the right a complex of star formation regions (W51). The aperture center used for pointing is marked with an orange dot which is independent of the roll angle.

Regarding the W40 observation, this pointing is to test a different potential spectroscopic use of the Roman telescope, the identification of low-mass brown dwarfs and free-floating planets in nearby star formation regions (with supporting deep imaging). Because of the proximity (~ 440 pc) and youth (~1 Myr) of W40, 66.7 minute exposures attain the necessary photometric depth to detect a 1 Jupiter mass object, given evolutionary models, and characterize the end of the IMF. Deep prism observations at a close to 90 degree roll will be sensitive to ~ 3 Jupiter mass objects and allow for the potential disentanglement of confused spectra in this relatively dense star-forming region. The prism overlaps significant water absorption features, shown in Fig. 1.6, that can place strong constraints on the membership and effective temperature of detected sources, and at the very least identify candidates for follow-up. The prism is favored over the grism since 1) the water absorption features are broad and do not require the grism to detect and 2) grism spectra would extend over a larger range of the detector resulting in an increase of source confusion especially within a dense star-forming region.

## 4.5 Sensitivity, saturation, extinction, and proper motion uncertainties

*Sensitivity*: In this section, we estimate the precision that the RGPS will obtain as a function of magnitude in its wide-area program for a number of key parameters. Specifically, we focus on the signal to noise ratio, the photometric precision, and the proper motion accuracy. The signal-to-noise ratios as a function of magnitude come from Pandeia. We use the following options in Pandeia:

- The "minzodi" background model with a "medium" background level. This is optimistic as the background in many RGPS fields will be dominated by unresolved stars.
- A 0.2" radius aperture. This is somewhat pessimistic relative to PSF photometry.
- The baseline Pandeia 60 s MA table. This will be updated as the MA table definitions evolve but should make only a small impact.
- No uncertainty in the background estimation. This uncertainty is subdominant to other terms and should have little effect.

**Table 4.4 Overview of Roman GPS Deep/Spectroscopic Targets**

| Deep/Spectroscopic Targets | Time[1] (hrs) | Percent (of total) | Area (deg$^2$) | $l$ | $b$ | $\langle A_K \rangle$[2] | F158 Source Density[3] ($\log_{10} N/\text{deg}^2$) | Diffuse Emission (est) | Observing mode[4] | Field notes[6] |
|---|---|---|---|---|---|---|---|---|---|---|
| NGC_3324_Carina | 1.79 | 0.3% | 0.28 | -73.80 | -0.20 | 0.6 | 6.9 | High | All filters deep, grism,prism | Stellar cluster and star formation bubble (d=2419±18 pc); JWST Early Release imaging field. |
| Acrux | 1.79 | 0.3% | 0.28 | -59.75 | -0.29 | 0.9 | 6.9 | Med | All filters deep, grism,prism | Alpha Crux stellar wind bowshock and 15-20 known members of Sco Cen Association (d=105.5±2.1 pc) |
| NGC_5269_5281 | 1.79 | 0.3% | 0.28 | -50.86 | -0.55 | 1.2 | 6.9 | Low | All filters deep, grism,prism | Pair of open clusters at d=1930±7 pc (age=200 Myr) and d=1471±3 pc (100 Myr), respectively. |
| Window_319.5-0.2 | 1.79 | 0.3% | 0.28 | -40.50 | -0.21 | 0.9 | 7.2 | Low | All filters deep, grism,prism | VVV extinction window for studies of outer Galactic disk. |
| G333 | 1.79 | 0.3% | 0.28 | -27.00 | -0.48 | 1.2 | 7.3 | Med | All filters deep, grism,prism | Major MW star forming region (SFR=56xOrion) with both high extinction and diffuse emission in inner Galaxy at d=2500±700 pc |
| ASCC_85 | 1.79 | 0.3% | 0.28 | -20.17 | -0.32 | 1.3 | 7.3 | Low | All filters deep, grism,prism | Nearby stellar cluster (d=852±2 pc) of moderate age (120 Myr) |
| Teutsch_84 | 1.79 | 0.3% | 0.28 | -15.56 | -0.46 | 1.4 | 7.4 | Low | All filters deep, grism,prism | An unusually old (0.4 Gyr), massive (~5000 $M_{sun}$), optically reddened, inner Galaxy cluster at distance d=2400±30 pc. |
| NGC_6357_Lobster | 1.79 | 0.3% | 0.28 | -6.86 | 0.85 | 1.5 | 7.8 | Very High | All filters deep, grism,prism | Large angular size (~deg) heavily studied MW star formation (SFR=32xOrion) at d=1672±7 pc along a molecular filament that also contains NGC 6334. Also targeted for time domain science. |
| Window_355.0-0.3 | 1.79 | 0.3% | 0.28 | -5.21 | -0.37 | 0.6 | 8.2 | Low | All filters deep, grism,prism | VVV extinction window for inner Galactic disk and bulge (AK=0.60±0.08) with red clump excess detected at 14 kpc and enhanced micolensing events (Saito et al 2020) |
| VVV-CL001_UKS_1 | 1.79 | 0.3% | 0.28 | 5.20 | 0.77 | 0.8 | 8.2 | Low | All filters deep, grism,prism | Pair of high extinction globular clusters at very different distances d=15400±700 pc (UKS1) and d=8230±1900 pc (VVV CL01); VVV CL01 is one of the most metal poor globular clusters in the Milky Way, currently crossing through the disk. |
| M17_Omega | 1.79 | 0.3% | 0.28 | 15.04 | -0.73 | 1.6 | 7.4 | Very High | All filters deep, grism,prism | Heavily studied nearby (d=1600±100 pc) star formation region (SFR=37x Orion); part of a kiloparsec-long structure also containing M16 (Eagle), M8 (Lagoon), and M20 (Triffid). (Kuhn et al 2020). |
| Trumpler_35 | 1.79 | 0.3% | 0.28 | 28.21 | -0.06 | 1.9 | 7.3 | Low | All filters deep, grism,prism | Young (22 Myr) cluster at d=2674±19 pc |
| W40 | 5.74 | 0.8% | 0.28 | 28.83 | 3.54 | 1.0 | 6.9 | High | Very Deep F129,F158,F213, prism[5] | One of the nearest (d=436±9 pc) high-mass star forming regions to the Sun embedded in the dark clouds of the Aquila Rift; at a similar distance as Orion Nebula, but with 20% of the star formation rate. |
| W44 | 1.79 | 0.3% | 0.28 | 34.70 | -0.40 | 1.7 | 7.1 | Med | All filters deep, grism,prism | Large angular size supernova remnant and gamma-ray source at d=2600±700 pc (Wang et al 2020) with star formation extending for a kiloparec in depth along the line of sight (Reid et al 2019). |
| W51 | 1.79 | 0.3% | 0.28 | 49.40 | -0.20 | 1.4 | 7.0 | High | All filters deep, grism,prism | Complex of several star forming regions with SFR=167x Orion extending from 5100±200 pc to 6100±700 pc (Reid et al 2019). Long thought to be the tangency direction of Sagittarius "spiral arm". |
| **TOTAL** | **30.80** | **4.4%** | **4.22** | | | | | | | |

[1] All deep fields have 1.43 hours of imaging and 0.36 hours of grism/prism observation except W40 (5.64 hrs imaging, 2.27 hrs prism).
[2] $A_K$ from Zucker et al (2025) [$l$<−65°], Zhang & Kainulainen (2023) [−65°<$l$<−10°], Surot et al (2020) [−10° < $l$ < 10°], Nidever et al (2012) [$l$ > 10°].
  $A_K$ for W40 from Comeron et al (2022).
[3] Estimated source count (with x4 "standard" exposure time) using TRILEGAL plus exintction from above.
[4] Single Roman pointings with 4x "standard exposure time (0.75 mag deeper) in all filters using BOXGAP8. Grism and prism exposures at two roll angles for 300 sec each.
[5] Two 30 min exposures in F129, F158, F213 only, with 60 min observations with prism.
[6] Star formation rates from Table 1 of Binder & Povich (2020). Cluster ages and distances from Hunt & Reffert (2023) except Acrux/ScoCen (Ratzenböck et al 2023),
  G333 (Ramírez-Tannus et al 2020), VVV CL001 (Fernández-Trincado et al 2021), UKS 1 (Alonso-Garcia et al 2025), M17 (Kuhn et al 2020), W51 (Reid et al 2019).

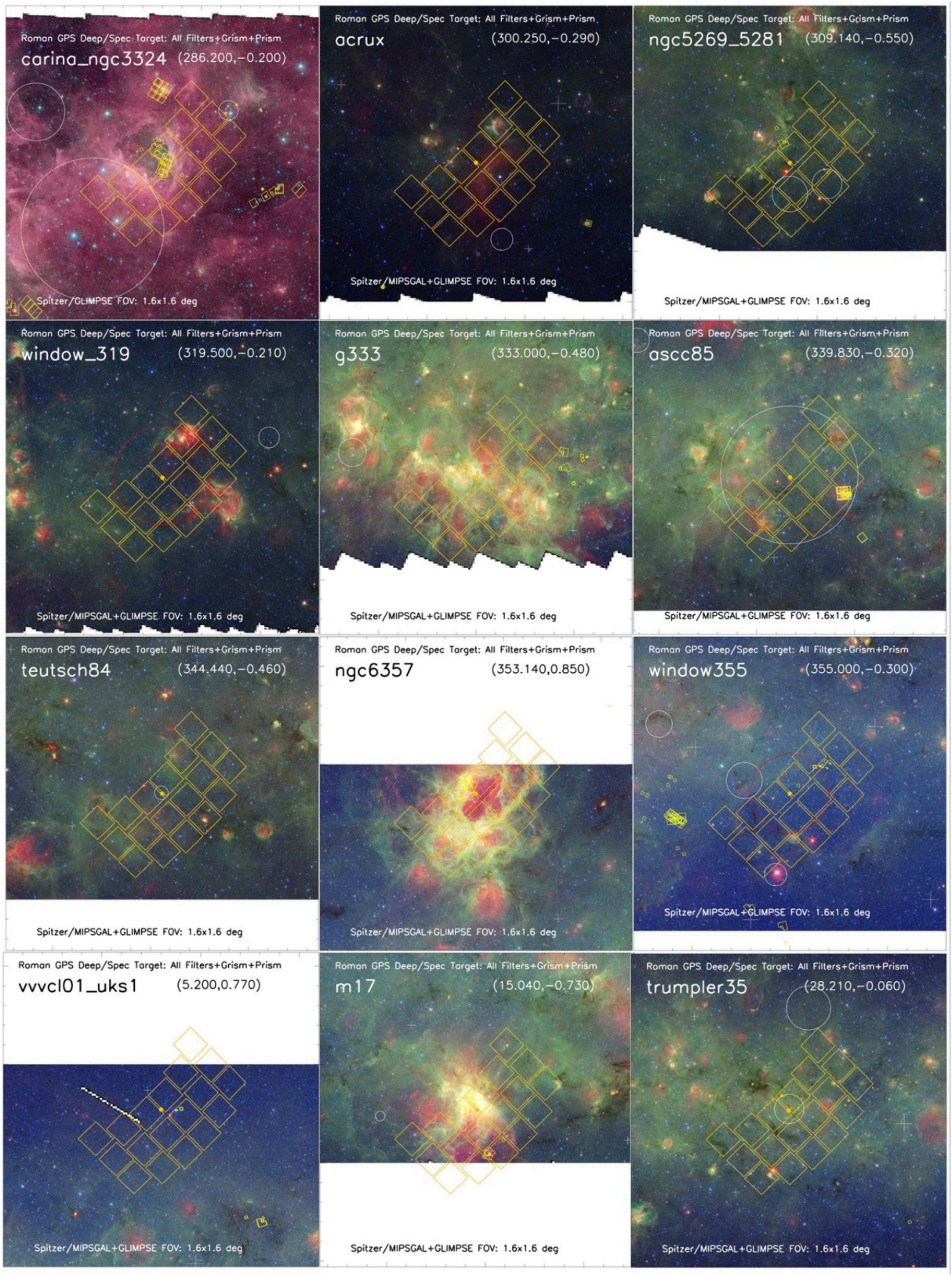

**Figure 4.6** Same as Fig 4.2, but showing single Roman pointings recommended for deep imaging in all filters plus grism and prism observations. See Table 4.4 for more information on these fields. The white-space occurs when the display of the targeted region extends beyond the region covered by either the *Spitzer* MIPSGAL or GLIMPSE surveys.

For proper motion estimates, we make the following modeling assumptions:
- The bright-end systematic uncertainty is 0.01 pixel per exposure, following Sanderson et al 2019. We note that JWST and HST have obtained systematic uncertainties roughly 2x better than this.
- Eight total exposures, with each field getting four exposures at each of two epochs separated by two years. The four exposures per epoch correspond to two exposures in each of two filters.

We assume that the sources are detected with equal SNR in each filter and use a fixed effective FWHM of 0.11" for simplicity. More accurate estimates need to build in the specific source spectrum and account for the modest variation in effective FWHM between filters. *Astrometric uncertainty:* The astrometric uncertainty in each exposure is therefore $\sigma_a = \sqrt{(1.1 mas)^2 + (110\ mas\ /\ SNR)^2}$ and the proper motion uncertainty is given by $2^{-3/2} \sigma_a\ mas/yr$. Figure 3.10 provides the performance of the RGPS as a function of magnitude using the above prescription. The proper motion uncertainty approaches 0.38 mas / yr at high SNR due to the assumed 0.01 pix astrometric uncertainty floor. This uncertainty is proportional to the astrometric uncertainty floor assumed.

Over two years, the typical proper motions are small and will not make cross-matching challenging for most sources. We did not do simulations of cross-band proper motions, but do not expect this to be a problem—because the PSF is well known in each band, one can model pixel-phase effects on the astrometry at the <0.01 pix level (and Hubble and Webb perform 2–3x better than this).

*Extinction:* To estimate the effects of extinction (Table 3.1), we used the extinction curve of Schlafly et al (2016) assuming $A_H / A_K = 1.73$ (Nishiyama et al. 2009). These extinction coefficients are appropriate for the central wavelengths of each Roman band and do not correspond to a real astronomical SED, but because the Roman filters are not very wide (excluding F146) this is a good approximation. Note that Roman's F129 filter is somewhat redder than the 2MASS J filter and Roman's F213 is somewhat bluer than the 2MASS K filter, so Roman's $A_{F129} / A_{F213}$ ratio tends to be smaller than its 2MASS analog.

*Saturation (imaging):* The saturation values given in Table 3.1 assumes that the first resultant will consist of a single read and that the second resultant will consist of two reads, following the current proposed microlensing MA table. It uses the time-to-saturation estimates from the Goddard Roman WFI technical pages. Note that the fluxes of even saturated sources may be estimated by fitting the PSF outside of the saturated core.

*Sensitivity (spectroscopy):* Spectroscopy in crowded fields is complicated by the large numbers of sources, whose spectra will overlap significantly. However, the 600 s exposures proposed here should enable good stellar parameters for red clump stars in the Galactic bulge with $A_K = 1$; this corresponds to 16.3 mag in the H band. To roughly estimate the SNR achievable on these fields, we take our nominal 10 min exposure time and assume that backgrounds are 60x nominal due to blending of starlight among the different sources in these crowded fields. The 60x number is based on H-band HST images of the bulge, artificially dispersed to resemble Roman slitless spectroscopic observations, and compared to the backgrounds measured on high Galactic latitude fields. This leads to signal-to-noise ratios per pixel of 170 for the prism and 25 for the grism, at 1.6 microns, for our nominal red clump bulge target at $A_K = 1$.

Our recommended strategy and sensitivity estimates are similar to the GBTDS—which recommended 5 x 300 sec observations of their fields—but in much wider range of Galactic environments. It is unclear how well extraction of stellar and diffuse emission spectra will work in crowded fields, so it is difficult to predict the scientific payoff. But it was clear after the RGPS February workshop that many community members felt that it was worth investing some time and the committee agreed.

*Saturation (spectroscopy):* Roman grism and prism images saturate at significantly brighter magnitudes because the light from stars is dispersed over 200–1000 pixels. We estimate the point at which the grism and prism saturate by extrapolating from the imaging saturation limits, with the following adjustments. First, the grism (prism) spreads the light of 1000x (200x) as many pixels as the imaging. The grism and prism bandpasses are roughly 4 times as wide as the imaging bandpasses, but the grism has roughly half the throughput of the imaging. Finally, the spectroscopic exposure times for the first and second resultants are 1.6 times longer than the equivalent values for the imaging due to the longer spectroscopic frame times. Combining these factors, we expect the grism (prism) to saturate 6.2 (3.7) mag brighter than the corresponding imaging magnitudes, allowing for spectral characterization of sources too bright for proper photometric extraction. Of course, depending on the source spectrum, grism and prism observations may saturate at some wavelengths and not others; we provide these rough estimates only to inform when saturation may be a concern.

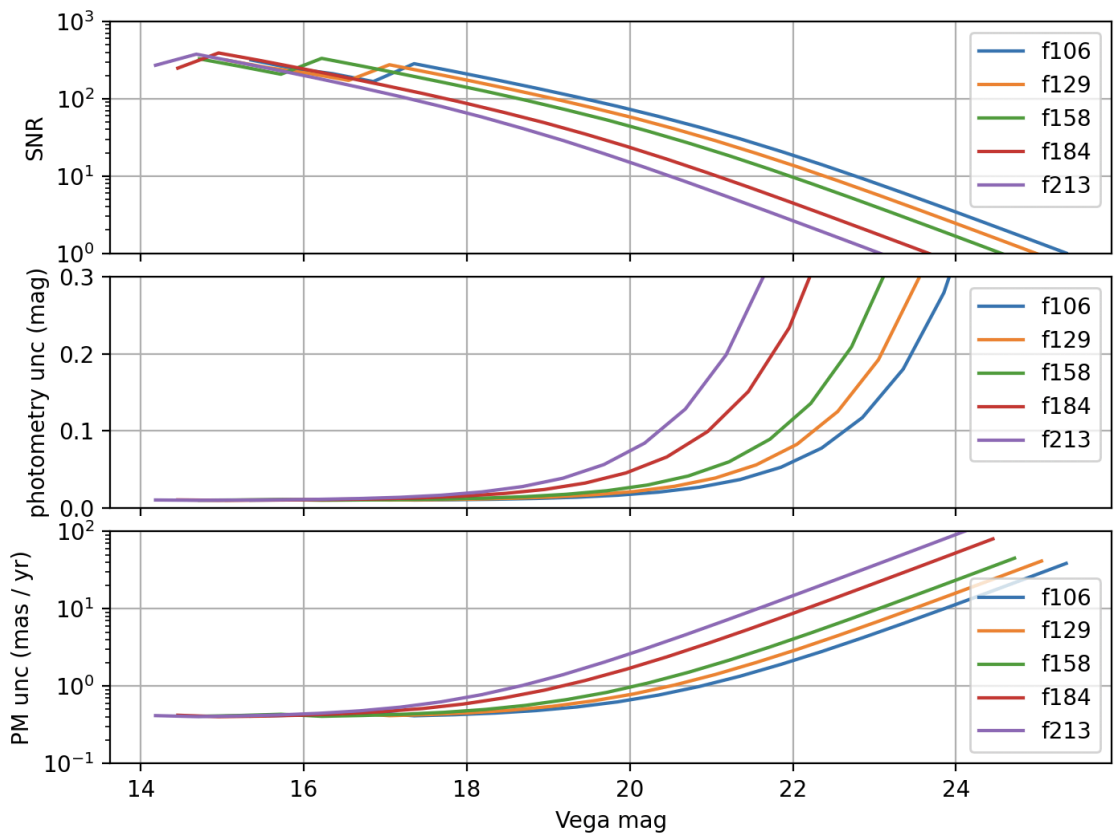

**Figure 4.6** [top] Estimated signal-to-noise ratio as function of (Vega) magnitude and Roman telescope/filter combination using assumptions described above, [middle] associated photometric uncertainty for the same parameters, [bottom] uncertainties in proper motion measurements for the same parameters.

# 5. Constraints, Verification, and Contingencies

In this final section, we discuss scheduling constraints for the RGPS program and the sensitivity of this program to the start date for science operations (§5.1), the importance of specific fields and specific roll angles (§5.2), the issue of gaps (§5.3), the feasibility of spectroscopic investigations (§5.4) and whether early scientific verification observations are needed (§5.5).

## 5.1 Scheduling constraints and sensitivity to start date

*Wide-field constraints:* Figure 5.1 shows the observability of the Galactic midplane ($b=0°$) as a function of Galactic longitude and day of the year. This shows the two of the three major constraints on the scheduling of the RGPS. The *first* constraint, determined by the orientation and orbit of the spacecraft, is that the Galactic plane from longitude range from l=+50° to -40° is only visible for two seasons of fewer than 90 days, with a minimum window size of ~74 days. Beyond these longitude bounds, the visibility window for the Galactic plane increases rapidly with distance from the Galactic center, becoming continuously observable for $l$ >78° and $l$<-65° (295°).

The *second* constraint is that the first three bulge seasons (dark black boxes in Fig 5.1) will be continuously occupied with GBTDS observations. If fully occupied, this would nearly preclude any observations in the longitude range $|l|$ < 8° for the first 1.5 years of Roman operations. Some small amount of time might remain due to the joint visibility constraint of both the Galactic center and bulge fields for the GBTDS program, but some of this time will also be needed for routine telescope operations.

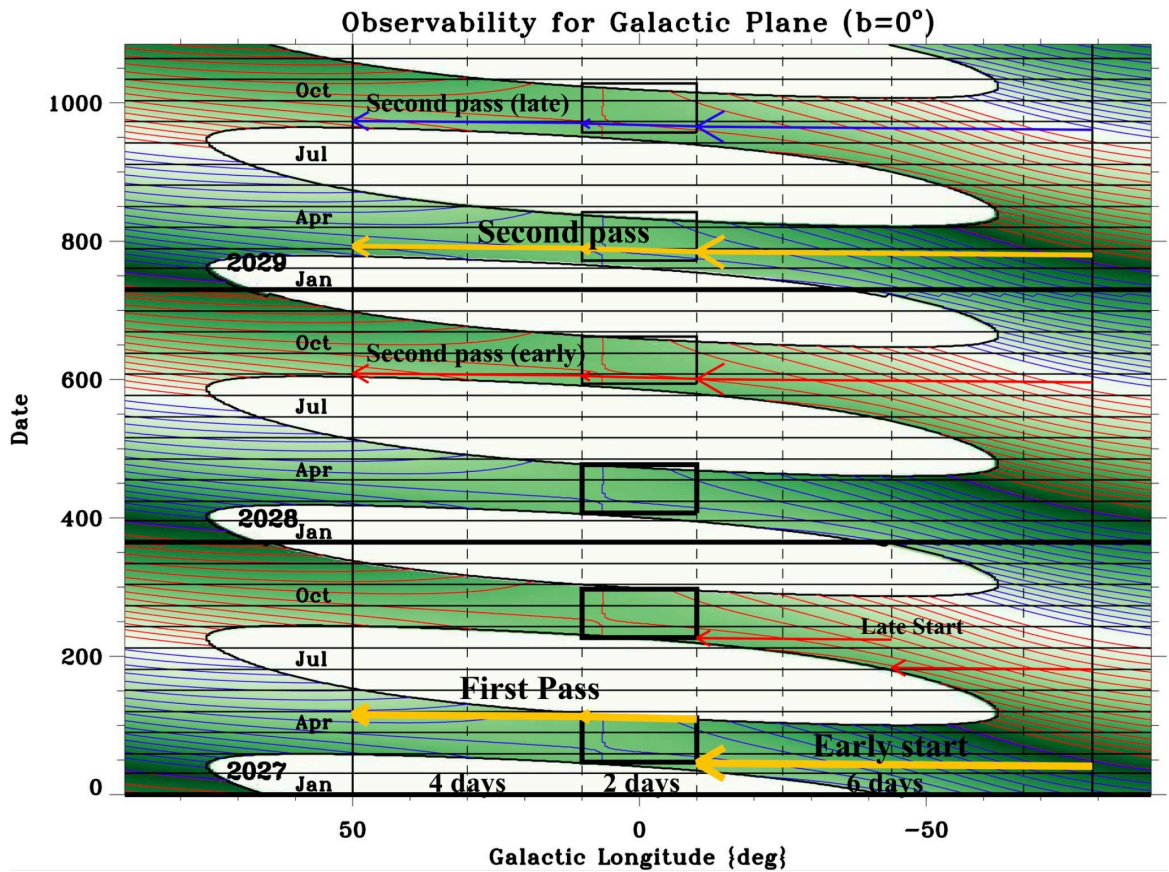

**Figure 5.1** Observability of the Galactic plane ($b=0º$) shaded in green with the roll angle of the telescope with red (positive roll angles in ecliptic coordinates) and blue (negative roll angle) contours spaced every 10 degrees. The duration of the GBTDS seasons is shown with black rectangles. The first three GBTDS seasons will be almost continuously occupied with observations for this program, with the possible exception of a two day window in GBTDS season 1 that may be reserved for early RGPS observations. The preferred scheduling for the wide-field section of the RGPS is shown with yellow/orange arrows and discussed in the text. Other, less desirable times, are shown with red or blue arrows.

The *third* constraint is based on the scientific value of getting RGPS F129 and F213 observations as early as possible in order to obtain scientifically valuable proper motions with a two-year baseline. With a mapping speed of 4.9 deg$^2$/hour per filter and two filters (F129 and F213), *the RGPS can map 14.7 degrees of longitude per day in the disk ($|b|<2º$), and 4.9 degrees of longitude per day in the bulge ($|b|<6º$)*. Taking advantage of the fact that some fraction of the bulge to be mapped by the RGPS can be covered outside of the GBTDS seasons, observations can be broken into three segments: the fourth Galactic quadrant ($l=-79º$ to $-10º$), bulge ($l=-10º$ to $+10º$), and first Galactic quadrant ($l=+10º$ to $+50º$) which will take six, two, and four days to map respectively.

*With an early start to operations, ideally the RGPS could map the fourth quadrant, bulge, and first quadrant over twelve days in F129 and F213, with the latest possible start for the fourth quadrant mapping being 11 Feb 2027.* Missing this early window for the fourth quadrant mapping would mean breaking up the first pass of the Galactic plane into the early bulge and first quadrant starting on 20 April 2027 (the end of the first GBTDS window) with the fourth quadrant mapping completed by 15 Aug 2027.

With the reduced demand for the bulge in the fourth, fifth and sixth bulge season, it should be less challenging to schedule the second pass of the RGPS to map the plane in F158 and F184. **Our preference is that this second pass takes place during the fifth bulge season, starting on or around 19 Feb 2028.** With an early start for the fourth quadrant observations in Feb 2026, this would provide a two year baseline to measure proper motions *and* these measurements would be made at the same parallax angle. Even if the first pass of the fourth quadrant is delayed until August 2027, giving a 15 month rather than 24 month time baseline for parts of the Galactic plane, the degradation in the proper motion uncertainties is outweighed by the timely completion of the survey. If the first full pass of the Galactic plane was completed much later than August 2026, then the second pass should be delayed until the sixth bulge season, the last bulge season before intensive GBTDS monitoring resumes.

We note that the call for early-definition science described observations that could be completed in the first two years of operations, while our preferred timing for the second pass of the Galactic plane lies outside this window. Using a two year window between start and end of the RGPS observations satisfies the spirit of this requirement, if not the letter, and we hope it will be allowed. But if not, the second pass of the Galactic plane would need to be completed during the fourth bulge season, as opposed to the fifth. This would produce degraded proper motions, but otherwise still satisfy all of the other science programs and would allow RGPS to provide an early delivery of the full survey to the community.

*Time domain science constraints*: Regarding the time-domain science fields, we are anticipating that these would all be scheduled towards the end of the two year survey timeframe. Certainly for the Galactic center fields and the NGC 6334/57 field this must be the case since they directly conflict with the GBTDS investigations. Moreover, we suspect that interleaving these time-domain investigations (which include hourly cadences) with wide-field mapping would be the most efficient use of observing time. On the other hand, the Carina and W43 programs could be done earlier in the mission without any scientific impact or loss of efficiency.

*Deep/spec field constraints:* There are no particular time-constraints on the timing of the Deep/Spectroscopic fields, and it would benefit many investigators, particularly those interested in the possibility of future spectroscopic programs to have access to this data as early as practical. We also note that if there is the need for "filler" observations between large slews that cross the Galactic plane that these targets could be considered. In addition, using the input provided from the community, the RGPS-DC could also provide an extended list of Deep/Spectroscopic targets along the full Galactic plane.

## 5.2. The importance of specific fields and specific roll angles

The largest element of the RGPS program is wide-field survey. The full program was driven by the desire to cover the Galaxy from about $l=+50º$ to $l=–80º$. We expect that, barring major problems, the survey will maintain these longitude ranges with four-filter coverage. If there is any need to significantly reduce the wide-field footprint, due to longer slewing times or the need to revise the dithering or frame overlap strategy, we recommend reducing the latitude range of the survey. We agreed that this reduction should be applied proportionally to disk, bulge and the Serpens South regions.

Regarding time-domain science, the highest priority fields are the two fields on either side of the Galactic center. But one goal of the time-domain programs, as designed, is to compare yields in different environments using the same observational strategies. These regions will also be the largest area panchromatic observations of the RGPS. We do not anticipate removing any of them from the program unless there are significant problems with maintaining the expected wide-field coverage.

Finally, for the deep/spectroscopic fields, these are a small time investment but have the potential to seed future Galactic programs for deeper or spectroscopic observations. Adding or removing a few of these would be fine just so long as the goal of testing Roman's capability in a diversity of environments is maintained. One small issue is that in an ideal world, most of these fields have roll angles that would maximize the number of interesting objects contained. But we felt that given the experimental nature of the pointings (and because they are all contained within the wide-field survey) the ease of scheduling outweighs the gain in science in specifying roll-angle constraints. Depending on when they are scheduled, we would hope to have the opportunity to slightly revise the pointing locations to best align the Roman field-of-view with objects of interest that might lie in those directions.

## 5.3 Gaps, dithering, and mitigation strategies

The LINEGAP2 dithering/gap-filling strategy we recommend would cover each part of the survey region thrice (3.7%), twice (67.2%), once (27.9%), or no times (1.2%). Although a survey with 1.2% of the area not covered (in each filter) is not ideal, it is not expected to compromise the science that could be obtained. A single-pass coverage fraction of 27.9% could represent more risk to some science programs, depending on the data quality. Currently, no significant problems with single-pass coverage are anticipated. Cosmic ray removal, for example, should be possible with single-pass observations using the up-the-ramp sampling.

Since any data quality issues associated with single-pass data would reduce the total area of the sky covered, it is the largest uncertainty in the planning of the RGPS. One possibility the committee has identified is to increase the overlap fraction along one axis of the detector. Increasing the overlap from 0% to 30% would reduce the single-pass fraction of the survey to below 10%, a more tolerable value. If such a modification to the survey proves necessary, we would recommend keeping the longitude extent of the survey fixed and reducing the latitude range.

Another implication of the LINEGAP2 strategy is the characterization of the PSF. For most science cases, the driving consideration was source detection over a wide area. Using the science pitch submission form, thirty-one (61%) of the science pitches received marked *Total Survey Covered* as "Very Important" for the proposed science investigation. In contrast, only sixteen of fifty-one (31%) of submitted science pitches marked the category *Subpixel Dithering* as "Very Important", and twelve of those sixteen also marked *Total Survey Covered* as "Very Important". To the extent that a full characterization of the PSF is needed, the RGPS data will have ~8 images of each part of the sky. The result is not well sampled in the technical sense, but it will mean that most stars/blends are observed at a variety of pixel phases. These data may be utilized to achieve roughly milli-arcsecond astrometric precision for a single PSF realization (e.g. [Anderson 2016](#)). Since the multiband analysis will have access to all of this information, we expect the main benefit of better sampling in each band to be depth rather than deblending performance. We also note that the time-domain science regions (19 deg$^2$)—which includes the two regions covering the Nuclear Stellar Disk—and the deep/spectroscopic fields (4 deg$^2$) use the BOXGAP8 dithering pattern, so the effects of using LINEGAP2 for most of the survey area can be characterized.

## 5.4 Feasibility of spectroscopic observations

We were asked to deliver an observing plan both with and without a grism/prism component due to concerns about guiding in crowded fields. The spectroscopic element of the recommended program is very small (only 1.06% of the total time) for pilot observations. If it turns out not to be possible to carry out these observations or if the resulting data looks unusable, the time devoted to spectroscopy will be used to increase the photometric depth of the deep/spec fields.

## 5.5 Early scientific verification programs

Because the RGPS will take place early during operations, there is little time available for early scientific verification observations. It would be desirable to obtain a small mosaic of a challenging region (high stellar density, high backgrounds) to test data quality and the dithering strategy before the RGPS commences . This could occur right at the start of science operations using a southern field that is observable year-round or using an inner Galaxy field just before the start of the first bulge season. We anticipate that such data will be available without any specific RGPS involvement, especially if there will be early observations of the Galactic plane or the Magellanic clouds. A subset of the Definition Committee will evaluate this data and recommend any changes in implementation of this program.

## 5.6 Efficiently observing time-domain fields with slow cadences

We note that the hourly-cadence time-domain observations of the Galactic center and Serpens South fields can be observed with modest overheads by interleaving them with the wide-field RGPS observations. The wide-field RGPS program maps 15° of longitude per day, and these fields feature hourly-cadence observations over a period of 3 days. The average slew from the RGPS region to the time-domain field would then be roughly 11°, meaning that it would take 8 minutes to slew to the time domain field and then return to the RGPS field. This is comparable to the 12 minutes required to observe the Galactic center fields and much smaller than the 54 minutes required to observe the Serpens South field, meaning these observations can be made with a modest hit to the survey efficiency.

We also recommend observations with a weekly cadence of the Galactic center fields. These cannot be effectively interleaved with RGPS observations and may require slews of 30+ minutes to obtain. This is a factor of few longer than the time spent observing the Galactic center fields, making these observations much more costly than other time-domain observations. However, these fields may still be able to be observed efficiently if their observations can be interleaved with other Roman programs.

# Appendix A. Community White Papers and Science Pitches

The following tables summarize all of the White Papers and Science Pitches received from the community.

| Reference | Lead Author | Title | Topics | Crossref. |
|---|---|---|---|---|
| WP01 | Guenzani | Coronagraph Angles as Proxy Theta Angles in the Galactic Plane for Protostar Dust Coupling Analysis | Coronography | |
| WP02 | Collins | Roman and the Search for ExtraTerrestrial Intelligence | SETI | |
| WP03 | Kupfer | The high-cadence Roman Galactic Plane survey | Compact binaries | |
| WP04 | Rich | A survey of the Galactic Plane Emphasizing the Central 2 kpc | Galactic structure, time domain | See also SP31 |
| WP05 | Benecchi | Characterization of the Deep, Extended Kuiper Belt in the Galactic Disk | KBO | |
| WP06 | Meli | Investigating the interplay between infrared observations, galactic magnetic fields, and cosmic rays | Gal. mag. field, Cosmic rays, SNR, Sgr A*, Galaxies | See also SP21 |
| WP07 | Kruszyńska | Synergies between Roman Galactic Plane Survey and other major surveys | Microlensing, Compact binaries, CV, RR Lyrae, Galactic structure, SFR, YSO, Neptune Trojans, GC | |
| WP08 | Paladini | The Milky Way as The z=0 template to investigate star and planet formation | Red Giant Stars, YSO, Extinction map, Ionized gas | |
| WP09 | Zari | Young stars in the obscured galactic disk | YSO | |

**Table A.1**: Summary of the community White Papers received

| Reference | Lead Author | Title | Topics | Crossref. |
|---|---|---|---|---|
| SP01 | Elowitz | Searching for Interstellar Objects of Aritifical Origin Using Roman Galactic Plane Survey Data | Interstellar Objects | |
| SP02 | Amrar | Oukaimeden observatory , morocco , physics stellar | Variable stars | |
| SP03 | Lodieu | High proper motion metal-poor brown dwarfs | Brown Dwarfs | |
| SP04 | Soam | Dust in higher latitude clouds (HLCs) | High latitude clouds | |
| SP05 | Hunt | Tracing the inner galaxy with star clusters in Roman | OC | |
| SP06 | Werner | Linking SPHEREx to the Roman Galactic Plane Survey | SFR | |
| SP07 | Bryant | Galactic | OC | |
| SP08 | Morihana | The synergy between ULTIMATE-Subaru and Roman of the Galactic Plane Survey | CV | |
| SP09 | Lian | Search for isolated stellar mass black holes | Microlensing | |
| SP10 | Anderson | Ionized Hydrogen in the Galaxy | Ionized gas | |
| SP11 | Lim | YSO Outflow hunting with Roman WFI Grism | YSO | |
| SP12 | D'Ammando | Characterizing jetted AGN in the Galactic Plane in synergies with other MWL facilities | AGN | |
| SP13 | Freeman | The metal-rich RR Lyrae stars of the Galactic disk | RR Lyrae | |
| SP14 | Bachelet | High-resolution imaging of all microlensing events detected in the Milky Way | Microlensing | |
| SP15 | Villaseñor | Unveiling the Multiplicity of Young Massive Stars in the Galactic Plane | Massive stars | |
| SP16 | Minniti | Free floating planets in nearby stellar associations | FFP | |
| SP17 | Smith | Hypervelocity stars in the central region of the Milky Way | Hypervelocity stars | |
| SP18 | Navarro | Long timescale microlensing events in the Galactic center region | Microlensing | |
| SP19 | De Furio | Pushing the Limits of Star Formation: An in Depth Exploration of Star-forming Regions in the Galactic Plane | SFR | |
| SP20 | Baravalle | Galaxies and Quasars in the Zone of Avoidance | Galaxies, AGN | |
| SP21 | Meli | Investigating the interplay between infrared observations, galactic magnetic fields, and cosmic rays | Gal. mag. field, Cosmic rays, SNR, Sgr A*, Galaxies | |
| SP22 | Morihana | The synergy between ULTIMATE-Subaru and Roman of the Galactic Plane Survey | CV | |
| SP23 | Wang | Toward a 3-D Panchromatic View of the Galactic Center | Galactic Center, Sgr A* | |
| SP24 | Lucas | Clusters and YSOs in the Galactic Plane Survey | YSO, OC | |
| SP25 | Lim | Tracing Sequentially triggered star formation with Roman | SFR, Molecular Clouds | |
| SP26 | Daylan | Searching the Galactic Plane for Free-Floating Planets using Roman Space Telescope | Microlensing | |
| SP27 | Bahramian | A deep and comprehensive view of X-ray binaries in the Galactic plane | X-ray binaries | |
| SP28 | Carey | Synergies with existing infrared surveys of the Galactic Plane | Survey overlap | |
| SP29 | Kuzuhara | Identifying Nearby Stars Accelerated by Substellar-Mass Companions | Hypervelocity stars | |
| SP30 | Carey | Roman Galactic Plane synergies with NEO Surveyor and other Planetary Science missions | NEO, Interstellar Objects | |
| SP31 | Rich | A survey of the Galactic Plane Emphasizing the Central 2 kpc | Galactic Center | See also WP04 |
| SP32 | Ivanov | Census of Milky Way star clusters | Star clusters | |
| SP33 | Sankrit | Identifying Tiny Low-extinction Windows | Extinction map | |
| SP34 | Sabin | Stellar Feedback: The Influence of Evolved Stars in the Galactic Plane | Evolved stars | |
| SP35 | Street | Windows to the other side of the Milky Way | Far side of Milky Way | |

**Table A.2:** Summary of the community science pitches received (part 1)

| Reference | Lead Author | Title | Topics | Crossref. |
|---|---|---|---|---|
| SP36 | Gallart | Deriving star formation histories from color-magnitude diagrams reaching the oldest main sequence turnoff in the Milky Way disk and bulge. | Stellar ages | |
| SP37 | Pascucci | Probing Accretion and Outflows: A Slitless Spectroscopic Survey of Off-Plane Star-Forming Regions | SFR | |
| SP38 | Ivanov | Spectral library | Stellar spectroscopy | |
| SP39 | Ivanov | Spectroscopic extinction map | Extinction map | |
| SP40 | Craig | Searching for Nova Counterparts with Roman | Novae | |
| SP41 | Craig | Examining Dusty Nova Shells with Roman | Novae | |
| SP42 | Kuhn | The Cygnus X and Carina Galactic Plane Regions | SFR | |
| SP43 | Saydjari | Extinction and Extinction Curve Estimates towards NIR DIB Targets | ISM | |
| SP44 | Hillenbrand | Don't Completely Neglect the Grism Opportunity | Stellar properties, Extinction map | |
| SP45 | Rojas | A Multi-Wavelength Study of the Galactic Bulge with Roman HLWAS and Radio Observations | | |
| SP46 | Gramze | Star formation and HII regions with the Roman Telescope: SHIIRT | SFR | |
| SP48 | Bonito | Young stellar objects: investigation of a diverse range of variability timescales | YSO, SFR | |
| SP49 | Minniti | Homogeneous ages for the innermost Galactic globular clusters and search for missing globulars with the Roman GPS | GC | |
| SP50 | Pal | Roman IR Stellar Spectral Library | Stellar spectroscopy | |
| SP51 | Saito | Expanding the Roman observations of the Milky Way bulge | Stellar populations, Galactic Center, Galactic structure, Stellar populations, Solar System Objects | |
| SP52 | Pardo | Detecting Gravitational Waves with Roman | GW | |

**Table A.2** (continued) Summary of the community science pitches received (part 2)

# Appendix B: Metrics

We agreed on a number of metrics, or statistical criteria used to evaluate key areas of survey performance. Although a comprehensive study would include metrics describing the specific scientific results for each of the proposed science cases, this was beyond the resources or time allowed for our process. The metrics therefore focus on describing aspects of the survey observational parameters that were reported to be essential to the scientific yield of all proposals. These metrics are described in Table B.1.

One of the most important metrics for Galactic science is simply whether a given desired sky region is included within the survey footprint. Tables B.2 and B.3 summarize the overlap between the deep/spectroscopic and time domain fields in the RGBTDS with the survey regions. proposed in the community science cases.

In designing a survey strategy that would benefit as many science cases as possible, we analyzed the filters requested by the community for different proposals. Figure 3.20 charts the bandpasses requested for all science cases, while Figure B.1 compares the filters requested for time domain science with those ultimately selected for the RGPS TDA fields.

For those fields which will receive time series observations in the RGPS, the temporal cadence can be represented by the number of visits to a field and the total interval, or duration, over which those visits are performed. Constraints on the maximum duration of the RGPS (2yrs) and on the time allocation necessarily restrict the time domain fields to relatively short durations in order to achieve the desired visit intervals. Figure B.2 takes the sky area overlap into account and shows the percentage of the requested observations to be provided by the survey for each science case. It is possible for data in this plot to have metric percentages higher than 100% - this indicates that a field received either more visits than a science case asked for, or that observations continued over a longer duration than specified. This is the case for the Galactic Center region, which was indicated as a high priority in a number of community proposals. In interpreting this plot, it should be remembered that the survey cadence is considered for each filter. For example, Bahramian requested observations of two large regions for X-ray binary studies in F087, F129, F146, F158, F184 and F213. While the total time required for time series observations in all these filters would be beyond the scope of the RGPS, time series observations are included in F213 for some smaller regions of overlap. Single exposures will be acquired in F062, F087, F106 and F129.

Table B.1 Metrics used to evaluate survey performance

| Table B.1 Metrics used to evaluate survey performance ||
|---|---|
| **M1: Survey footprint coverage** | **How well do the survey fields defined cover the survey footprint requested in White Papers/Science Pitches?** |
| The sky regions requested by each science case, and covered by each survey design, are converted to HEALpixel sky maps to allow for irregularly-shaped regions. <br><br> This allows regions of interest to be combined by co-adding the sky maps and indicating the number of science cases requesting each HEALpixel as a measure of its scientific value. This can be a somewhat crude indicator since a high priority field might be requested only once by a White Paper with many authors interested for a range of different science projects. As a result, the software also provides the means to assign priorities to different science regions, and these can also be co-added to produce a sky map. <br><br> Regions of interest are generally combined on a per-filter basis, since observations of a particular region may only be useful at selected wavelengths. For example, proposals to observe highly extincted regions usually requested data using the longest wavelength filters. The metric calculates the percentage of the requested region included in survey design's footprint per filter. ||
| **M2: Star counts** | **How many stars are included in the survey fields?** |
| The total number of stars in a given HEALpixel is estimated from a map of stellar densities across the Galactic Plane derived from the Trilegal galactic model [Girardi et al. 2005] in the Roman passbands. Trilegal's default extinction model was used to generate these data. A revised approach to extinction, using published infrared extinction is described in <br><br> The metric value return is the total number of stars included. ||
| **M3: Extended object counts** | **How many known targets of different types are included in the survey fields?** |
| Catalogs are available for known examples of a range of different targets of interest, such as Open and Globular Clusters. <br> These catalogs, which provide both the target locations and their apparent radial extent, can be used to generate a sky map of a set of desired survey regions. This metric calculates what percentage of the targets in the catalog lie fully or partially within the designed survey footprint. <br> Catalogs are included for Open Clusters, Globular Clusters, Local Star Forming Regions, and Active Galactic Nuclei ||
| **M4: Proper Motions** | **How well can PM be measured ?** |
| We can measure the number of visits / field and % of survey region that receive >2 observations at some required interval . This metric aims to estimate the proper motion measurement uncertainty, σ, from the expression: σ = 1 mas / T [years] / N [exposures per epoch]. The recommended threshold for "detection" is ~1 mas / yr, since this precision makes it possible to measure quantities like Galactic rotation and bulge dynamics. <br><br> This is calculated for all HEALpixels within all region for a given survey design, summing visits over all filters chosen. The metric represents the percentage of the desired survey region that meets the precision threshold. ||
| **M5: Color measurements** | **Sky region observed in multiple bandpasses** |
| This metric evaluates the total area of sky to receive observations in each optical element individually, and in selected combinations of the filters, as a proxy for color measurements. <br> Note that this does NOT require contemporaneous observations in different filters to derive color measurements and is therefore unsuitable to evaluate color measurements of variable objects. <br><br> The filter combinations evaluated are drawn from those requested by the science cases, and the metric results are returned as area in square degrees. ||
| **M6: Sky area number of visits** | **Sky region to receive the requested number of visits** |
| This metric evaluates whether regions requested for time domain observations receive the repeat observations requested in each filter. The metric produces three values for each desired survey region and survey strategy: <br> • the percentage of the region to receive time series visits in any filter. <br> • the percentage of the requested number of visits received. <br> • the percentage of the requested total duration of observations received. ||

| Deep/spectroscopic target | Lead authors of community contributions | Contribution codes | Topics of proposals with desired footprints overlapping target field |
|---|---|---|---|
| Deep_ASSC_85 | Sankrit,Baravalle,Bahramian,Hillenbrand,Rich,Gallart,Kruszynska,Lucas,De_Furio,Meli,DAmmando,Carey1,Anderson,Bachelet,Ivanov,Hunt,Paladini,Werner,Sabin,Craig,Zari | SP33,SP20,SP27,SP44,SP31, WP04,SP36,WP07, SP24,SP19,SP21,SP12,SP28,SP10,SP14,SP39, SP05,WP08,SP06,SP47,SP40,WP09 | X-ray binaries, Galactic structure, Symbiotic stars, Ionized gas, SNR, Stellar ages, YSO, Stellar properties, CV, AGB, Red Giant Stars, GC, Cosmic rays, Extinction map, Novae, AGN, Overlap with other surveys, Planetary nebulae, SFR, Galaxies, Open Clusters, RR Lyrae, Compact binaries, Neptune Trojans, Microlensing, Galactic magnetic field |
| Deep_Acrux | Sankrit,Baravalle,Bahramian,Hillenbrand,Rich,Kruszynska,Lucas,Meli,Carey1,Anderson,Bachelet,Ivanov,Hunt,Paladini,Werner,Sabin,Craig,Minniti,Zari | SP33,SP20,SP27,SP44,SP31, WP04,WP07,SP24, SP21,SP28,SP10,SP14,SP39,SP05,WP08,SP06, SP47,SP40,SP16,WP09 | X-ray binaries, Galactic structure, Symbiotic stars, Ionized gas, SNR, YSO, Stellar properties, CV, AGB, Red Giant Stars, Globular Clusters, Cosmic rays, Extinction map, Novae, AGN, Overlap with other surveys, Planetary nebulae, SFR, Galaxies, Open Clusters, RR Lyrae, Compact binaries, Neptune Trojans, Microlensing, Galactic magnetic field |
| Deep_G333 | Sankrit,Villasenor,Baravalle,Bahramian,Rich,Gallart,Lucas,De_Furio,Meli,Carey1,Anderson,Bachelet,Ivanov,Hunt,Paladini,Werner,Sabin,Craig,Zari | SP33,SP15,SP20,SP27,SP31, WP04,SP36,SP24, SP19,SP21,SP28,SP10,SP14,SP39,SP05,WP08, SP06,SP47,SP40,WP09 | X-ray binaries, Symbiotic stars, SNR, Stellar ages, YSO, CV, Massive stars, AGB, Red Giant Stars, Cosmic rays, Extinction map, Galactic magnetic field, Novae, Overlap with other surveys, Planetary nebulae, SFR, Galaxies, Open Clusters, Microlensing, Ionized gas |
| Deep_M17_Omega | Saydjari,Sankrit,Villasenor,Baravalle,Bahramian,Lim3_grism_targets,Rich,VVV_keyholes_2nd_priority,Gallart,Kruszynska,Lucas,Meli,Ivanov2,Carey1,Craig1,Anderson,Bachelet,Ivanov,Hunt,Craig2,Paladini,Wern | SP43,SP33,SP15,SP20,SP27,SP11,SP31, WP04, SP35,SP36,WP07,SP24,SP21,SP39,SP28,SP40, SP10,SP14,SP39,SP05,SP41,WP08,SP06,SP47, SP40,SP38,WP09 | X-ray binaries, Keyholes, Galactic structure, Symbiotic stars, Ionized gas, Stellar spectroscopy, SNR, Stellar ages, YSO, CV, Massive stars, AGB, Red Giant Stars, Globular Clusters, ISM, Cosmic rays, Extinction map, Novae, AGN, Overlap with other surveys, Planetary nebulae, SFR, Galaxies, Open Clusters, RR Lyrae, Compact binaries, Neptune Trojans, Microlensing, Galactic magnetic field |
| Deep_NGC3324_Carina | Villasenor,Baravalle,Bahramian,Rich,Kruszynska,Lian,Lucas,Meli,Anderson,Bachelet,Lim2,Bonito,Werner,Craig,Minniti,Zari | SP15,SP20,SP27,SP31, WP04,WP07,SP09,SP24, SP21,SP10,SP14,SP25,SP48,SP06,SP40,SP16, WP09 | X-ray binaries, Ionized gas, Molecular Clouds, SNR, YSO, CV, Massive stars, Globular Clusters, Cosmic rays, Novae, AGN, SFR, Galaxies, Open Clusters, RR Lyrae, Compact binaries, Neptune Trojans, Microlensing, Galactic magnetic field |
| Deep_NGC5269_5281 | Sankrit,Baravalle,Bahramian,Rich,Kruszynska,Lucas,Meli,Carey1,Anderson,Bachelet,Ivanov,Hunt,Paladini,Werner,Sabin,Craig,Minniti,Zari | SP33,SP20,SP27,SP31, WP04,WP07,SP24,SP21, SP28,SP10,SP14,SP39,SP05,WP08,SP06,SP47, SP40,SP16,WP09 | X-ray binaries, Galactic structure, Symbiotic stars, Ionized gas, SNR, YSO, CV, AGB, Red Giant Stars, Globular Clusters, Cosmic rays, Extinction map, Novae, AGN, Overlap with other surveys, Planetary nebulae, SFR, Galaxies, Open Clusters, RR Lyrae, Compact binaries, Neptune Trojans, Microlensing, Galactic magnetic field |
| Deep_NGC6357_Lobster | Saito,Minniti_GCs,Sankrit,Villasenor,Baravalle,Bahramian,Gramze,Rich,Gallart,Kruszynska,Lucas,De_Furio,Meli,Carey1,Navarro,Anderson,Bachelet,Ivanov,Hunt,Paladini,Werner,Sabin,Craig,Zari | SP51,SP49,SP33,SP15,SP20,SP27,SP45,SP31, WP04,SP36,WP07,SP24,SP19,SP21,SP28,SP18, SP10,SP14,SP39,SP05,WP08,SP06,SP47,SP40, WP09 | X-ray binaries, Galactic structure, Symbiotic stars, Ionized gas, SNR, Stellar ages, Stellar populations, SNR, Stellar ages, YSO, CV, Massive stars, AGB, Red Giant Stars, Globular Clusters, Cosmic rays, Extinction map, Novae, AGN, Overlap with other surveys,Planetary nebulae, SFR, Galaxies, Open Clusters, RR Lyrae, Compact binaries, Neptune Trojans, Microlensing, Galactic magnetic field |
| Deep_Teutsch_84 | Sankrit,Villasenor,Baravalle,Bahramian,Rich,Gallart,Kruszynska,Lucas,De_Furio,Meli,Carey1,Anderson,Bachelet,Ivanov,Hunt,Paladini,Werner,Sabin,Craig,Zari | SP33,SP15,SP20,SP27,SP31, WP04,SP36,WP07, SP24,SP19,SP21,SP28,SP10,SP14,SP39,SP05, WP08,SP06,SP47,SP40,WP09 | X-ray binaries, Galactic structure, Symbiotic stars, Ionized gas, SNR, Stellar ages, YSO, CV, Massive stars, AGB, Red Giant Stars, Globular Clusters, Cosmic rays, Extinction map, Novae, AGN, Overlap with other surveys, Planetary nebulae, SFR, Galaxies, Open Clusters, RR Lyrae, Compact binaries, Neptune Trojans, Microlensing, Galactic magnetic field |
| Deep_Trumpler_35 | Sankrit,Bahramian,Rich,Gallart,Kruszynska,Lucas,De_Furio,Meli,Carey1,Anderson,Bachelet,Morihana1,Morihana2,Ivanov,Hunt,Paladini,Werner,Sabin,Craig,Zari | SP33,SP27,SP31, WP04,SP36,WP07,SP24,SP19, SP21,SP28,SP10,SP14,SP22,SP08,SP39,SP05, WP08,SP06,SP47,SP40,WP09 | X-ray binaries, Galactic structure, Symbiotic stars, Ionized gas, SNR, Stellar ages, YSO, CV, AGB, Red Giant Stars, Globular Clusters, Cosmic rays, Extinction map, Novae, Overlap with other surveys, Planetary nebulae, SFR, Galaxies, Open Clusters, RR Lyrae, Compact binaries, Neptune Trojans, Microlensing, Galactic magnetic field |
| Deep_VVV_CL001_UKS_1 | Saito,Minniti_GCs,Sankrit,Baravalle,Bahramian,Kupfer,Gramze,Rich,Gallart,Kruszynska,Lucas,Meli,Carey1,Navarro,Smith,Anderson,Bachelet,Ivanov,Hunt,Paladini,Werner,Sabin,Craig,Carey2,Zari | SP51,SP49,SP33,SP20,SP27,WP03,SP45,SP31, WP04,SP36,WP07,SP24,SP21,SP28,SP18,SP17, SP10,SP14,SP39,SP05,WP08,SP06,SP47,SP40, SP30,WP09 | X-ray binaries, Galactic structure, Symbiotic stars, Ionized gas, Solar System Objects, Stellar populations, SNR, Stellar ages, YSO, CV, AGB, Red Giant Stars, Globular Clusters, Cosmic rays, Extinction map, Novae, Interstellar Objects, Hypervelocity stars, AGN, Overlap with other surveys, Planetary nebulae, SFR, Galaxies, Open Clusters, NEO, RR Lyrae, Compact binaries, Neptune Trojans, Microlensing, Galactic magnetic field |
| Deep_W40 | Sankrit,Villasenor,Bahramian,Gallart,Paladini2,De_Furio,Meli,Bachelet,Lim2,Paladini,Werner,Sabin,Craig | SP33,SP15,SP27,SP36,WP08,SP19,SP21,SP14, SP25,WP08,SP06,SP47,SP40 | X-ray binaries, Symbiotic stars, Exoplanets, Molecular Clouds, SNR, Stellar ages, YSO, CV, Massive stars, AGB, Red Giant Stars, Cosmic rays, Novae, Extinction map, Galactic magnetic field, Novae, Planetary nebulae, SFR, Galaxies, Microlensing, Ionized gas |
| Deep_W44 | Sankrit,Villasenor,Bahramian,Rich,Gallart,Lucas,Meli,Carey1,Anderson,Bachelet,Morihana1,Morihana2,Ivanov,Hunt,Paladini,Werner,Sabin,Craig,Zari | SP33,SP15,SP27,SP31, WP04,SP36,SP24,SP21, SP28,SP10,SP14,SP22,SP08,SP39,SP05,WP08, SP06,SP47,SP40,WP09 | X-ray binaries, Symbiotic stars, SNR, Stellar ages, YSO, CV, Massive stars, AGB, Red Giant Stars, Cosmic rays, Extinction map, Galactic magnetic field, Novae, Overlap with other surveys, Planetary nebulae, SFR, Galaxies, Open Clusters, Microlensing, Ionized gas |
| Deep_W51 | Sankrit,Villasenor,Bahramian,Lim3_grism_targets,Rich,Gallart,Lucas,Meli,Carey1,Anderson,Bachelet,Morihana1,Morihana2,Ivanov,Hunt,Paladini,Werner,Sabin,Craig,Zari | SP33,SP15,SP27,SP11,SP31, WP04,SP36,SP24, SP21,SP28,SP10,SP14,SP22,SP08,SP39,SP05, WP08,SP06,SP47,SP40,WP09 | X-ray binaries, Symbiotic stars, SNR, Stellar ages, YSO, CV, Massive stars, AGB, Red Giant Stars, Cosmic rays, Extinction map, Galactic magnetic field, Novae, Overlap with other surveys, Planetary nebulae, SFR, Galaxies, Open Clusters, Microlensing, Ionized gas |
| Deep_Window_319.5-0.2 | Sankrit,Baravalle,Bahramian,Hillenbrand,Rich,VVV_keyholes_2nd_priority,Gallart,Kruszynska,VVV_keyholes_1st_priority,Lucas,De_Furio,Meli,Carey1,Anderson,Bachelet,Ivanov,Hunt,Paladini,Werner,Sabin,Craig | SP33,SP20,SP27,SP44,SP31, WP04,SP35,SP36, WP07,SP35,SP24,SP19,SP21,SP28,SP10,SP14, SP39,SP05,WP08,SP06,SP47,SP40,SP16,WP09 | X-ray binaries, Galactic Keyholes, Galactic structure, Symbiotic stars, Ionized gas, SNR, Stellar ages, YSO, Stellar properties, CV, AGB, Red Giant Stars, Globular Clusters, Cosmic rays, Extinction map, Novae, AGN, Overlap with other surveys, Planetary nebulae, SFR, Galaxies, Open Clusters, RR Lyrae, Compact binaries, Neptune Trojans, Microlensing, Galactic magnetic field |
| Deep_Window_355.0-0.3 | Saito,Minniti_GCs,Sankrit,Villasenor,Baravalle,Bahramian,Gramze,Rich,VVV_keyholes_2nd_priority,Gallart,Kruszynska,VVV_keyholes_1st_priority,Lucas,Meli,DAmmando,Carey1,Navarro,Smith,Anderson,Bachelet,Ivanov,Hunt,Paladini,Werner,Sabin,Craig,Carey2,Zari | SP51,SP49,SP33,SP15,SP20,SP27,SP45,SP31, WP04,SP35,SP36,WP07,SP35,SP24,SP21,SP12, SP28,SP18,SP17,SP10,SP14,SP39,SP05,WP08, SP06,SP47,SP40,SP30,WP09 | X-ray binaries, Galactic Keyholes, Galactic structure, Symbiotic stars, Ionized gas, Solar System Objects, Stellar populations, SNR, Stellar ages, YSO, CV, Massive stars, AGB, Red Giant Stars, Globular Clusters, Cosmic rays, Extinction map, Novae, Interstellar Objects, Hypervelocity stars, AGN, Overlap with other surveys, Planetary nebulae, SFR, Galaxies, Open Clusters, NEO, RR Lyrae, Compact binaries, Neptune Trojans, Microlensing, Galactic magnetic field |

**Table B.2** Overlap between the deep/spectroscopic fields and the regions of sky requested by different community proposals, giving the science proposed by each paper.

| Deep/spectroscopic targets | Lead authors of community contributions | Contribution codes | Topics of proposals with desired footprints overlapping target field |
|---|---|---|---|
| TDS_Carina | Villasenor, Baravalle, Bahramian, Rich, VVV_keyholes_2nd_priority, Kruszynska, VVV_keyholes_1st_priority, Lian, Lucas, Meli, Kuhn, Anderson, Bachelet, Lim2, Bonito, Werner, Craig, Minniti, Zari | SP15, SP20, SP27, SP31, WP04, SP35, WP07, SP35, SP09, SP24, SP21, SP42, SP10, SP14, SP25, SP48, SP06, SP40, SP16, WP09 | SFR, Keyholes, Globular Clusters, Cosmic rays, Microlensing, Gal. mag. field, AGN, Compact binaries, SNR, X-ray binaries, Massive stars, CV, Open Clusters, Molecular Clouds, Novae, RR Lyrae, Free-floating planets, Galaxies, Galactic structure, Neptune Trojans, YSO, Ionized gas |
| TDS_Galactic_Center_Q1 | Saito, Minniti_GCs, Sankrit, Baravalle, Bahramian, Freeman, Gramze, Rich, VVV_keyholes_2nd_priority, Gallart, Kruszynska, Lucas, Meli, Carey1, Navarro, Smith, Anderson, Wang, Bachelet, Ivanov, Hunt, Paladini, Werner, Sabin, Craig, Carey2, Rich2, Zari | SP51, SP49, SP33, SP20, SP27, SP13, SP45, SP31, WP04, SP35, SP36, WP07, SP24, SP21, SP28, SP18, SP17, SP10, SP23, SP14, SP39, SP05, WP08, SP06, SP47, SP40, SP30, SP31, WP04, WP09 | SFR, Keyholes, Red Giant Stars, Globular Clusters, Cosmic rays, Microlensing, Gal. mag. field, AGN, Symbiotic stars, Compact binaries, SNR, Interstellar Objects, X-ray binaries, CV, Open Clusters, Novae, AGB, Near-Earth Objects, RR Lyrae, Stellar populations, Galaxies, Solar System Objects, Galactic structure, Hypervelocity stars, Extinction map, Planetary nebulae, Neptune Trojans, Stellar ages, YSO, Ionized gas, Overlap with other surveys |
| TDS_Galactic_Center_Q4 | Saito, Minniti_GCs, Sankrit, Baravalle, Bahramian, Freeman, Gramze, Rich, VVV_keyholes_2nd_priority, Gallart, Kruszynska, Lucas, De_Furio, Meli, DAmmando, Carey1, Navarro, Smith, Anderson, Wang, Bachelet, Ivanov, Hunt, Paladini, Werner, Sabin, Craig, | SP51, SP49, SP33, SP20, SP27, SP13, SP45, SP31, WP04, SP35, SP36, WP07, SP24, SP19, SP21, SP12, SP28, SP18, SP17, SP10, SP23, SP14, SP39, SP05, WP08, SP06, SP47, SP40, SP30, SP31, WP04, WP09 | SFR, Keyholes, Red Giant Stars, Globular Clusters, Cosmic rays, Microlensing, Gal. mag. field, AGN, Symbiotic stars, Compact binaries, SNR, Interstellar Objects, X-ray binaries, CV, Open Clusters, Novae, AGB, NEO, RR Lyrae, Stellar populations, Galaxies, Solar System Objects, Galactic structure, Hypervelocity stars, Extinction map, Planetary nebulae, Neptune Trojans, Stellar ages, YSO, Ionized gas, Overlap with other surveys |
| TDS_NGC6334_6357 | Saito, Minniti_GCs, Sankrit, Villasenor, Baravalle, Bahramian, Gramze, Rich, Gallart, Kruszynska, Lucas, De_Furio, Meli, Carey1, Navarro, Anderson, Bachelet, Ivanov, Hunt, Paladini, Werner, Sabin, Craig, Zari | SP51, SP49, SP33, SP15, SP20, SP27, SP45, SP31, WP04, SP36, WP07, SP24, SP19, SP21, SP28, SP18, SP10, SP14, SP39, SP05, WP08, SP06, SP47, SP40, WP09 | SFR, Red Giant Stars, Globular Clusters, Cosmic rays, Microlensing, Gal. mag. field, AGN, Symbiotic stars, Compact binaries, SNR, Massive stars, X-ray binaries, CV, Open Clusters, AGB, Novae, RR Lyrae, Stellar populations, Galaxies, Solar System Objects, Galactic structure, Planetary nebulae, Extinction map, Neptune Trojans, Stellar ages, YSO, Ionized gas, Overlap with other surveys |
| TDS_Serpens_W40 | Sankrit, Villasenor, Bahramian, Gallart, Kruszynska, Paladini2, De_Furio, Meli, Anderson, Bachelet, Lim2, Hunt, Paladini, Werner, Sabin, Craig, Zari | SP33, SP15, SP27, SP36, WP07, WP08, SP19, SP21, SP10, SP14, SP25, SP05, WP08, SP06, SP47, SP40, WP09 | SFR, Red Giant Stars, Globular Clusters, Cosmic rays, Microlensing, Gal. mag. field, Symbiotic stars, Compact binaries, SNR, X-ray binaries, Massive stars, CV, Molecular Clouds, Open Clusters, AGB, Novae, RR Lyrae, Galaxies, Galactic structure, Planetary nebulae, Extinction map, Neptune Trojans, Stellar ages, Exoplanets, YSO, Ionized gas |
| TDS_W43 | Sankrit, Villasenor, Bahramian, Hillenbrand, Rich, Gallart, Kruszynska, Lucas, De_Furio, Meli, Carey1, Anderson, Bachelet, Morihana1, Morihana2, Ivanov, Hunt, Paladini, Werner, Sabin, Craig, Zari | SP33, SP15, SP27, SP44, SP31, WP04, SP36, WP07, SP24, SP19, SP21, SP28, SP10, SP14, SP22, SP08, SP39, SP05, WP08, SP06, SP47, SP40, WP09 | SFR, Red Giant Stars, Globular Clusters, Cosmic rays, Stellar properties, Microlensing, Gal. mag. field, Symbiotic stars, Compact binaries, SNR, X-ray binaries, Massive stars, CV, Open Clusters, AGB, Novae, RR Lyrae, Galaxies, Galactic structure, Planetary nebulae, Extinction map, Neptune Trojans, Stellar ages, YSO, Ionized gas, Overlap with other surveys |

**Table B.3** Overlap between the time domain fields and the regions of sky requested by different community proposals, indicating the science cases proposed.

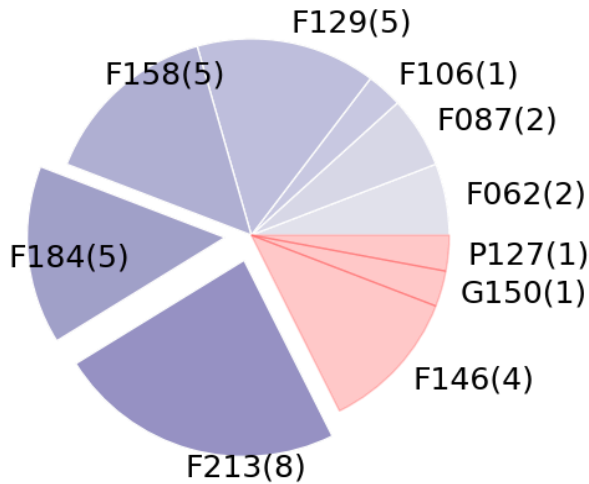

**Figure B.1** Breakdown of the filters requested for time domain science in community proposals, compared with the observations included for RGPS time domain fields. Note that many science cases requested the majority of their observations in a single filter, supported by less frequent observations in other filters. The separated wedges indicate those used for multi-epoch observations in the RGPS time domain fields, which also receive single epoch observations in almost all other filters. Wedges shaded in red indicate filters that are not included in RGPS time domain field observations.

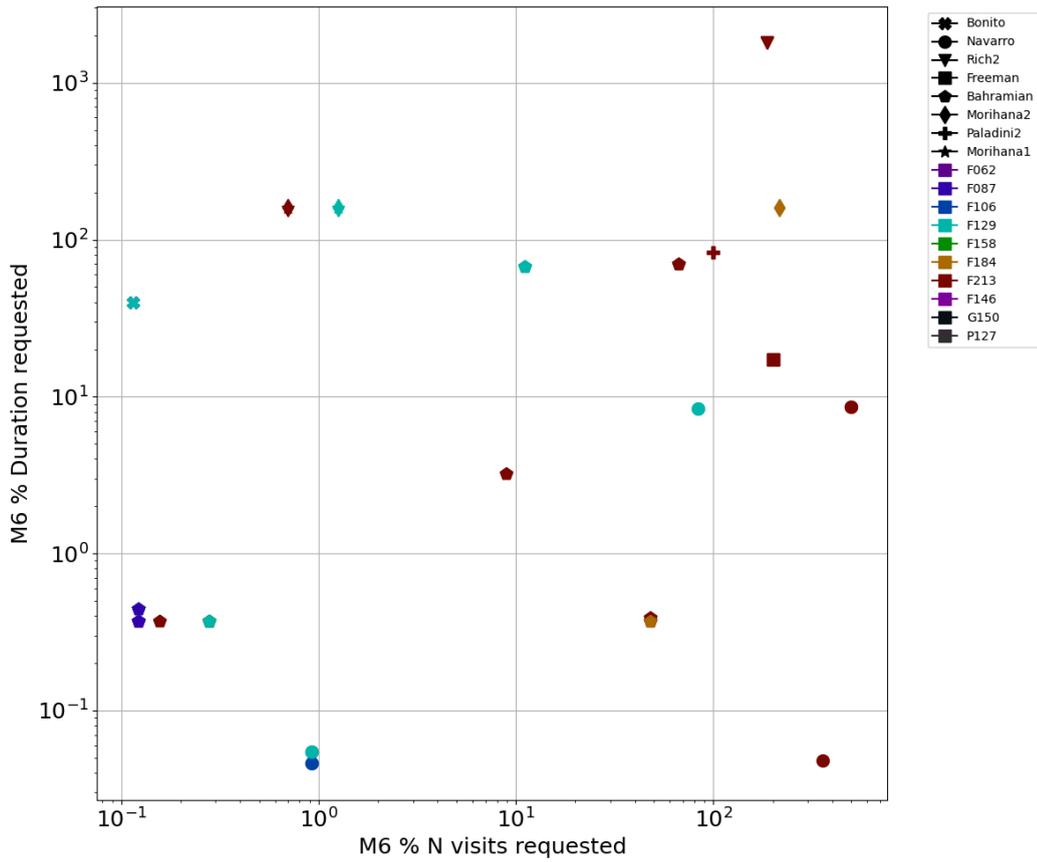

**Figure B.2** Comparison of the percentage of the requested cadence parameters (number of visits, duration) for time domain science cases to be provided by the time domain fields in the RGPS, taking into account spatial overlap between the regions requested and surveyed.

# Appendix C: Modeling and Calculations

## C.1 Extinction modeling

One chief advantage of the RGPS, particularly in comparison to optical coverage of the Galactic plane, is the greatly reduced extinction in the infrared (Schlafly et al 2016, Table 3.1). For $A_{5420 Å}$=20 mag (a transmission factor of $10^{-8}$), the corresponding extinctions in the four longest wavelength Roman filters are $A_{F129}$=4.2 mag (2.1%), $A_{F158}$=2.7 mag (8.0%), $A_{F184}$=2.0 mag (16.5%), and $A_{F213}$=1.5 mag (25%). Figure C.1 [left] shows how the extinction and the transmission fraction vary with filter and assumed $A_{5420}$ (or $A_K$), while Figure C.1 [right] shows the fraction of survey area or the fraction of sources (within |b|<1º) expected to have extinctions greater than some threshold.

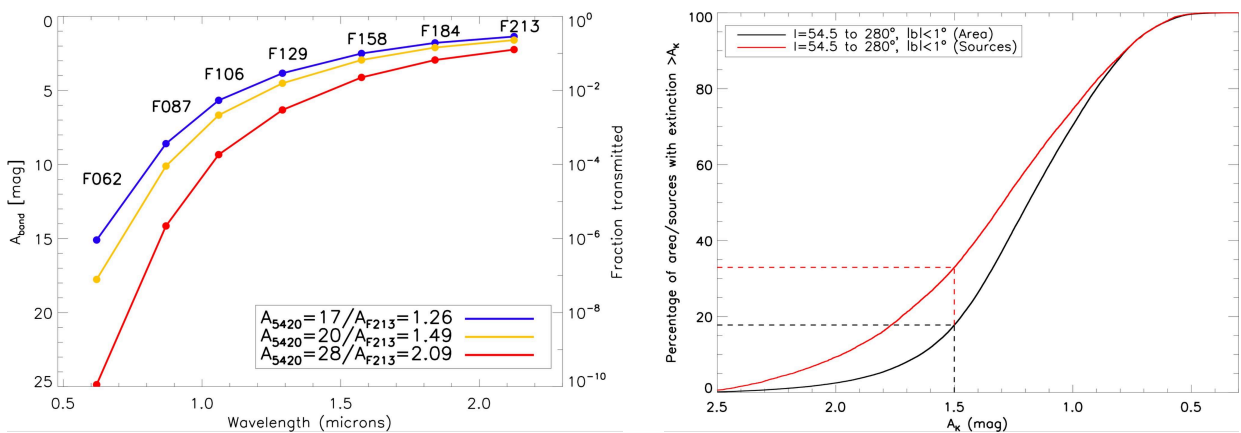

**Figure C.1** [left] Extinction (left axis) and transmission fraction (right axis) as a function of Roman filter [right] Fraction of survey area (or sources) with extinction greater than a threshold $A_{F213}$.

For the many science programs which rely on large sample sizes, a good characterization of this extinction is paramount for estimating the number of sources and correcting for incompleteness in samples. Estimating extinction in the Galactic plane, where the sources are inter-mixed with the absorption, requires three-dimensional dust extinction maps. But most available 3D dust maps are limited by the use of optical photometric data or optically measured Gaia parallaxes. This makes them less valuable for estimating infrared extinction since the inner Galaxy becomes opaque for many directions in the inner Galaxy. Three exceptions are the maps of Marshall et al (2006) (shown in Fig C.2), Nidever et al (2012), and Chen et al (2019) which were based solely on infrared data, but have angular resolution of several arcminutes, and do not use Gaia to refine stellar distance estimates.

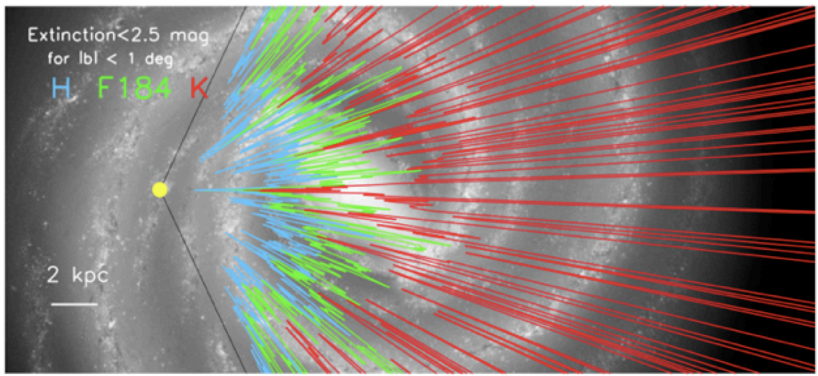

**Figure C.2** From Stauffer et al (2018) The extinction "horizon" in the H, F184 and K band, i.e. the distance where $A_{band}$ >2.5, as a function of Galactic longitude.. For each longitude, the distance corresponding to the maximum extinction (Marshall et al 2006) for any latitude value is plotted.

For estimating Roman starcounts, we use instead the much higher angular resolution 2D infrared extinction maps drawn from three sources—Zhang & Kainulainen (2023) [-65º<*l*<–10º], Surot et al (2020) [ –10º < *l* < 10º] , and Nidever et al (2012) red clump maps [*l* > 10º]—a small section of which is shown in Figure C.3. As documented in Fig 16 of Zhang & Kainulainen (2023), the Surot et al (2020) extinction values are systematically lower than the other two sources; we multiply the Surot map extinction values by 1.25 for consistency.

The benefit of these 2D maps is that they capture the highly structured extinction of the sky at higher angular resolution; a principal uncertainty is the distance probed. (See Figure 12 of Nidever et al (2012) for an illustration of how the distance range depends on stellar populations.) Typically the three maps we have used measure the extinction out to ~8 kpc, so that all of the extinction foreground to the bulge/bar should be accounted for.

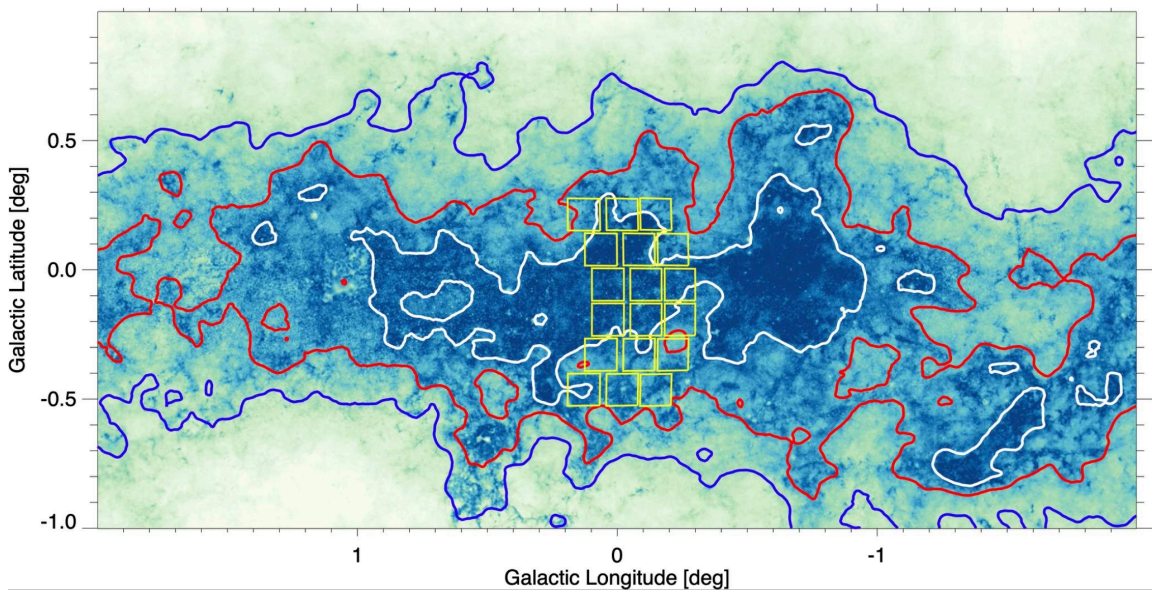

**Figure C.3** 2D extinction map with 7.2 arcsec pixel size towards inner Galaxy from Surot et al (2020), multiplied by 1.25 to make this map consistent with Nidever et al (2012) and Zhang & Kainulainen (2023). Contours show $A_K$=1.2 (blue), 1.75 (red), and 2.2 (white). The Galactic Center Roman pointing of the GBTDS is shown in yellow.

Although sources observed with the F213 filter will suffer less extinction, the Roman sensitivity is better in the shorter wavelength filters. The cross-over point for three filters is at $A_{F213}$=1.75, where the *combination* of sensitivity and extinction for F158, F184, and F213 is nearly identical. For a *b*=±1º latitude strip—the dustiest section of the RGPS— only 7% of the area and 18% of the sources exceed this threshold. Therefore, for much of the survey area, F158 and F184 will detect more sources. However, in the direction of the Central Molecular Zone (|*l*|<2º, |*b*|<0º.5), 59% of the area and 56% of sources (using the star count model in the next section) exceed $A_{F213}$=1.75. So in regions of high foreground extinction, as well as for embedded sources or those with significant circumstellar absorption, the F213 can be expected to uncover the largest number of sources.

## C.2 Star-count modeling and source densities

There are several models available for estimating Galactic starcounts as a function of photometric band, limiting magnitude, direction, and Galaxy model, including the Besançon model (Robin et al 2003), *TRILEGAL* (Girardi et al 2005), *Galaxia* (Sharma et al 2011), and *SynthPop* (Klüter et al 2025). The LSST survey has been modelled with a modified version of *TRILEGAL* (Dal Tio et al 2022), while the GBTDS used the "Huston2024" model within the Synthpop framework. This last model, unlike the others, includes the contribution of a nuclear stellar disk which is an important consideration in calculating microlensing event rates, but does not focus on matching the details of the disk of the Galaxy.

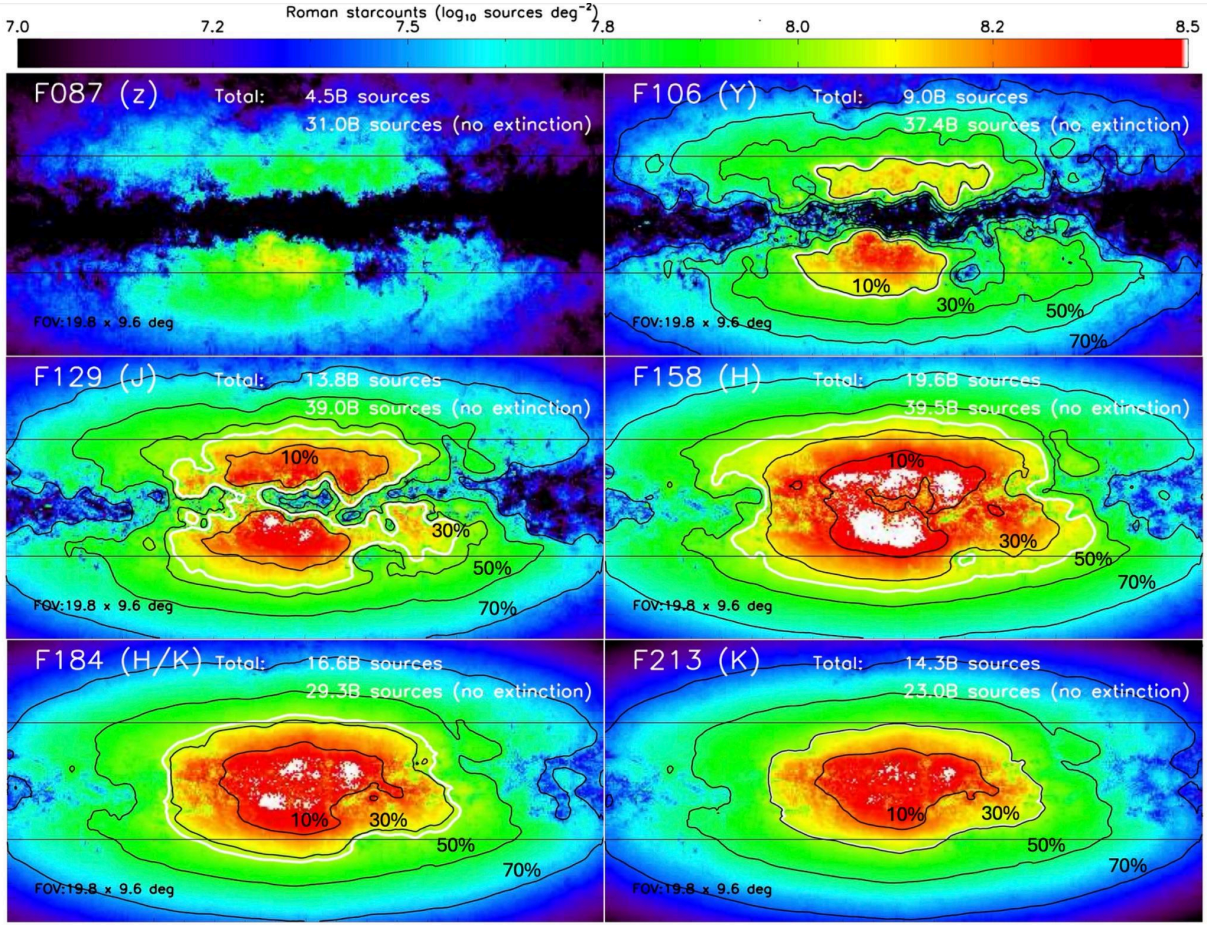

**Figure C.4** 2D starcount maps from TRILEGAL (Girardi et al 2005) simulations applying a foreground extinction described in §C.1. The total number of detectable sources in the full survey area (using sensitivities from Table 3.1) are shown. Black contours show the fraction of total sources within a given contour, with 10% of all sources within the innermost contour. The white contour shows a source density of one source per 10 pixel, which occurs at the 30% to 40% contours for the longest wavelength filters. The horizontal line shows $b=\pm 1º$. The total number of sources that would be detected in the full survey area (both with and without extinction) is report.

We used the web version of the *TRILEGAL* starcount model with the default options for the geometry, star-formation rate, and age-metallicity relationship for the thin disk, thick disk, bulge, and halo, and no extinction. We sampled 0.001 deg² areas spaced every degree in Galactic latitude, and every five degrees in Galactic longitude out l=±30º, and every 10º from |l|=30º to 90º. Source counts were linearly interpolated between these points and binned into seven cubes (one for each filter) with a resolution of 0.05 deg (longitude) x 0.05 deg (latitude) x 0.05 mag. For each spatial bin, the 2D extinction from the previous section was applied and the source counts were totalled down to the sensitivity limits given in Table 3.1. The results for the inner Galaxy— |l|<9º.8, |b|<4º.8 —is shown in Figure C.4. This figure shows the estimated total number of detectable sources in the *full* survey area: approximately 20 billion in F158, 17 billion F184, 14 billion in F213 and F129, and 9 billion sources in F106. Some of these numbers differ from the numbers in Table 3.1, since only part of the full survey region will be covered by F087, F106, and F184.

These values can be compared to the number of sources in the DECaPS2 (3.3 billion) or VVV (0.9 billion) programs, obtained with lower angular resolution and sensitivity than the RGPS. To check the calculations presented here, we used the survey footprint and filter/sensitivity of the DECaPS2 survey predicting a total of 2.3 billion sources as compared to the 3.3 billion sources observed. Approximately 70% of the sources are within ±10º of the Galactic center, and approximately 30% are within ±5º of the Galactic center. So approximately one third of the sources are predicted to lie in a region of the sky where the source densities will exceed one billion sources per square degree, equivalent to one source for every ten pixels for Roman. In these regions, source confusion becomes a serious consideration.

Most of the assumptions made to generate these estimates are on the conservative side. The extinction is assumed to be all foreground, even though much of the 2D extinction is expected to arise a distance of three to four kiloparsecs (in the "Molecular Ring") and we have scaled up the extinction maps to match the highest estimates. We also neglect the "Long Bar", an extension of the inner bulge/bar that is approximately 15% the mass of the inner bulge/bar (Sormani et al 2022). As part of the efforts to model the LSST survey, Dal Tio et al 2022 find that the star-counts in the Baade's window field are under-predicted by TRILEGAL (using a parameterized 3D extinction model) by a factor of two as compared to DECaPS data. It is not clear whether this difference is due to the predicted source density or the extinction law adopted. However, it is clear that Roman's high angular resolution and infrared sensitivity will make the RGPS the new gold standard survey for testing star-count models.

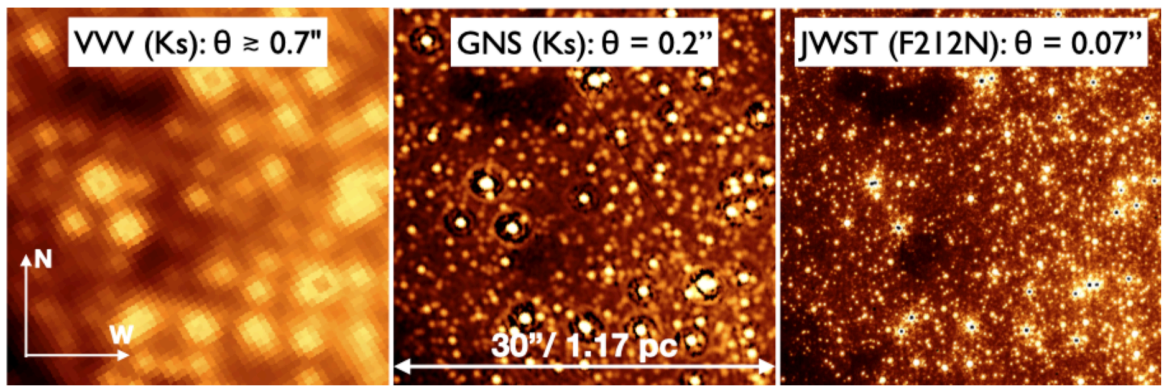

**Figure C.5** (From Schodel et al 2023) A field 25" northeast of Sgr A* with size 1.17 pc at the distance of Galactic Center as observed by VVV, GALACTIC- NUCLEUS and JWST (Proposal 1939, PI J. Lu). The PSF FWHM of *Roman* is 0.11 to 0.16".

However, some of our assumptions may be optimistic. Local extinction, especially in star formation regions, can be much higher than the average values adopted here and although dust clouds beyond the Galactic center will be distant and small, they may still block a significant fraction of the stars on the far side of the Galactic disk. In addition, the treatment of crowding is important for developing realistic estimations of star counts. To probe confusion closer to the Galactic center or in high density environments at higher angular resolution (Figure C.5), near-IR data for selected regions is available from the ESO-VLT GALACTICNUCLEUS Program (Nogueras-Lara et al 2018), HST (Clarkson et al 2008, Brown et al 2010, Calamida et al 2015, Hosek et al 2019, Terry et al 2020), JWST (Crowe et al 2025) and Euclid (Massari et al 2025). Estimates of source crowding were also made as part of the GBTDS CCS report; see their Figure 16 and associated text.

All of these programs demonstrate the scientific value of high angular resolution investigations in the inner Galaxy, and that the magnitude limit at which source crowding will limit photometric measurements is very sensitive to the 3D distribution of stars and dust, and the density of the dust clouds along the line of sight. Even with the minimum integration time, the RGPS will reach the confusion limit. Deeper coverage of the time-domain science and deep/spectroscopic fields, along with the ultra-deep GBTDS observations will provide data to further constrain this crowding limit.

## C.3 Measurements of stellar parameters and source extinction

Figure 3.9 shows an extinction-free color-absolute magnitude diagram from a TRILEGAL simulation (described above) of a 0.001 deg² field towards l=5°, b=0°, applying the sensitivity limits from Table 3.1. An example with a more complex star formation history in the nuclear stellar disk (NSD) (Nogueras-Lara et al 2020; Schödel et al 2023) is provided in Figure C.6, which shows a model with a bar foreground and four different populations in the NSD. A key step in the analysis is the ability to correct the measurements for extinction in order to recover the true CMD. In this figure, panel 3 shows what might be achieved using JWST mid-infrared observations in F480M, since the F213-F480M colors of all sources are similar and F480M suffers far less extinction. An open question is whether a combination of Roman-only filters could achieve a similar fidelity of extinction correction and the depth of the observations needed to achieve different scientific goals.

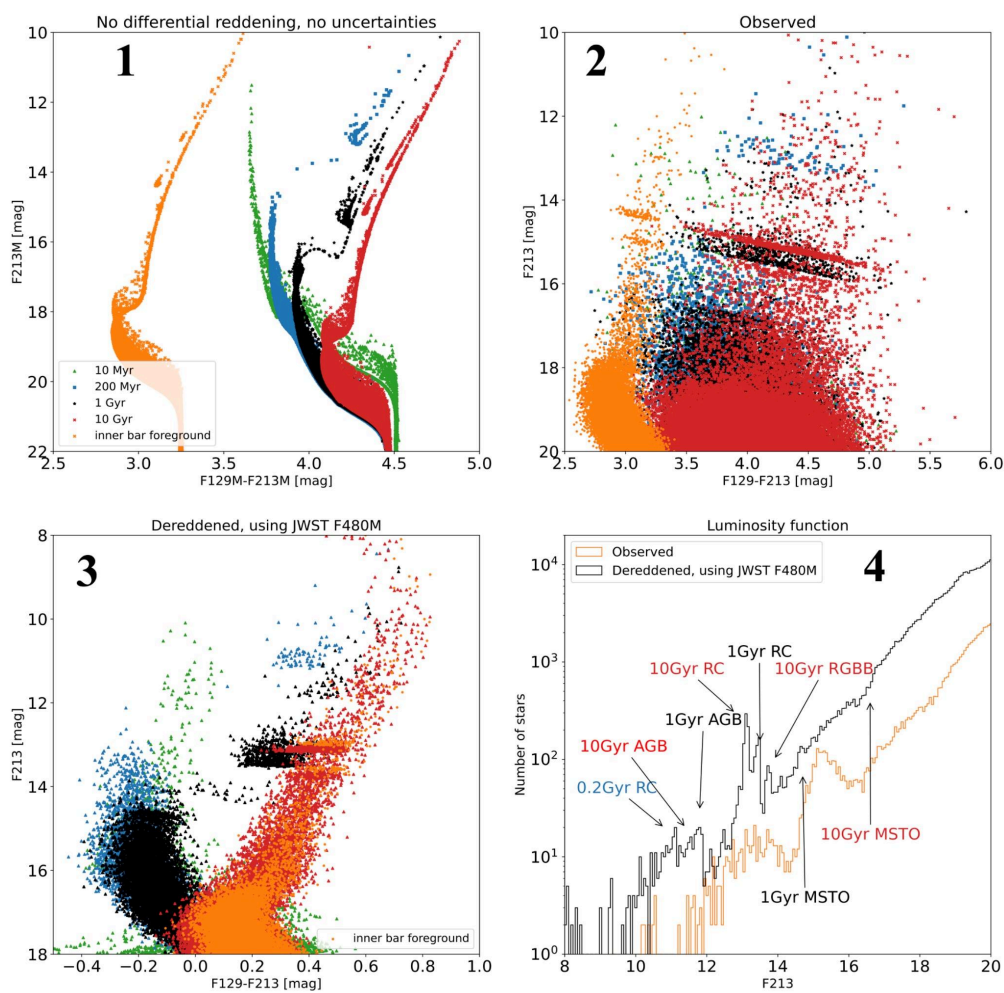

**Figure C.6** (from R. Schödel) **[Panel 1]** Line of sight model towards the CMZ/NSD with a bar foreground (at $A_K$=1.3) and four epochs of star formation in the NSD ($A_K$=2); **[Panel 2]** photometric uncertainties and differential reddening of $\sigma_{AK}$=0.05 (foreground) and 0.2 mag for observed CMD; **[Panel 3]** extinction-corrected CMD using supplemental JWST 480M observations; **[Panel 4]** detection of diagnostic features of red giant branch and main sequence turn-off (MSTO) red clump (RC), asymptotic giant branch (AGB), bump) and MSTO of different aged stellar population.

Other science cases, particularly mapping the spatial distribution of dust, require using the measured SED to estimate the extinction on a star-by-star basis. In principle, one might expect that the filter set with the largest wavelength range would yield the best SED reconstruction, but both because many sources become undetectable or have lower precision measurements in the F106 and bluer filters and because the F184 filter contains some prominent spectral features (water absorption, Paschen alpha emission), the best filter combination for source identification is not clear.

Figure C.7 shows some preliminary work on how stellar parameter recovery (stellar type, distance, and extinction) may vary based on filter choice. By generating synthetic spectra as a function of stellar type and convolving these spectra with the Roman filter curves, it is possible to predict intrinsic absolute magnitudes of stars in the Roman photometric bands. These intrinsic magnitudes can then be reddened to simulate Roman photometry as a function of Galactic environment (distance and extinction). By fitting noisy versions of this simulated photometry using established Bayesian stellar modeling codes (Speagle et al. 2025; Zucker, Saydjari, & Speagle et al. 2025, Cargile et al. 2020), one can evaluate which filter combinations minimize uncertainties on stellar parameters of interest by generating full synthetic posteriors on distance, extinction, and stellar type.

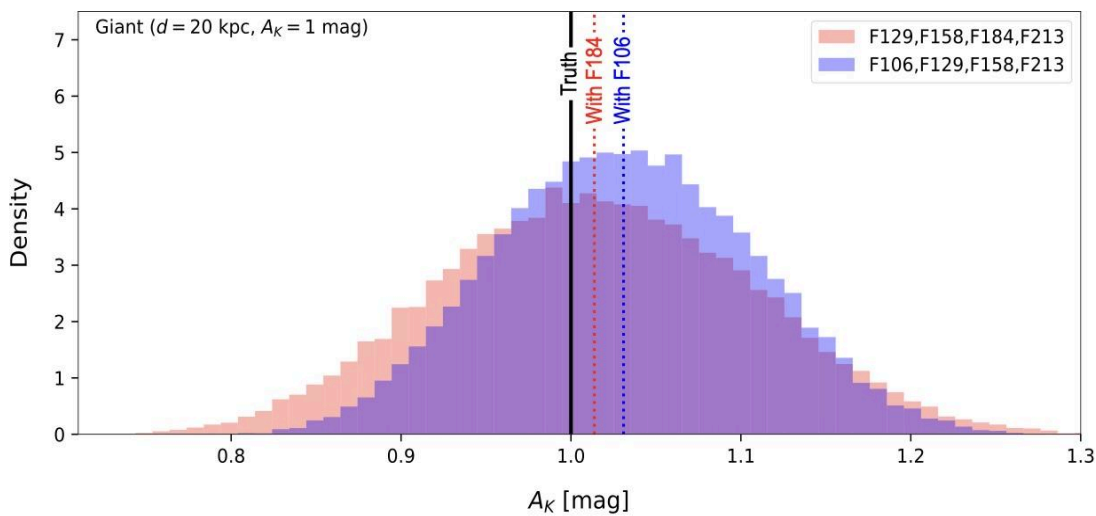

**Figure C.7** Preliminary results on how filter sets affect the recovery of extinction for a reddened ($A_K$=1), distant (20 kpc) giant (M=0.75 $M_o$) star. Black line shows the true extinction of the star; blue and red histograms and lines show the distribution and averages of inferred extinction values given different Roman filter sets. *Figure credit:* Catherine Zucker, Phillip Cargile, Andrew Saydjari, Adam Wheeler

Figure C.7 shows preliminary results on the recovered extinction posterior samples for a highly reddened ($A_K$=1), distant (d=20 kpc), giant star (M = 0.75 $M_\odot$) given two Roman filter sets. For this example, we also incorporate a parallax measurement (providing some constraint on the star's distance) even though most stars in the RGPS will not have a parallax. Both filter combinations provide reasonable recovery of the "true" extinction ($A_K$=1), suggesting that a longer wavelength baseline does not necessarily improve the stellar inference given the same number of bands. More study of filter choice is needed over a broader range of stellar types and environments to evaluate degeneracies inherent to photometric stellar modeling codes as well as maximize parameter recovery over all targets of interest in the RGPS. It will be fruitful to combine the RGPS filters with complementary filters (e.g. LSST), since the stacked depths of LSST in the Galactic plane will be similar to the single exposure RGPS depths.

## C.4 Proper motion measurements and modeling

A variety of science cases require proper motion accuracies better than a few tenths of a mas/yr, for example using Galactic rotation and bar streaming and dispersion to identify different dynamical populations. The RGPS survey will robustly measure proper motions better than 1 mas / yr to an F158 magnitude of roughly 21.5, which will allow for mapping the rotation of the Galaxy to the other side of the Galaxy's disk (Figure C.8).

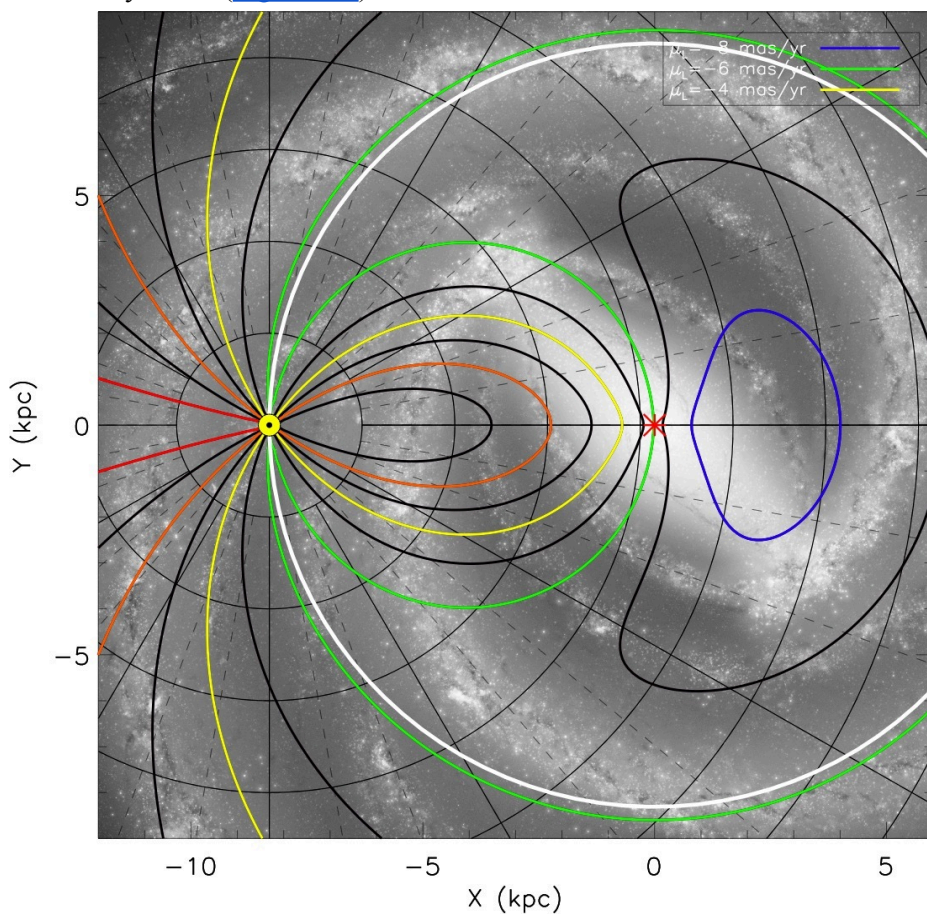

**Figure C.8** Map of proper motion as a function of Galactic position using the rotation curve of Reid et al (2019), with the solar motion with respect to Local Standard of rest removed; colored contours are used every 2 mas/yr, with red (0 mas/yr), orange (-2 mas/yr), and so on, with black contours marking the odd values.. The white line indicates the solar radius. . Note that the region of highest proper motion lies just beyond the Galactic center, traditionally one of the most difficult regions in the entire Galaxy to map.

The Galactic bulge has a velocity dispersion of about 2 mas / yr; RGPS proper motion estimates will be more accurate than this value to an F158 magnitude of roughly 22.5, providing detailed measurements of the motions of millions of individual bulge stars. The RGPS will also robustly measure the motions of high-velocity stars being ejected from the Milky Way, which can have proper motions in excess of 10 mas / yr. Finally, Roman will be able to make strides studying the motions of stars in nearby star forming regions, with proper motions > 0.4 mas/yr within 0.5 kpc (e.g., the distance of W40), especially for the most embedded, red stars where Gaia proper motions are not possible.

## C.5 Calculating yields for time-domain science observations

*Hourly cadence:* For the hourly cadences, we have adapted the strategy of the Spitzer YSOVAR program ([Rebull et al 2014](#)) and recommend reobserving the Serpens South region so the results of Roman observations can be calibrated against this previous effort. Although the number of Roman visits is much less than this program, sampling the Spitzer light curves of 1500 YSOVAR sources with the proposed Roman sampling yields 30 (2%) variable objects down to K=15 (Rebull 2025, priv comm). With the much greater depth of the Roman observations, we expect a large sample of variable sources will be obtained. The same sampling strategy will then be applied to the full Central Molecular Zone.

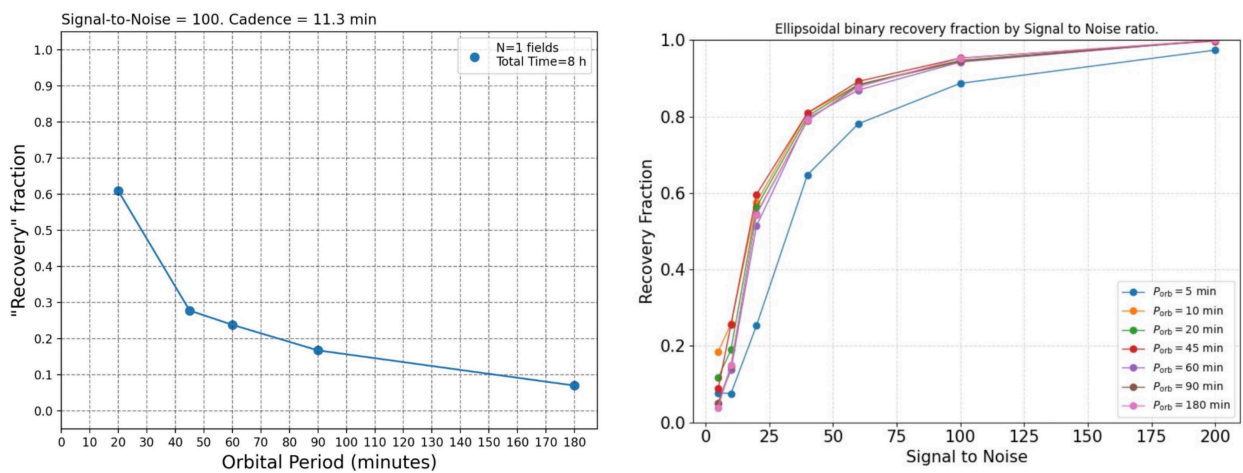

**Figure C.9** [left] Recovery fraction of eclipsing double white dwarf systems as a function of orbital period for the recommended high-cadence observations. [right] Recovery fraction of ellipsoidal variables of different periods as a function of signal-to-noise.

*High cadence (11.3 min):* The optimization of the Roman Space Telescope for high cadence observations for the GBTDS program opens up the opportunity to use a similar observing strategy in different environments. These observations may detect a large range of phenomena, including stellar flares, microlensing by free-floating planets, etc. Here we focus on the recovery of compact binaries. As discussed in §1.10, the detection rate of compact binaries will provide important constraints on the nature of gravitational wave sources. Figure C.9 (left) shows the recovery fraction of eclipsing double white dwarf systems as a function of orbital period for the recommended high cadence observations. For short orbital periods (< 20 mins) the recovery fraction is ~60%, dropping to ~10% for double white dwarfs with orbital periods of ~3 hours (due to the smaller number of eclipses). To recover a source, the light curve must have one of the following: (1) at least three 3-sigma photometric outliers due to eclipses, or (2) at least two 4-sigma photometric outliers due to eclipses, or (3) at least one 6-sigma photometric outliers due to eclipses. This is a very conservative assumption and so the given recovery fraction is likely a lower limit. Eclipsing double white dwarfs which show only eclipses with duty cycles less than 10% are most difficult to detect in photometric time domain data. Any other period source with additional variability including pulsators will have a significantly improved recovery fraction. As a second example, the right panel of Fig C.9 shows the recovery fraction of ellipsoidal variables with sinusoidal variability which shows a significantly higher recovery fraction across a large range of orbital periods. By observing in the Galactic center and bulge direction, this program should maximize the yields of both types of systems.